\documentclass{aa}  

\usepackage{graphicx}
\usepackage{txfonts}
\usepackage{multirow} 
\usepackage{diagbox} 
\usepackage{setspace}
\usepackage{float} 

\usepackage{blindtext}
\usepackage{gensymb}
\usepackage{multirow}
\usepackage[font=small,labelfont=bf]{caption}

\usepackage{enumitem}

\usepackage{caption}
\usepackage{subcaption}

 \usepackage{tabularx}             
    \usepackage{array}           
\usepackage{ amssymb }

\usepackage{natbib,twoopt}
\usepackage[colorlinks,allcolors=blue]{hyperref} 
\usepackage{verbatim}
\usepackage{soul}

\usepackage[normalem]{ulem}
\usepackage{color}

\begin{document} 

\newcommand{\doce}{$^{12}$CO }
\newcommand{\docep}{$^{12}$CO}
\newcommand{\trece}{$^{13}$CO }
\newcommand{\trecep}{$^{13}$CO}
\newcommand{\dosuno}{$J$\,=\,2\,$-$\,1 }
\newcommand{\dosunop}{$J$\,=\,2\,$-$\,1}
\newcommand{\unocero}{$J$\,=\,1\,$-$\,0 }
\newcommand{\unocerop}{$J$\,=\,1\,$-$\,0}
\newcommand{\kms}{\,km\,s$^{-1}$ }
\newcommand{\kmsp}{\,km\,s$^{-1}$}
\newcommand{\ms}{\,M$_{\odot}$ }
\newcommand{\msp}{\,M$_{\odot}$}
\newcommand{\pagb}{post-AGB }
\newcommand{\pagbp}{post-AGB}
\newcommand{\pagbs}{post-AGBs }
\newcommand{\pagbsp}{post-AGBs}

\newcommand{\alls}{AC\,Her, 89\,Her, IRAS\,19125+0343, and R\,Sct }
\newcommand{\allsp}{AC\,Her, 89\,Her, IRAS\,19125+0343, and R\,Sct}
\newcommand{\on}{89\,Her }
\newcommand{\onp}{89\,Her}
\newcommand{\iras}{IRAS\,19125+0343 }
\newcommand{\irasp}{IRAS\,19125+0343}
\newcommand{\ac}{AC\,Her }
\newcommand{\acp}{AC\,Her}
\newcommand{\rs}{R\,Sct }
\newcommand{\rsp}{R\,Sct}

\newcommand{\nir}{NIR-excess }
\newcommand{\nirp}{NIR excess}

\newcommand{\fig}{Fig.\,}
\newcommand{\figs}{Figs.\,}
\newcommand{\tab}{Table\,}
\newcommand{\eq}{Eq.\,}
\newcommand{\eqs}{Eqs.\,}
\newcommand{\sect}{Sect.\,}
\newcommand{\sects}{Sects.\,}
\newcommand{\app}{Appendix\,}
\newcommand{\secp}{\mbox{\rlap{.}$''$}}

\newcommand{\x}{\,$\times$\,}
\newcommand{\xd}[1]{10$^{#1}$}

\newcommand{\mm}{\,$\pm$\,}

\newcommand{\lsim}{\raisebox{-.4ex}{$\stackrel{\sf <}{\scriptstyle\sf \sim}$}}
\newcommand{\gsim}{\raisebox{-.4ex}{$\stackrel{\sf >}{\scriptstyle\sf \sim}$}}

\newcommand\sz{1.02}

   \title{Keplerian disks and outflows in post-AGB stars: \\ AC\,Herculis, 89\,Herculis, IRAS\,19125+0343, and R\,Scuti\,\thanks{Based on observations with IRAM\,NOrthern Extended Millimeter Array (NOEMA). IRAM   is supported by INSU/CNRS (France), MPG (Germany), and IGN (Spain).} \thanks{Final datacubes are available at the CDS via anonymous FTP to \url{cdsarc.u-strasbg.fr} (130.79.128.5) or via \url{http://cdsweb.u-strasbg.fr/cgi-bin/gcat?J/A+A/}}}
   \author{I. Gallardo Cava\,\inst{1,4}, M. Gómez-Garrido\,\inst{1,2}, V. Bujarrabal\,\inst{1}, A. Castro-Carrizo\,\inst{3}, J. Alcolea\,\inst{4}, and H. Van Winckel\,\inst{5}}

   \institute{Observatorio Astronómico Nacional (OAN-IGN), Apartado 112, 28803, Alcalá de Henares, Madrid, Spain\\ \email{i.gallardocava@oan.es}
         \and
         Centro de Desarrollos Tecnológicos, Observatorio de Yebes (IGN), 19141, Yebes, Guadalajara, Spain 
         \and
         Institut de Radioastronomie Millimétrique, 300 rue de la Piscine, 38406, Saint-Martin-d'Hères, France   
        \and
        Observatorio Astronómico Nacional (OAN-IGN), Alfonso XII 3, 28014, Madrid, Spain
        \and
        Instituut voor Sterrenkunde, KU Leuven, Celestijnenlaan 200B, 3001, Leuven, Belgium
 }

\titlerunning{Keplerian disks and outflows in post-AGB stars: AC\,Her, 89\,Her, IRAS\,19125+0343, and R\,Sct}
\authorrunning{Gallardo Cava, I. et al.}

   \date{}
   \date{Received 5 October 2020 / Accepted 4 March 2021}
 
  \abstract
   {There is a class of binary \pagb stars with a remarkable near-infrared excess that are surrounded by Keplerian or quasi-Keplerian disks and extended outflows composed of gas escaping from the disk. The Keplerian dynamics had been well identified in four cases, namely the Red\,Rectangle, AC\,Her, IW\,Car, and IRAS\,08544$-$4431. In these objects, the mass of the outflow represents $\sim$\,10\% of the nebular mass, the disk being the dominant component of the nebula.}
   {We aim to study the presence of rotating disks in sources of the same class in which the outflow seems to be the dominant component.}
   {We present interferometric NOEMA maps of \doce and \trece \dosuno in 89\,Her and \doce \dosuno in \acp, \irasp, and \rsp. Several properties of the nebula are obtained from the data and model fitting, including the structure, density, and temperature distributions, as well as the dynamics. We also discuss the uncertainties on the derived values.}
   {The presence of an expanding component in \ac is doubtful, but thanks to new maps and models, we estimate an upper limit to the mass of this outflow of $\lsim$\,3\x\xd{-5}\msp, that is, the mass of the outflow is $\lsim$\,5\% of the total nebular mass. For \onp, we find a total nebular mass of 1.4\x\xd{-2}\msp, of which $\sim$\,50\% comes from an hourglass-shaped extended outflow. In the case of \irasp, the nebular mass is 1.1\x\xd{-2}\msp, where the outflow contributes $\sim$\,70\% of the total mass. The nebular mass of \rs is 3.2\x\xd{-2}\msp, of which $\sim$\,75\% corresponds to a very extended outflow that surrounds the disk.}
   {Our results for \iras and \rs lead us to introduce a new subclass of binary \pagb stars, for which the outflow is the dominant component of the nebula. Moreover, the outflow mass fraction found in \ac is smaller than those found in other disk-dominated binary \pagb stars. \on would represent an intermediate case between both subclasses.}

   

   \maketitle
%

\begin{table*}[h]
\caption{Binary \pagb stars observed in this paper.}
\small
\vspace{-5mm}
\begin{center}

\begin{tabular*}{\textwidth}{@{\extracolsep{\fill\quad}}llcccccll}
\hline \hline
\noalign{\smallskip}

\multicolumn{2}{c}{Source} & \multicolumn{2}{c}{Observed coordinates}   & $V_{LSR}$ & $d$ &  $P_{orb}$ & \multirow{2}{*}{Sp.\,Type} & \multirow{2}{*}{Comments}\\
GCVS name & IRAS name & \multicolumn{2}{c}{J2000}   & [km\,s$^{-1}$] & [pc]  &  [d] &  & \\
\hline
\\[-2ex]
\vspace{1mm} 
AC\,Herculis  & 18281+2149 &  18:30:16.24  &  +21:52:00.6    & $-$9.7 & 1100 &  1188.9 & F2\,$-$\,K4\,I & RV\,Tauri variable \\

\vspace{1mm} 
89\,Herculis  & 17534+2603 & 17:55:25.19  &  +26:03:00.0 & $-$8.0 & 1000 & 289.1 & F2\,I &   Semiregular variable \\

\vspace{1mm}
$-$ & 19125+0343 & 19:15:01.18  &  +03:48:42.7  & 82.0 & 1500  & 519.7 & F2 &  RV\,Tauri variable \\ 

R\,Scuti & 18448$-$0545 & 18:47:28.95  & $-$05:42:18.5  & 56.1  & 1000 & $-$ & G0\,$-$\,K0\,I &  Peculiar RV\,Tauri variable \\

\hline
\end{tabular*}

\end{center}
\small
\vspace{-1mm}
\textbf{Notes.} Distances and spectral types are adopted from \citet{bujarrabal2013a}. Velocities are derived from our observations. Orbital periods of the binary systems are taken from \citet{oomen2018}.

\label{prop}
\end{table*}

\section{Introduction}
\label{introduccion}

Most of the protoplanetary (or pre-planetary) and planetary nebulae (pPNe and PNe) show fast bipolar outflows (30\,$-$\,100\kmsp) with clear axial symmetry. These outflows are responsible for a good fraction of the total mass ($\sim$\,0.1\msp) and carry very large amounts of linear momentum \citep{bujarrabal2001}. The immediate precursor of the \pagb stars, the asymptotic giant branch (AGB) stars, present spherical circumstellar envelopes, which are in isotropic expansion at moderate velocities, around 10\,$-$\,20\kms \citep{castrocarrizo2010}.
The spectacular evolution from AGB circumstellar envelopes to \pagb nebulae takes place in a very short time ($\sim$\,1000\,yr). The accepted scenario to explain this evolution implies that material is accreted by a companion from a rotating disk, followed by the launching of very fast jets, in a process similar to that in protostars \citep{soker2002, frankblackman2004, blackmanlucchini2014}.
However, the effect of binarity in \pagb stars is still poorly understood \citep{demarcoizzard2017}.

There is a class of binary \pagb stars that systematically show evidence of the presence of disks \citep{vanwinckel2003,ruyter2006,bujarrabal2013a,hillen2017} and low initial mass \citep{alcoleabujarrabal1991}. The observational properties of these objects were recently reviewed by \citet{vanwinckel2018}. All of them present a remarkable near-infrared (NIR) excess and the narrow CO line profiles characteristic of rotating disks. Their spectral energy distributions (SEDs) reveal the presence of hot dust close to the stellar system, and its disk-like shape has been confirmed by interferometric IR data \citep{hillen2017, kluska2019}.
These disks must be stable structures, because their IR spectra reveal the presence of highly processed grains \citep{gielen2011, jura2003, sahai2011}.
Observations of \doce and \trece in the \dosuno and \unocero lines have been well analyzed in sources with such a NIR excess \citep{bujarrabal2013a} and they show line profiles formed by a narrow single peak and relatively wide wings.  
These line profiles are similar to those in young stars surrounded by a rotating disk made of remnants of interstellar medium and those expected from disk-emission modelling \citep{bujarrabal2005,guilloteau2013}. These results indicate that the CO emission lines of our sources come from Keplerian or quasi-Keplerian disks. The systematic detection of binary systems in these objects \citep{oomen2018} strongly suggests that the angular momentum of the disks comes from  the stellar system.

The study of Keplerian disks around \pagb stars requires high angular- and spectral-resolution observations because of the relative small size of the rotating disks. To date, there are only four resolved cases of Keplerian rotating disks: the Red\,Rectangle \citep{bujarrabal2013b, bujarrabal2016}, AC\,Her \citep{bujarrabal2015}, IW\,Car \citep{bujarrabal2017}, and IRAS\,08544$-$4431 \citep{bujarrabal2018}. All four have been very well studied through single-dish and interferometric millimeter(mm)-wave maps of CO lines.
These four studied sources show CO spectra with narrow line profiles characteristic of rotating disks and weak wings. This implies that most of the material of the nebula is contained in the rotating disk. However, according to the mm-wave interferometric
maps there is a second structure  surrounding the disk that is less massive, contains $\sim$\,10\% of the total mass, and is in expansion. This outflow is probably a disk wind consisting in material escaping from the rotating disk \citep[see extensive discussion of the best studied source by][]{bujarrabal2016}.

In the present work, we present NOEMA maps of \doce and \trece \dosuno emission lines in \acp, \onp, and \iras ---which are confirmed binaries---, and \rsp. \rs is a different object, it is a very bright star in the visible and is a RV\,Tau variable, but its SED is different from those of the other sources in our sample, with a less prominent \nirp. Also, its binary nature is not yet confirmed either. We discuss this source in \sect\ref{secrsctprev}. On the other hand, \acp, \onp, and \iras are confirmed binaries that also belong to the same class of binary \pagb stars with remarkable \nirp.
Our results of \onp, \irasp, and \rs show that the extended outflowing component is even more massive than the rotating disk. This suggests that they are part of a new subclass: the outflow-dominated nebulae around \pagb stars.

The paper is organized as follows. We present our \pagb star sample in \sect\ref{fuentes}. Technical information of our observations is given in \sect\ref{reduccion}. We discuss the mm-wave interferometric maps in \sect\ref{observaciones}. Results from our best-fit models are shown in \sect\ref{modelos}. Finally, we present our main conclusions in \sect\ref{conclusiones}.


\begin{table*}[h]
\caption{Observational parameters.}
\small
\tiny

\vspace{-5mm}
\begin{center}

\begin{tabular*}{\textwidth}{@{\extracolsep{\fill\quad}}lllccccccc}
\hline \hline
\noalign{\smallskip}

\multirow{2}{*}{Source}   & \multirow{2}{*}{Project} & \multirow{2}{*}{Observation dates} &  Baselines & Obs.\,time  & \multirow{2}{*}{Beam size}  & Sp. Resol.  & Noise \\
  &   & & [m]   &  [h]  &  & [km\,s$^{-1}$] & [mJy\,beam$^{-1}$]\\
\hline
\\[-2ex]
\vspace{1mm} 
AC\,Herculis      & W14BU & Dec.\,14, Apr.\,15, and Mar.\,16 &  17\,$-$\,760 & 21 & 0\secp35\x0\secp35  & 0.2 & 4.8\x\xd{-3}\\

\multirow{2}{*}{89\,Herculis}    & X073  & Jan.\,14 to Mar.\,14 & \multirow{2}{*}{15\,$-$\,760}  & \multirow{2}{*}{22}  & \multirow{2}{*}{0\secp74\x0\secp56}  & \multirow{2}{*}{0.2} & \multirow{2}{*}{4.9\x\xd{-3}}\\
\vspace{1mm}
 &  W14BT & Dec.\,14 and Feb.\,15 to Apr.\,15  &  &  &    & \\

\vspace{1mm}
\irasp    & WA7D &  Feb.\,13 to Mar.\,13 &  55\,$-$\,760 & 12 &  0\secp70\x0\secp70 &  1.0 & 5.4\x\xd{-3} \\

R\,Scuti & S15BA & Summer\,15 and Autumn\,15 & 18\,$-$\,240 & 12 &  3\secp12\x2\secp19  & 0.5 & 2.2\x\xd{-2} \\

\hline
\end{tabular*}

\end{center}
\small
\vspace{-1mm}
\textbf{Notes.} This table collects the main parameters of NOEMA maps of \doce\dosuno of \acp, \irasp, and \rsp. It also includes the main details of NOEMA maps of \trece\dosuno of \onp.

\label{obs}
\end{table*}


\section{Source descriptions and previous results}
\label{fuentes}

We produced interferometric maps of four sources (\tab\ref{prop}) using the IRAM NOrther Extended Millimeter Array (NOEMA). All of them are identified as binary \pagb stars (binary system including a \pagb star) with low gravity, high luminosity, FIR-excess indicative of material ejected by the star, and the mentioned remarkable NIR excess, and have been previously studied in CO by means of single-dish observations.

We adopted the same distances used in \citet{bujarrabal2013a}. We chose these distances to facilitate the comparison with the results derived in that work and because, in the case of binary stars, the estimation of distances via parallax measurements is very delicate \citep{dominik2003}.
We discarded other options because they present large uncertainties.
The best values are likely those derived from the not-yet-available \textit{Gaia} Data Release 3 (\textit{Gaia}\,DR3), which considers astrometry for binary systems.
As we discuss in detail in \sect\ref{incertidumbres}, we will easily be able to scale the derived values of our best-fit models for some parameters, including size and mass, if and when a new and better value for the distance to the sources is determined.

\subsection{\acp}
\label{secacherprev}

\ac is a binary \pagb star \citep{oomen2019}; although some authors have suggested that it could rather be a post-RGB star \citep{hillen2015}.
The CO line profiles and interferometric maps of \ac are very similar to the ones found in the Red\,Rectangle. According to this, \ac must be a \pagb star whose nebula is dominated by a Keplerian disk.

Previous mm-wave interferometric observations confirm that \ac presents a clear disk with Keplerian dynamics, which dominates the nebula  \citep{bujarrabal2015,hillen2015}. It presents a mass $\sim$\,1.5\x10$^{-3}$\msp, densities of 10$^{6}$\,$-$\,10$^{4}$\,cm$^{-3}$, and temperatures of 80\,$-$\,20\,K. These results were obtained adopting a distance of 1600\,pc.
Approximately \,40\% of the total flux was filtered out by interferometric observations. However, no outflowing material was detected in these papers.
Due to the large inner radius (30\,$-$\,35\,AU) and the low gas-to-dust ratio, the Keplerian disk is very likely to be in an evolved state and its mass was probably larger before \citep{hillen2015}.

\subsection{\onp}

\on is a binary \pagb star with a remarkable \nir that implies the presence of hot dust \citep{ruyter2006}. It also shows significant far-infrared (FIR) emission. The presence of large grains was proposed by \citet{shenton1995}.

Single-dish observations show narrow CO lines that are very similar to the ones detected in the Red\,Rectangle, but with more prominent wings, which suggests a significant contribution of the outflow \citep{bujarrabal2013a}.
The source was studied in detail by \citet{bujarrabal2007}; see also \citet{alcoleabujarrabal1995} and \citet{fong2006}. \citet{bujarrabal2007} discovered two features from NOEMA maps in \doce\dosuno in the nebula around \onp: an extended hourglass-shaped structure and a central clump, which probably  corresponds to an unresolved disk. 
Near-infrared observations strongly suggest that in \on the binary system is surrounded by a compact and stable circumbinary disk in Keplerian rotation, where large dust grains form and settle to the midplane \citep{hillen2013, hillen2014}.

\subsection{\irasp}
\label{secirasprev}

\iras is a binary \pagb star \citep{gielen2008}. It also belongs to this class of binary \pagb stars with remarkable \nir \citep{oomen2018}, implying the presence of hot dust.
For this source, the assumed value for the distance (\tab\ref{prop}) is highly questionable because some authors adopt very different values, such as 1800\mm 400\,pc in \citet{gielen2008} or 4131$^{+905}_{-645}$\,pc in \citet{bailerjones2018}.

Its CO lines are narrow \citep{bujarrabal2013a}, and so the hot dust must be located in a compact rotating disk, but they also have prominent wings. The circumbinary disk could be surrounded by an outflow containing most of the mass. The single-dish analysis yielded a nebular mass of 1.3\x\xd{-2}\msp.

\subsection{\rsp}
\label{secrsctprev}

While the first three sources mentioned above are spectroscopically confirmed binaries, this is not the case for \rsp. This bright RV\,Tauri star shows very irregular pulsations with variable amplitude \citep[see][]{kalaeehasazadeh2019}, and only a small IR excess \citep{kluska2019}, meaning that the SED is not clearly linked to the presence of a circumbinary disk. Hence, the star was labeled ``uncertain'' in the classification of \pagb stars surrounded by disks in \citet{gezer2015}.
The high amplitude of the pulsations is also reflected in the radial-velocity amplitude \citep{pollard1997}. It is worth mentioning that  a magnetic field has been detected; see e.g. \citet{tessore2015}.
Moreover, \citet{matsuura2002} suggest that it could be an AGB star in the helium-burning phase of the thermal pulse cycle. 

However, it is known that \rs is an RV\,Tauri type variable source and hence probably a \pagb star. Its CO line profiles are different from those of other binary \pagb stars studied by \cite{bujarrabal2013a}. Following this, it might be compatible with the presence of a Keplerian disk, and as Keplerian disks are only detected around binaries, it could well be binary as well, but this needs to be confirmed. 
This hypothesis is reinforced by interferometric data in the \textit{H}-band showing a very compact ring \citep{kluska2019}. Therefore, we tentatively consider \rs as a binary \pagb star, like the other three sources. The nebular mass derived from the single-dish studies is 5\x\xd{-2}\msp, where $\sim$\,14\% would correspond to the disk.


\begin{figure*}[h]
\centering
\includegraphics[width=\textwidth]{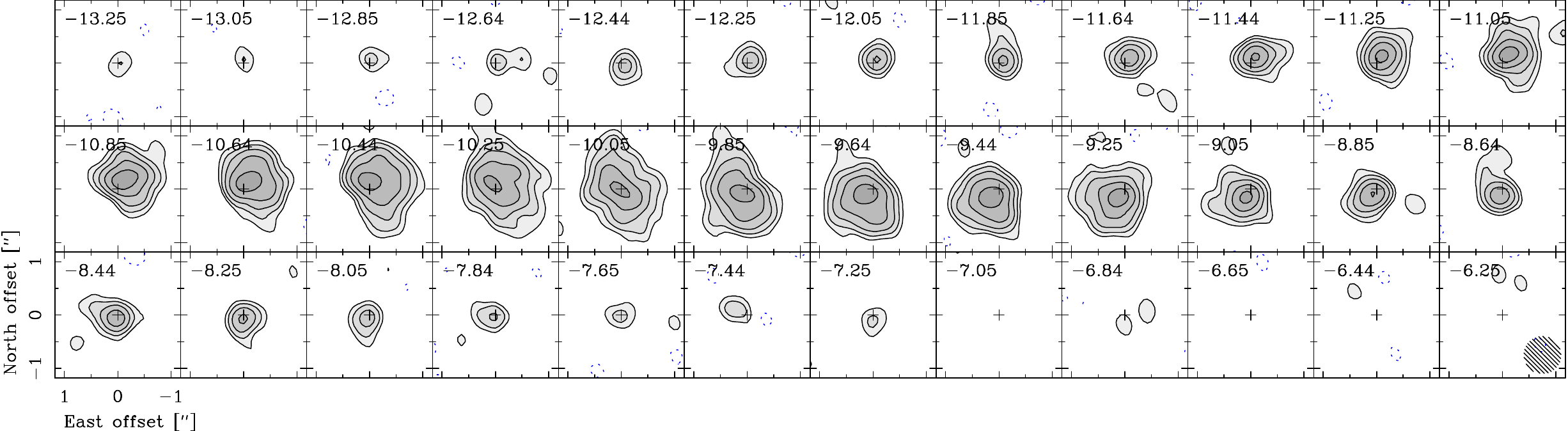}
 
\caption{\small NOEMA maps per velocity channel of \doce \dosuno emission from \acp. The beam size (HPBW) is 0\secp35\x0\secp35. The contour spacing is logarithmic: \mm9, 18, 36, 76, and 144\,mJy\,beam$^{-1}$ with a maximum emission peak of 230\,mJy\,beam$^{-1}$.
   The LSR velocity is indicated in the upper left corner of each velocity-channel panel and the beam size is shown in the last panel.}
    \label{fig:ac12mapas}  
\end{figure*}

\begin{figure*}[h]
        \centering
        \begin{minipage}[b]{0.48\linewidth}
                \includegraphics[width=\sz\linewidth]{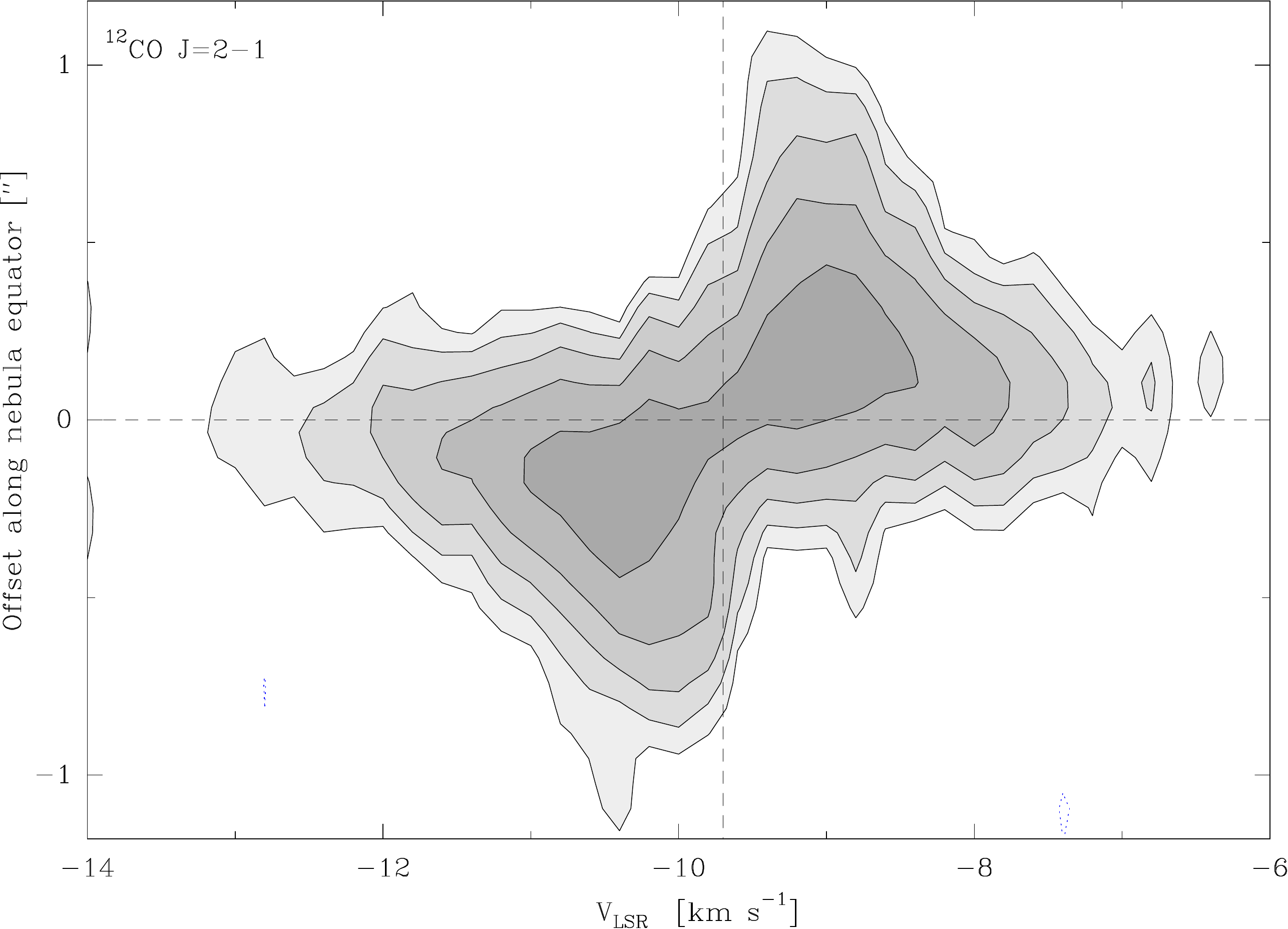}
        \end{minipage}
        \quad
        \begin{minipage}[b]{0.48\linewidth}
                \includegraphics[width=\sz\linewidth]{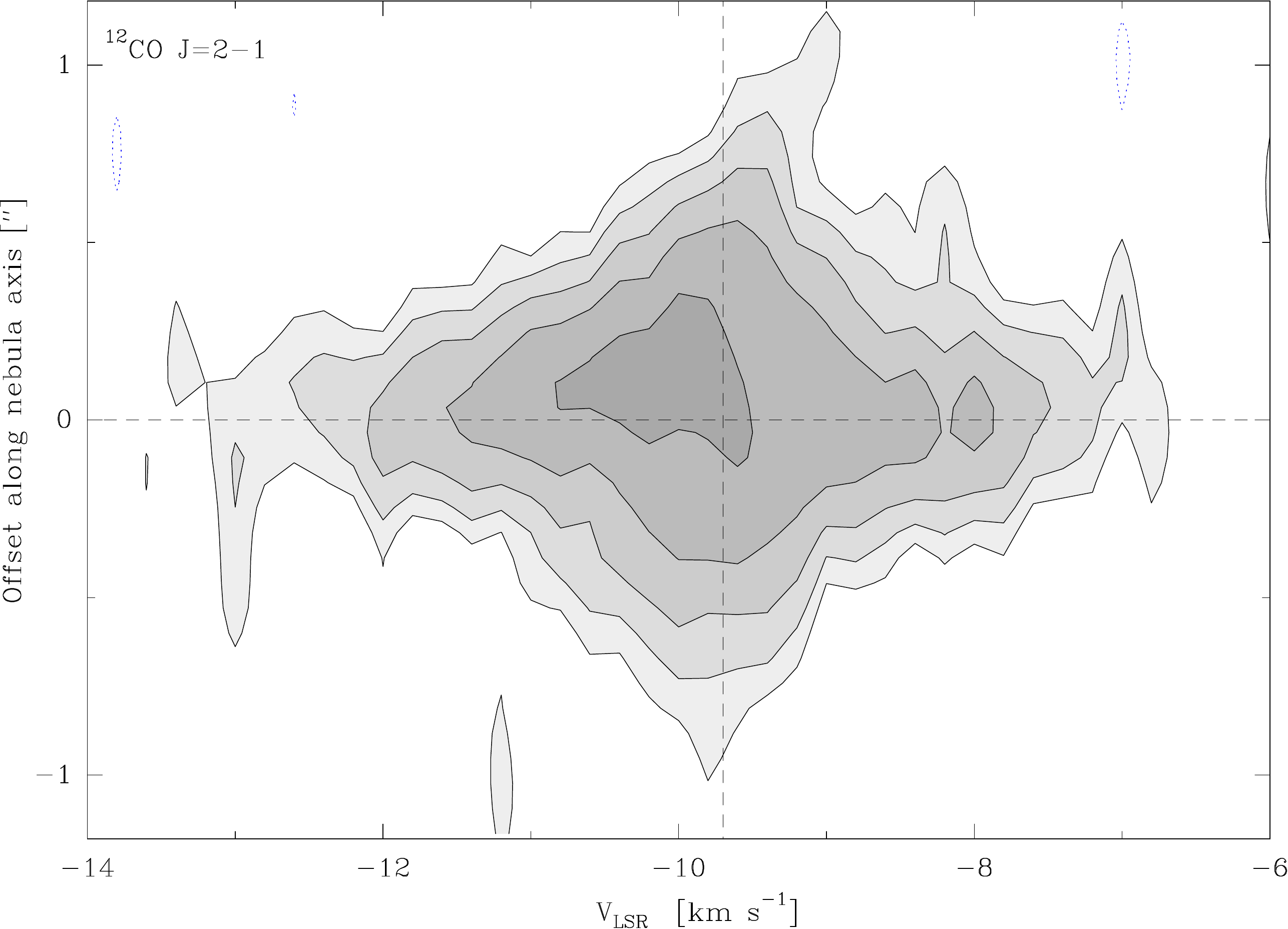}
        \end{minipage}
        \caption{\small \textit{Left:} Position--velocity diagram from our NOEMA maps of \doce \dosuno in \ac along the direction $PA=136.1\degree$, corresponding to the nebula equator. The contour spacing is logarithmic: \mm9, 18, 36, 76, and 144\,mJy\,beam$^{-1}$ with a maximum emission peak of 230\,mJy\,beam$^{-1}$. The dashed lines show the approximate central position and systemic velocity. \textit{Right}: Same as in \textit{Left} but along the perpendicular direction $PA=46.1\degree$.}
        \label{fig:acher12pv}
\end{figure*}

\begin{figure*}[h]
\centering
        \includegraphics[width=\textwidth]{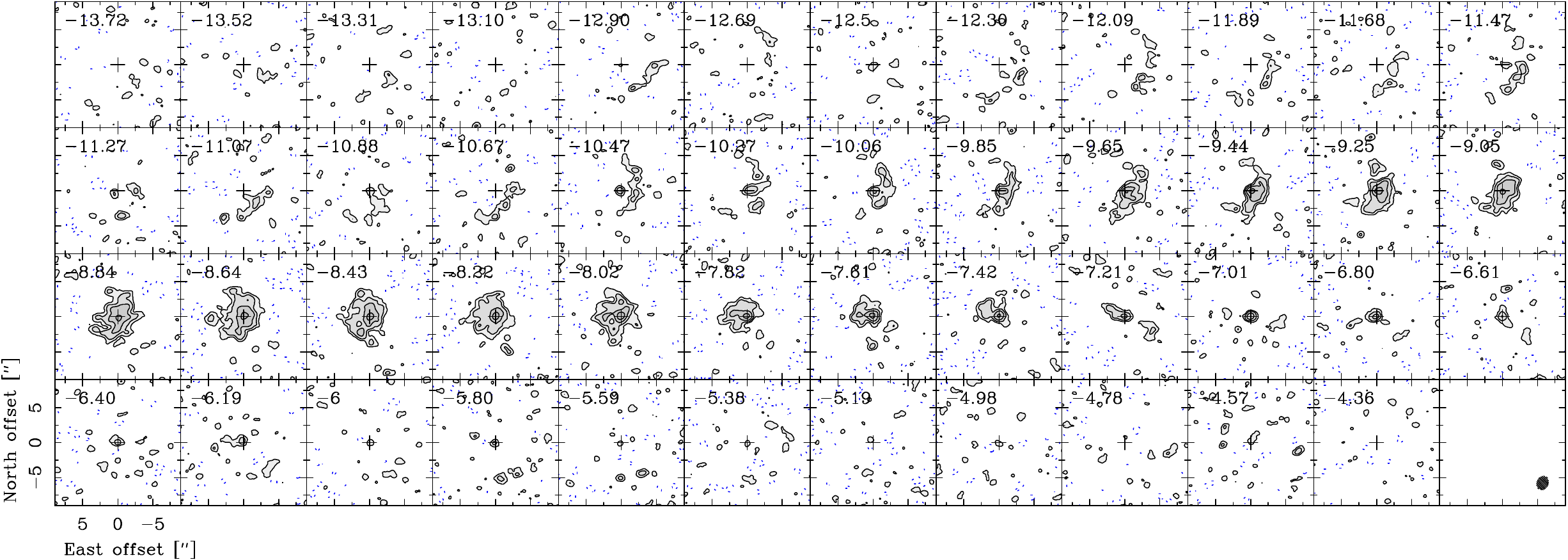}
        \includegraphics[width=\textwidth]{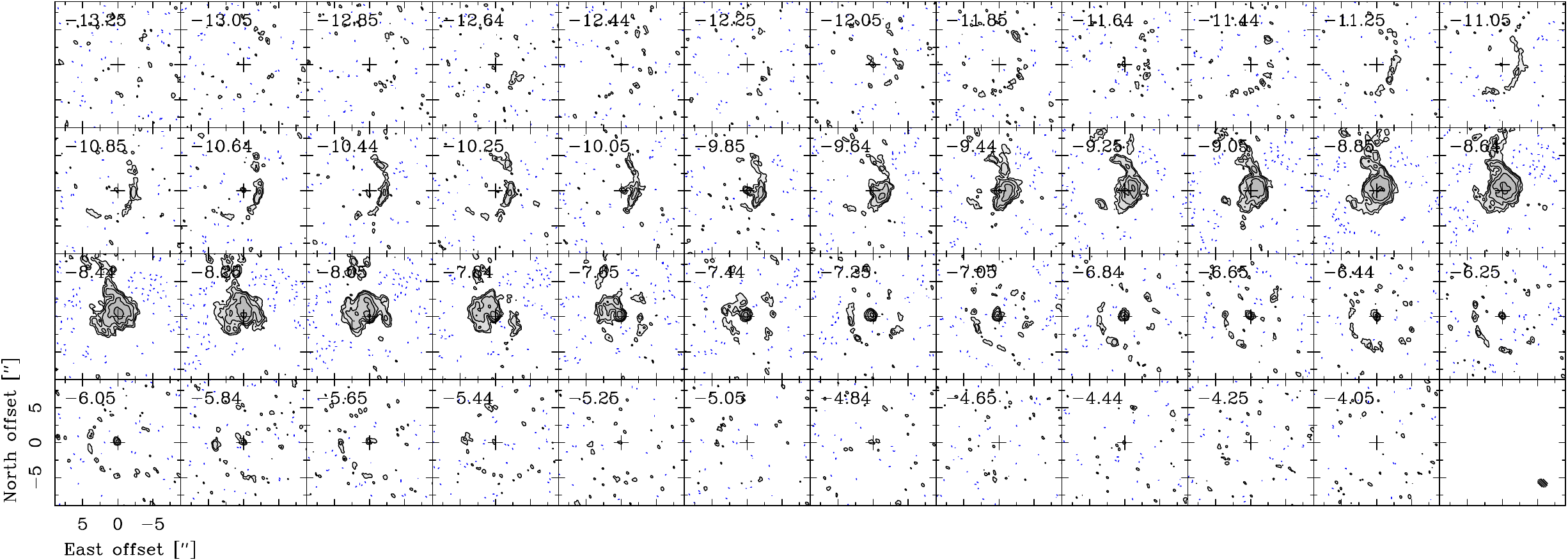}
   \caption{\small \textit{Top}: NOEMA maps per velocity channel maps of \doce\dosuno from 89\,Her. The beam size (HPBW) is 1\secp02\x0\secp83, the major axis being oriented at $PA=115\degree$. The contour spacing is logarithmic: $\pm$\,70, 140, 280, and 560\,mJy\,beam$^{-1}$ with a maximum emission peak of 870\,mJy\,beam$^{-1}$.
   \textit{Bottom}: Same as in \textit{Top} but for \trece\dosunop. The beam size (HPBW) is 0\secp74\x0\secp56, the major axis being oriented at $PA=28\degree$. The contour spacing is logarithmic: $\pm$\,11, 22, 44, 88, and 144\,mJy\,beam$^{-1}$ with a maximum emission peak of 225\,mJy\,beam$^{-1}$.
    The LSR velocity is indicated in the upper left corner of each velocity-channel panel and the beam size is shown in the last panel.}
    \label{fig:89hermapas}  
\end{figure*}

\begin{figure*}[h]
        \centering
        \begin{minipage}[b]{0.48\linewidth}
                \includegraphics[width=\sz\linewidth]{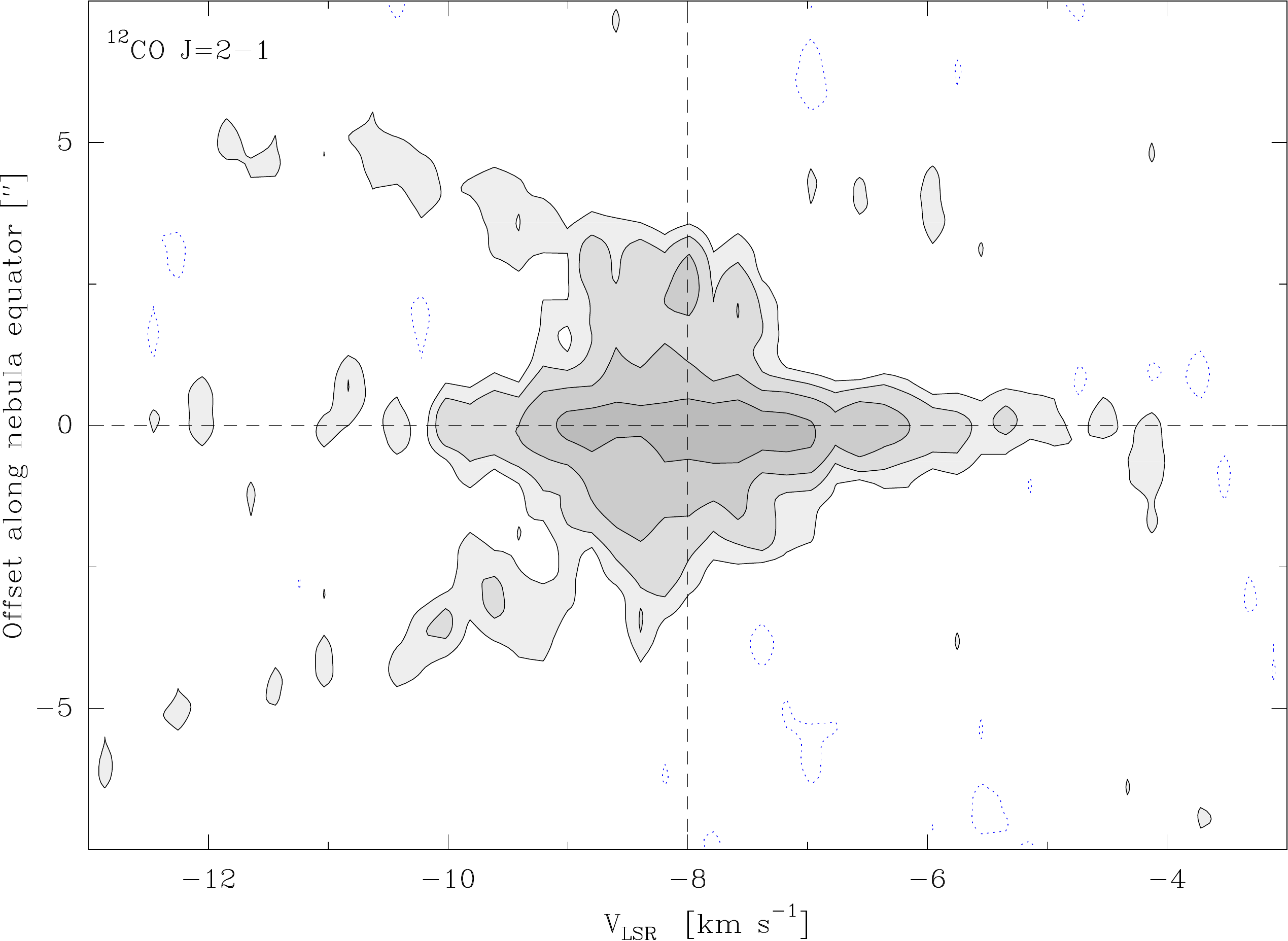}
        \end{minipage}
        \quad
        \begin{minipage}[b]{0.48\linewidth}
                \includegraphics[width=\sz\linewidth]{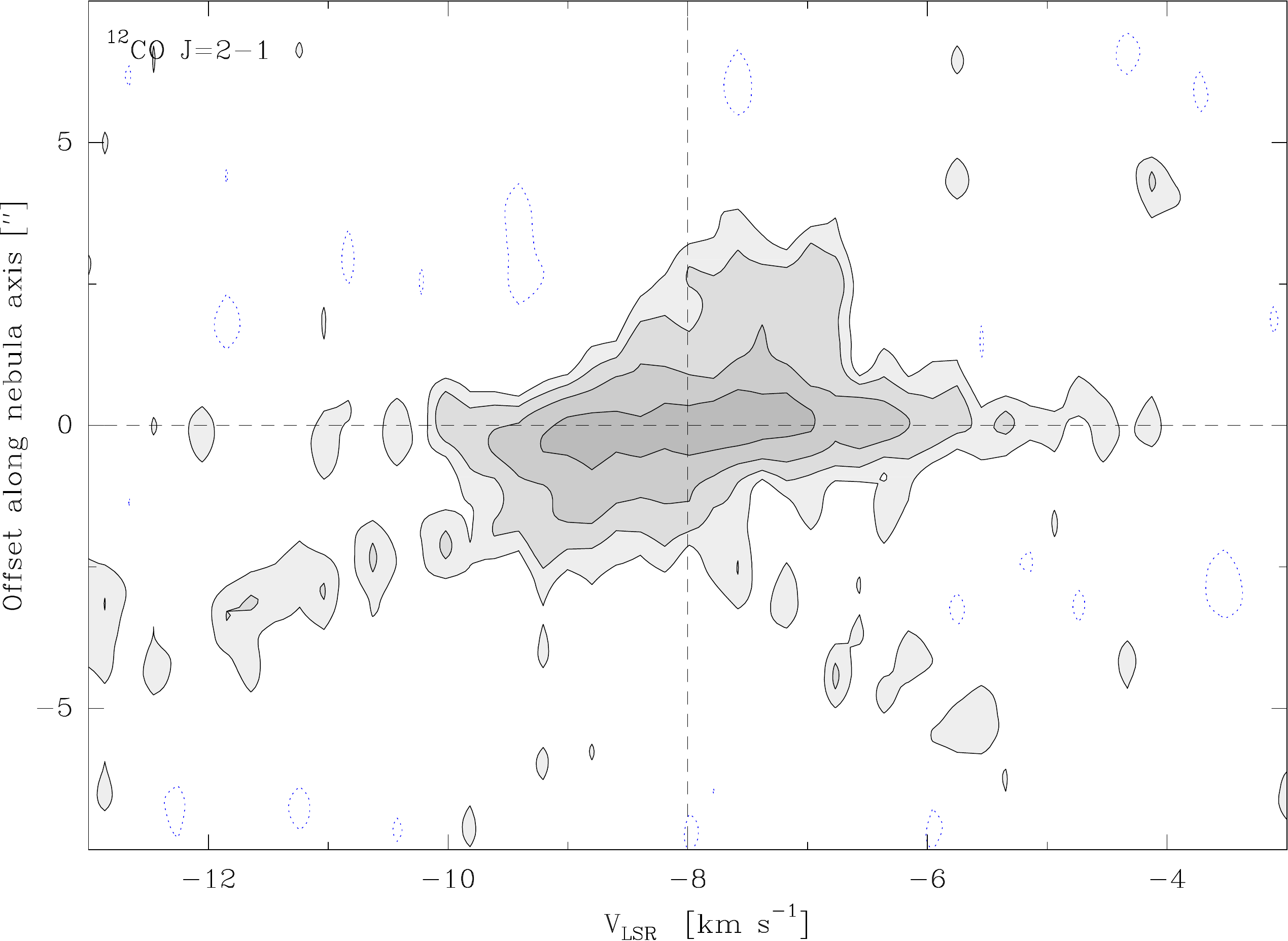}
        \end{minipage}
        \caption{\small \textit{Left:} Position--velocity diagram from our NOEMA maps of \doce \dosuno in \on along the direction $PA=150\degree$, corresponding to the nebula equator. The contour spacing is logarithmic: $\pm$\,70, 140, 280, and 560\,mJy\,beam$^{-1}$  with a maximum emission peak of 870\,mJy\,beam$^{-1}$. The dashed lines show the approximate central position and systemic velocity. The beam is represented in the last panel. \textit{Right}: Same as in \textit{Left} but along $PA=60\degree$.}
        \label{fig:89her12pv}
\end{figure*}

\begin{figure*}[h]
        \centering
        \begin{minipage}[b]{0.48\linewidth}
                \includegraphics[width=\sz\linewidth]{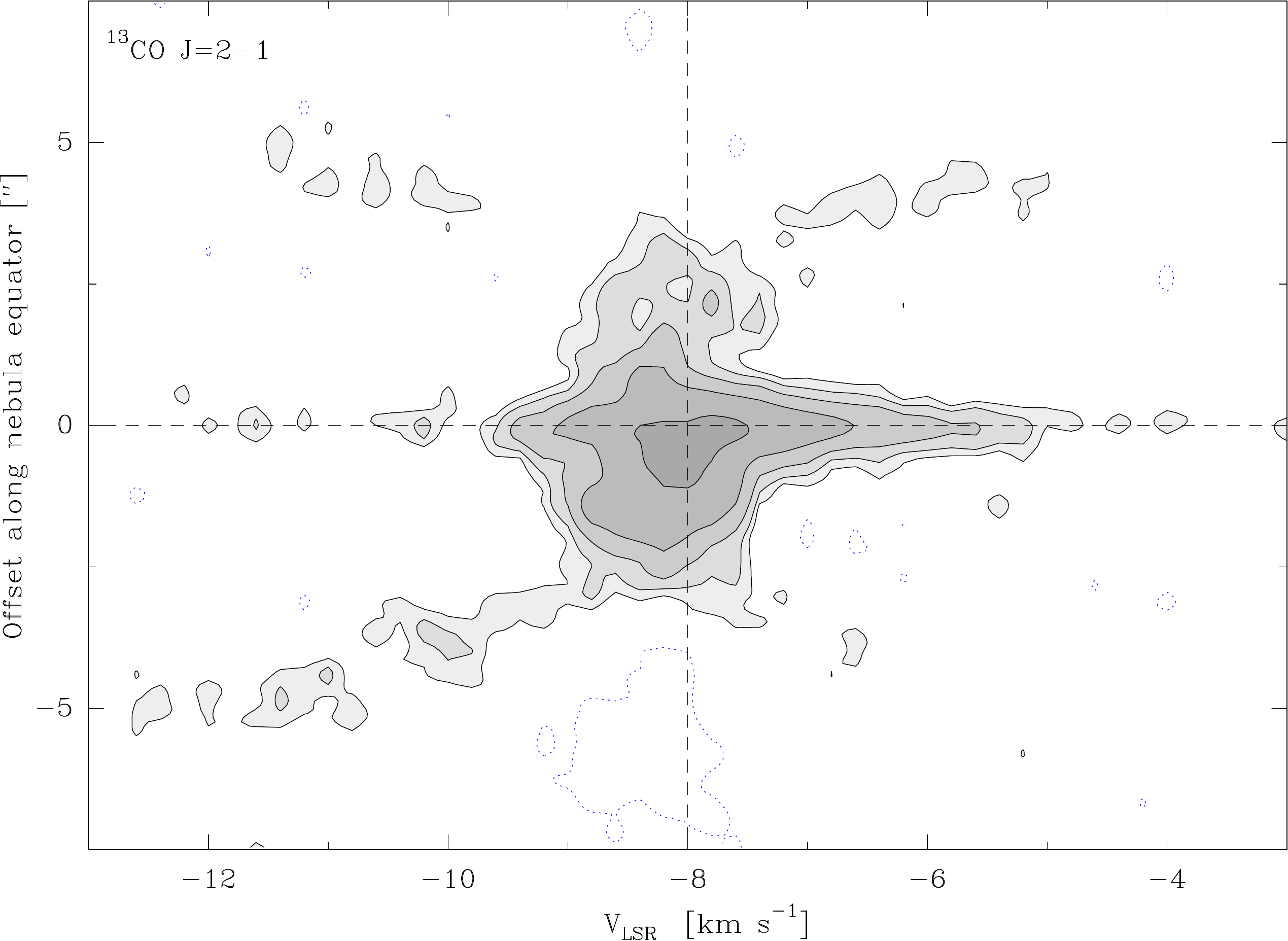}
        \end{minipage}
        \quad
        \begin{minipage}[b]{0.48\linewidth}
                \includegraphics[width=\sz\linewidth]{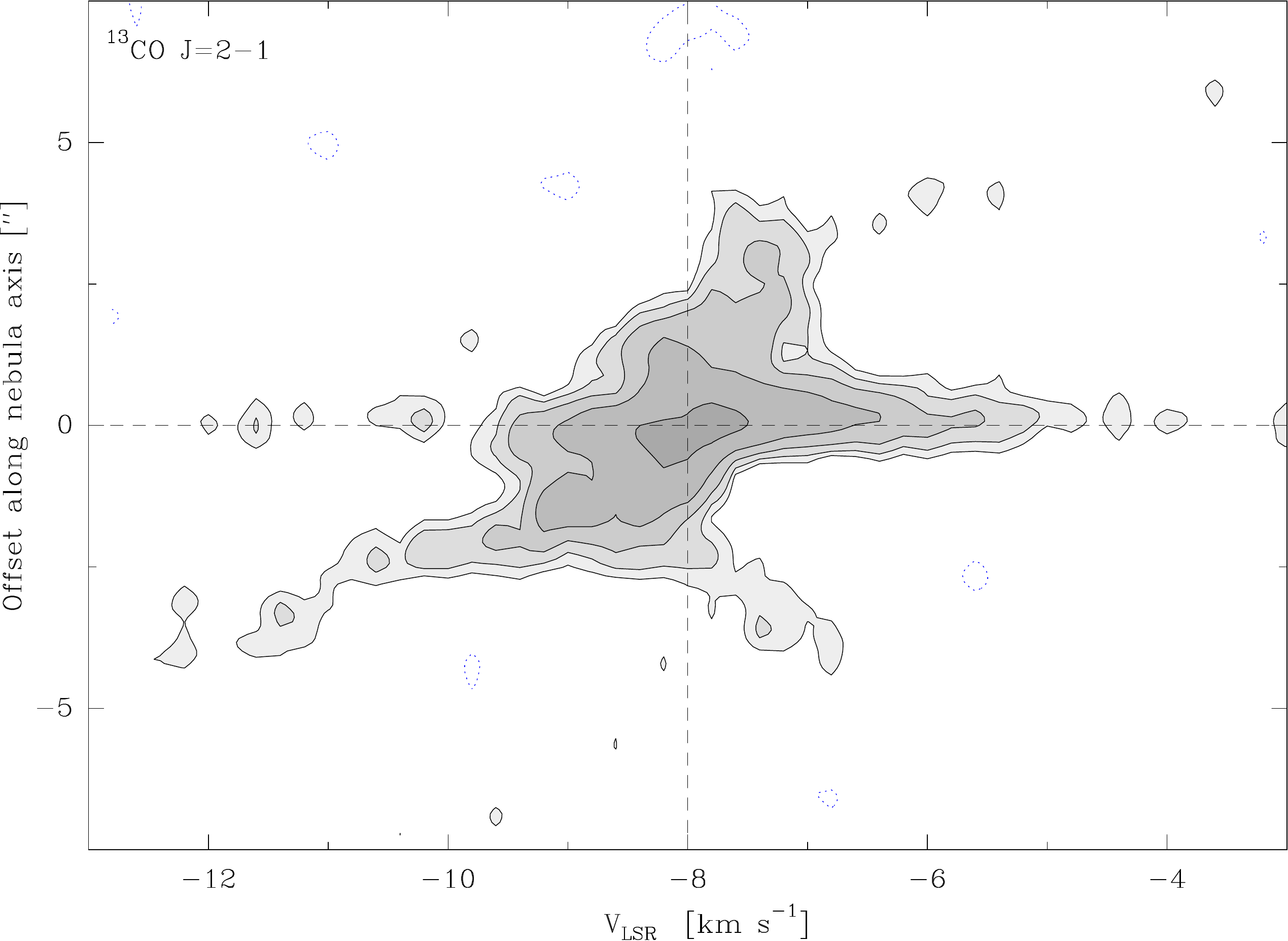}
        \end{minipage}
        \caption{\small \textit{Left:} Same as \fig\ref{fig:89her12pv} (\textit{Left}) but for \trece\dosuno emission. The contours are  $\pm$\,11, 22, 44, 88, and 144\,mJy\,beam$^{-1}$ with a maximum emission peak of 225\,mJy\,beam$^{-1}$. The dashed lines show the approximate centroid in velocity and position. \textit{Right}: Same as in \textit{Left} but along the perpendicular direction $PA=60\degree$.}
        \label{fig:89her13pv}
\end{figure*}

\begin{figure*}[h]
\includegraphics[width=\textwidth]{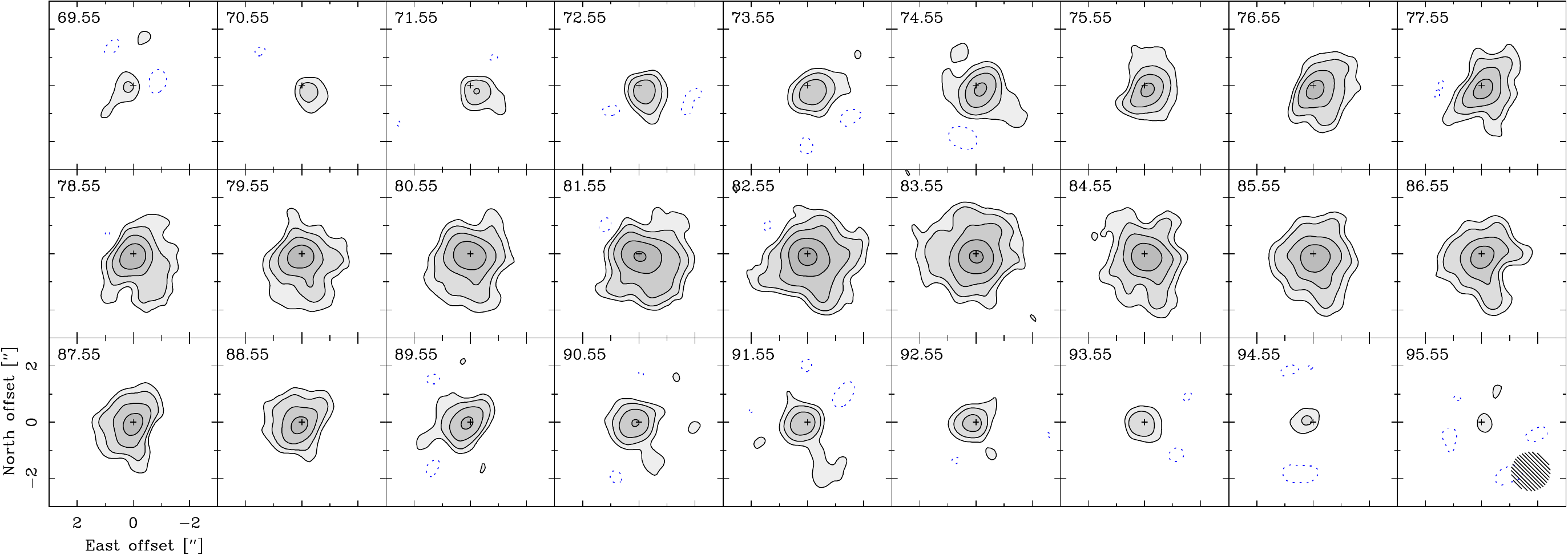}
 
\caption{\small Maps per velocity channel of \doce \dosuno emission from \irasp. The beam size (HPBW) is 0\secp70\x0\secp70. The contour spacing is logarithmic: $\pm$\,20, 40, 80, 160, and 320\,mJy\,beam$^{-1}$ with a maximum emission peak is 413\,mJy\,beam$^{-1}$.
   The LSR velocity is indicated in the upper left corner of each velocity-channel panel and the beam size is shown in the last panel.}
    \label{fig:iras12mapas}  
\end{figure*}

\begin{figure*}[h]
        \centering
        \begin{minipage}[b]{0.48\linewidth}
                \includegraphics[width=\sz\linewidth]{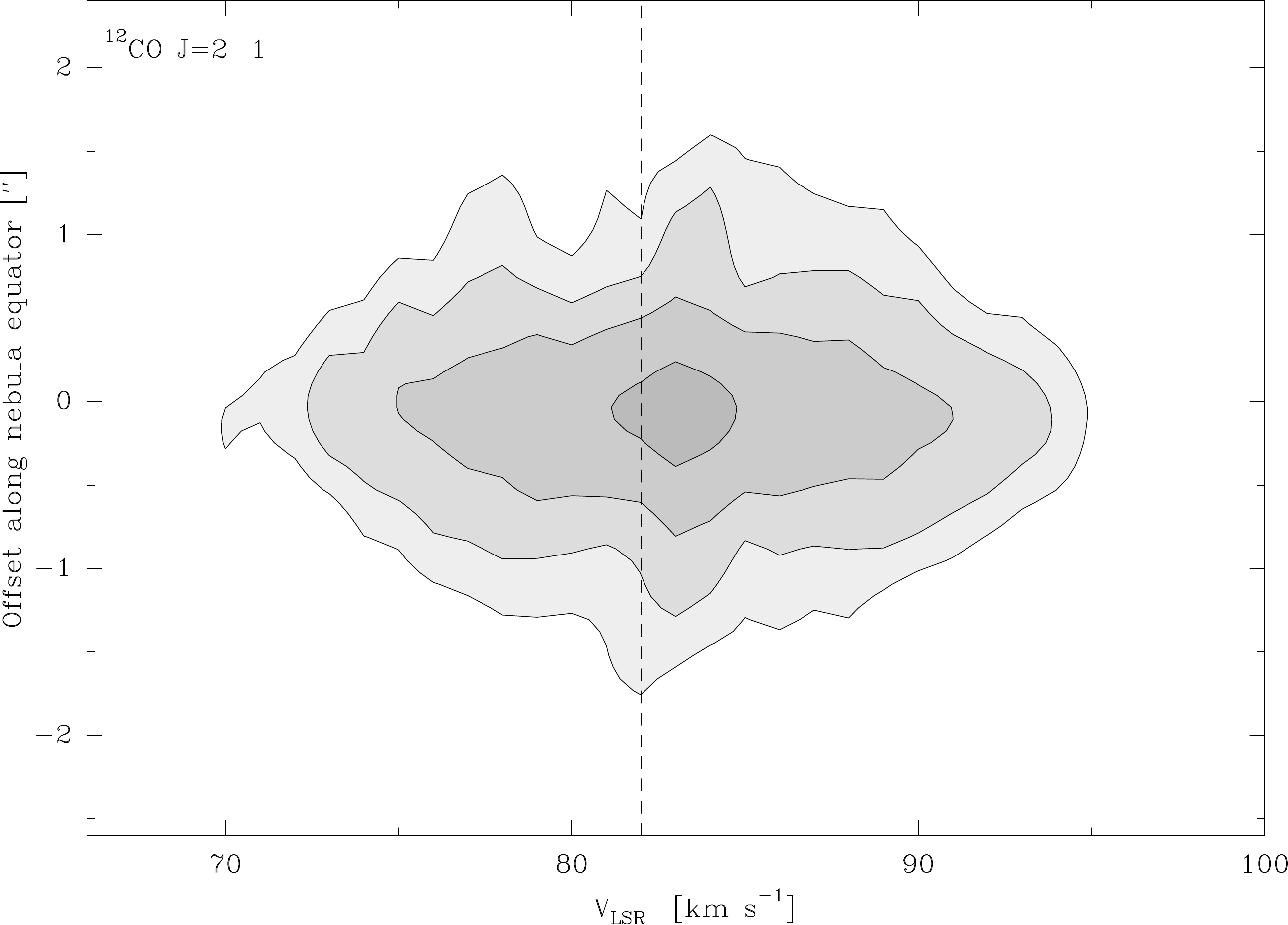}
        \end{minipage}
        \quad
        \begin{minipage}[b]{0.48\linewidth}
                \includegraphics[width=\sz\linewidth]{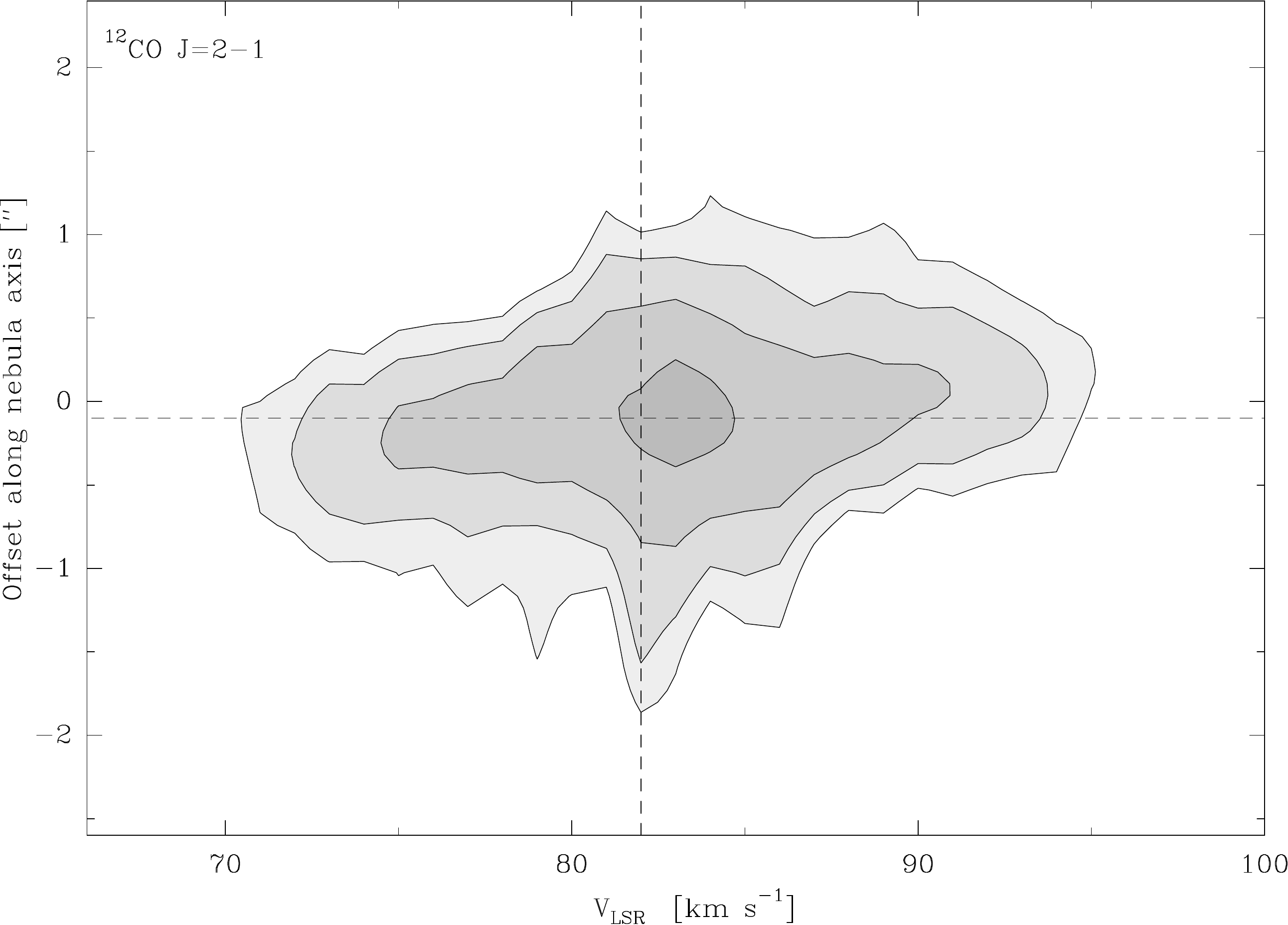}
        \end{minipage}
        \caption{\small \textit{Left:} Position--velocity diagram from our maps of \doce \dosuno in \iras along the direction $PA=-40\degree$, corresponding to the nebula equator. The contour spacing is logarithmic: $\pm$\,40, 80, 160, and 320\,mJy\,beam$^{-1}$ with a maximum emission peak of 413\,mJy\,beam$^{-1}$. The dashed lines show the approximate central position and systemic velocity. \textit{Right}: Same as in \textit{Left} but along the perpendicular direction  $PA=50\degree$.}
        \label{fig:iras12pv}
\end{figure*}

\begin{figure*}[h]
\centering
\includegraphics[width=\textwidth]{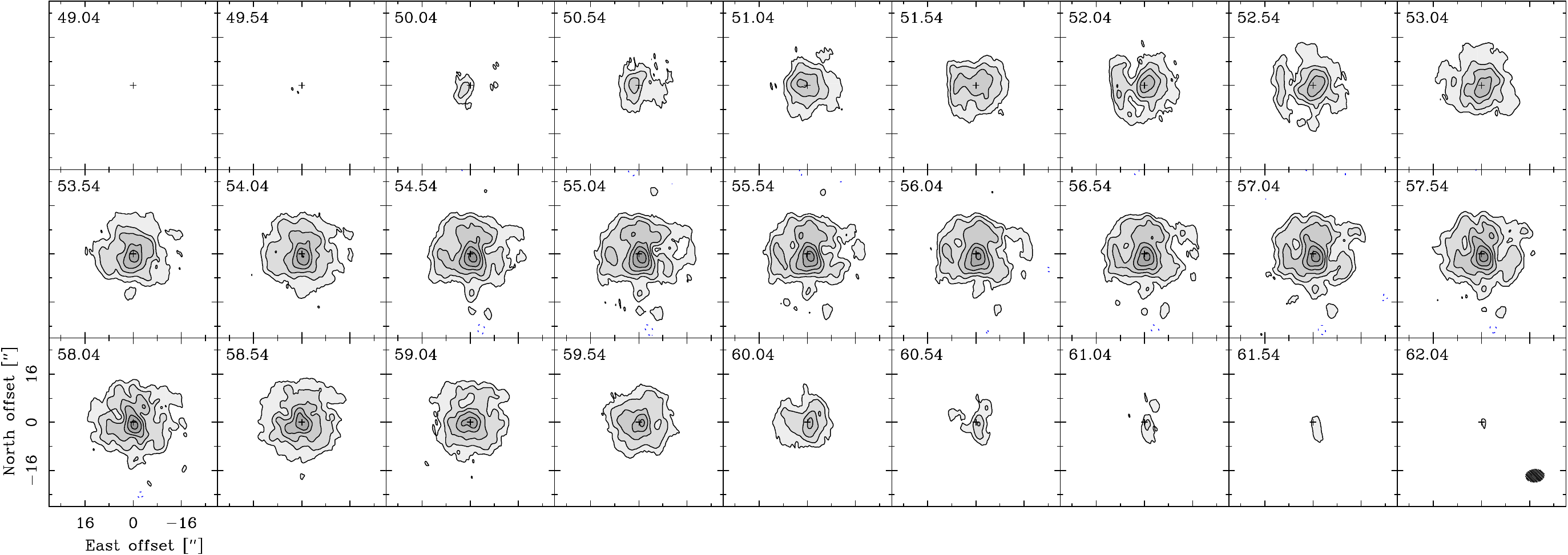}
 
\caption{\small Maps per velocity channel of \doce \dosuno emission from \rsp. The beam size (HPBW) is 3\secp12\x2\secp19, the major axis oriented at $PA=-185\degree$. The contour spacing is logarithmic: $\pm$\,50, 100, 200, 400, 800 and 1600\,mJy\,beam$^{-1}$ with a maximum emission peak of 2.4\,Jy\,beam$^{-1}$.
   The LSR velocity is indicated in the upper left corner of each velocity-channel panel and the beam size is shown in the last panel.}
    \label{fig:rs12mapas}  
\end{figure*}

\begin{figure*}[h]
        \centering
        \begin{minipage}[b]{0.48\linewidth}
                \includegraphics[width=\sz\linewidth]{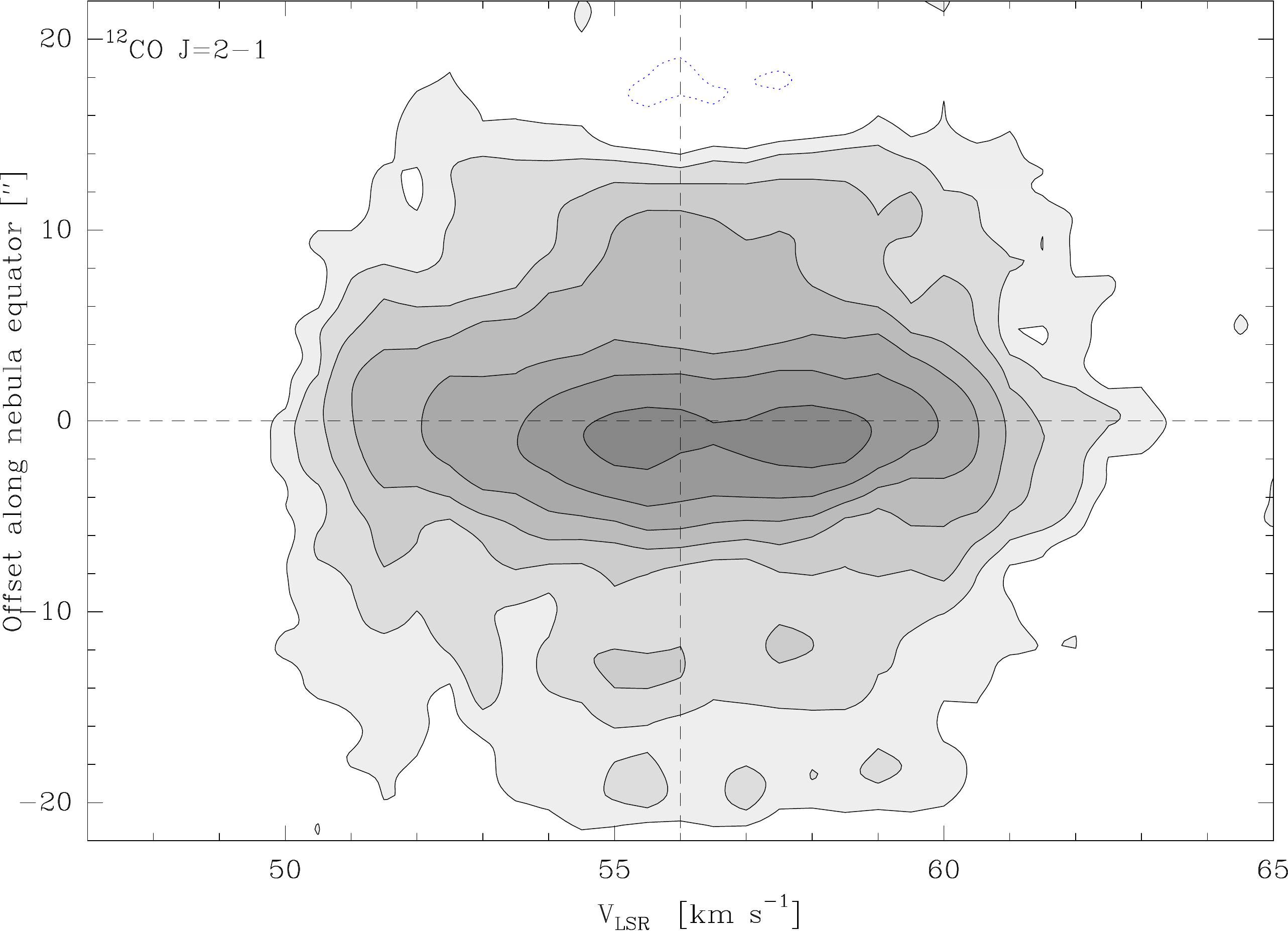}
        \end{minipage}
        \quad
        \begin{minipage}[b]{0.48\linewidth}
                \includegraphics[width=\sz\linewidth]{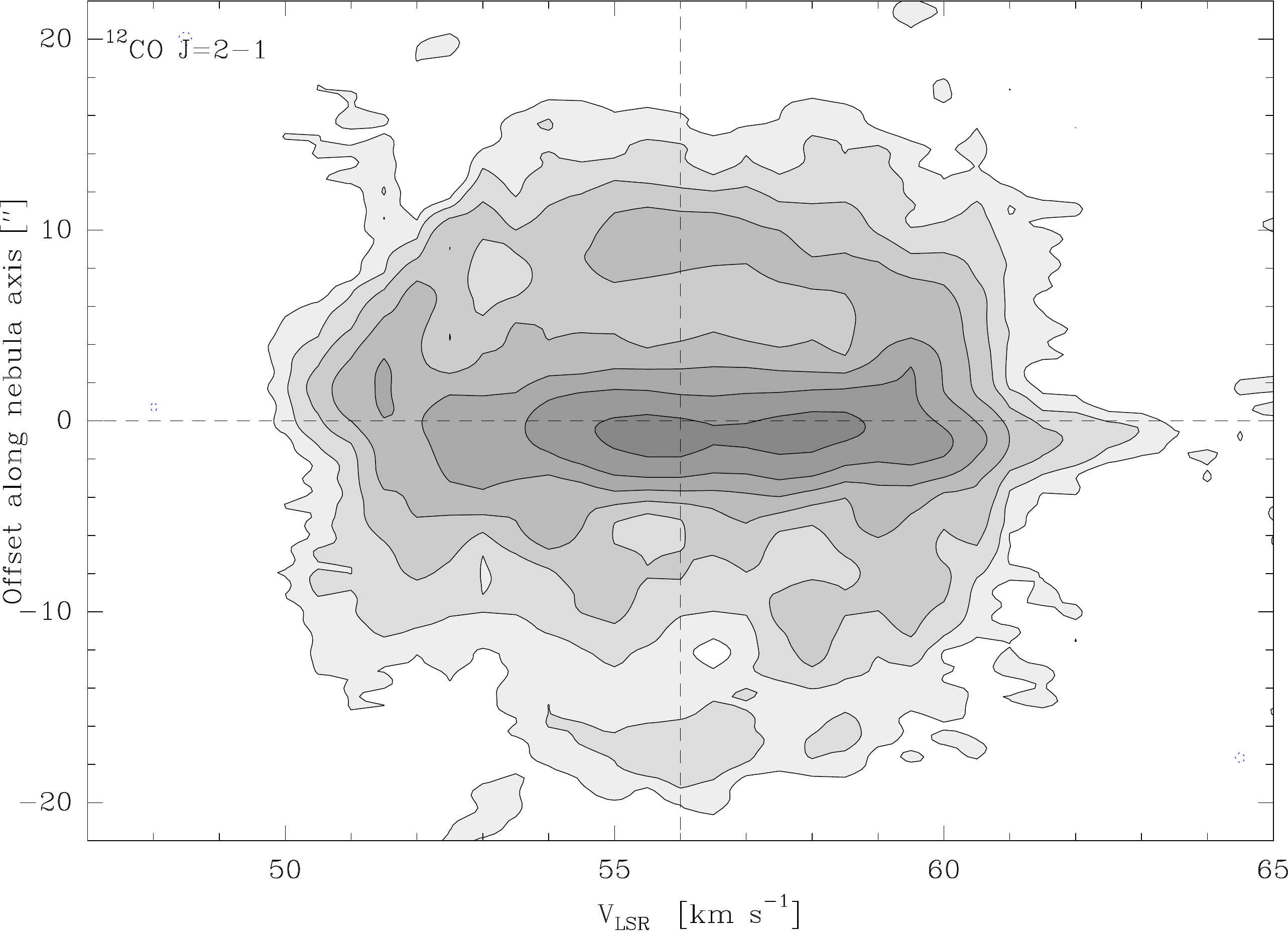}
        \end{minipage}
        \caption{\small \textit{Left:} Position--velocity diagram from our maps of \doce \dosuno in \rs along the direction $PA=0\degree$, corresponding to the nebula equator. The contour spacing is logarithmic: $\pm$\,25, 50, 100, 200, 400, 800, and 1600\,mJy\,beam$^{-1}$ with a maximum emission peak of 2.4\,Jy\,beam$^{-1}$. The dashed lines show the approximate central position and systemic velocity. \textit{Right}: Same as in \textit{Left} but along the perpendicular direction $PA=90\degree$.}
        \label{fig:rsct12pv}
\end{figure*}


\section{Observations and data reduction}
\label{reduccion}

Observations of the \doce\dosuno rotational transition at 230.53799\,GHz were carried out towards \alls with the IRAM NOEMA interferometer at Plateau de Bure (France) with six antennas. Data of the \trece\dosuno transition were also obtained for \onp. 
The data calibration was performed with the CLIC software (GILDAS package). In all the observing sessions, measurements with high signal-to-noise ratio were obtained on a bright calibration source in order to calibrate the
instrumental RF spectral response. Calibration sources close to each source were cyclically observed, and their data used to perform a first standard gain calibration. Phase self-calibration was performed later on the source continuum emission. In all the sessions, the calibration source MWC\,349 was observed as the primary flux reference, with an adopted flux of 1.91\,Jy at 230.5\,GHz. The data four all the observed line emissions were obtained with a spectral resolution of 0.05\kmsp. Additional bandwidth with a channel spacing of 2\,$-$\,2.5\,MHz was delivered to map continuum emission, the available bandwidth being different for the different observations. More details can be found below and a summary of observational parameters can be found in \tab\ref{obs}. MAPPING, also part of GILDAS, was used for data analysis and image synthesis. Final maps were obtained after continuum-emission subtraction and analyzed following image synthesis using natural and robust weightings of the visibilities. For the relatively low resolution, all sources present continuum images that are compact and unresolved, and are only used in the reduction of the data (see \sect\ref{observaciones}).

\ac was observed with observatory project names XB74 and W14BU. The data of project XB74 were already presented by \citet{bujarrabal2015}. In this work, we add to the previous data those of new project W14BU, which were aimed to increase the sensitivity. The new data were obtained in December\,2014, April\,2015, and March\,2016, with baselines ranging from 17 to 760\,m. Considering the addition of the two projects, a total of 21\,h were obtained on source, with an available bandwidth for continuum from 228.2 to 231.8\,GHz. 
In \fig\ref{fig:ac12mapas}, we present channel maps of \doce\dosuno emission with a resampled spectral resolution of 0.2\kms and a synthetic beam of 0\secp35\x0\secp35 in size, obtained using natural weighting. Relevant position--velocity diagrams are shown in \fig\ref{fig:acher12pv}.

\on was observed for 11\,h in November\,2005 and January\,2006 under project name P05E. In addition to the line emission, data were delivered for continuum in a bandwidth of 600\,MHz. These data were published by \citet{bujarrabal2007} where more details can be found. A new uv-data processing was performed using tapering in order to reduce small-scale variations in the crude map at the expense of spatial resolution. New channel maps are presented in the top panel of \fig\ref{fig:89hermapas}, which were obtained with a synthetic beam of 1\secp02\x0\secp83 in size 
with a spectral resolution of 0.2\kmsp , the major axis being oriented at $PA=115\degree$. 
In addition, observations of the \trece\dosuno line emission in \on were performed under the project names X073 and W14BT. X073 data were obtained with the most extended configuration between January and March\,2014, those for W14BT in December\,2014, and from February to
April\,2015 for more compact B and D configurations, in order to attain  baselines ranging from 15 to 760\,m. Acquisitions were obtained on source for a total of 22\,h, with a continuum bandwidth from 218.1 to 221.7\,GHz with 2\,MHz of channel spacing. MWC\,349 was observed in
all the tracks as in the primary flux calibrator, with an adopted flux of 1.86\,Jy. 
In the bottom panel of \fig\ref{fig:89hermapas}, we present channel maps of the \trece\dosuno line emission with a spectral resolution of 0.2\kms and a synthetic beam of 0\secp74\x0\secp56 in size, the major axis being oriented at $PA=28\degree$, obtained with natural weighting. In \fig\ref{fig:89her12pv} we present relevant position--velocity diagrams.

Observations of \iras were made between February and March\,2013 under the project name WA7D. A total of 12\,h were obtained on source with extended array configuration, baselines ranging from 55 to 760\,m. The interferometric visibilities were merged with zero-spacing data obtained with the 30\,m\,IRAM telescope, which guarantees that no flux is missed in final maps. We present channel maps of the \doce\dosuno line emission with a channel spacing of 1\kms and a synthetic beam of 0\secp7\x0\secp7 in size (\fig\ref{fig:iras12mapas}). We chose a circular beam in order to best compare with models, while the original data yield a synthetic beam of 0\secp7\x0\secp4 with natural weighting. The spectral configuration was the same as described for \acp. In \fig\ref{fig:iras12pv} we present relevant position--velocity diagrams.

Observations of \rs were performed in the summer and autumn of 2015 with the array in the most compact configurations (project name S15BA). A total of 12\,h were obtained on source. The spectral configuration was the same as that described for \acp. As \rs line emission is considerably extended, we added to the NOEMA visibilities short-spacing pseudo-visibilities obtained from on-the-fly maps with the 30\,m\,IRAM telescope to compensate the extended component filtered out by the interferometry.
In \fig\ref{fig:rs12mapas}, we present channel maps with a spectral resolution of 0.5\kms and a synthetic beam of 3\secp12\x2\secp19 in size, with the major axis oriented at $PA=-185\degree$.
We show relevant position--velocity diagrams in \fig\ref{fig:rsct12pv}.

\section{Interferometric observational results}
\label{observaciones}

In this section, we present the results directly obtained from the observations. We show maps per velocity channel and position--velocity (PV) diagrams along the assumed two major perpendicular directions of the nebula: along the equatorial rotating disk and along the revolution symmetry axis of the nebula.

In the cases of \onp, \irasp, and \rs the central disk is not well resolved. However, we underline that: (a) their CO single-dish profiles are similar to those of other binary \pagb stars \citep{bujarrabal2013a}, in which we know that there are central disks and that their contribution to the profiles is significant, (b) PV diagrams show hints of the typical PV diagram structure of disks with Keplerian dynamics, (c) the high-velocity dispersion observed in the central regions is the same as that expected from an unresolved Keplerian disk, and (d) when longer baselines are favored and inner regions are selected (even at the cost of losing flux), we find a line profile with two peaks characteristic of rotating disks (see CO line profiles of \on and \rs in \app\ref{dospicos}).
For these reasons, we think that these three nebulae probably harbor a rotating disk at their center and we include such a structure in our modeling (\sect\ref{modelos}).

\subsection{\acp}
\label{secacherobs}

The \doce\dosuno mm-wave interferometric results are presented in \fig\ref{fig:ac12mapas}. These observations are of higher resolution than those previously published \citep{bujarrabal2015}. In addition, we also include shorter baselines to reduce the flux loss (no significant amount of flux was missed in the interferometric data; see \fig\ref{fig:acher_flujo12co}) and to improve the sensitivity for the detection of the expanding component. A Keplerian disk was already detected in the published data; here we focus on the detection of the outflow.

In \app\ref{outflow_acher} we present an analysis of the emission of the outflow along different $PA$ (position angle, measured from north to east). Thanks to our detailed study, we determined that the direction better showing the Keplerian dynamics is  
$PA=136.1\degree \pm 1.4\degree$ (see Appendix\,\ref{outflow_acher} for details); this direction is termed the ``equatorial direction'' hereafter. 

As we can see in the left panel of \fig\ref{fig:acher12pv}, by comparison with results in \app\ref{outflow_acher}, the PV diagram along this equatorial direction very nicely shows the characteristic signature of Keplerian rotation. The investigation of the PV diagram along the nebula axis (i.e., perpendicular to the equatorial direction) should help us to detect the presence of an axially expanding outflow.
In the PV along the nebula axis in the right panel of \fig\ref{fig:acher12pv}, we can see a strong emission from the Keplerian disk in the central regions.
The theoretical PV diagram along the nebula axis in pPNe in the presence of just a rotating disk shows emission with a form similar to a diamond or rhombus with similar emission in all four PV diagram quadrants. On the contrary, we see how the emission at central velocities is slightly inclined, which could be explained by the presence of a low-mass outflow surrounding the Keplerian disk. 
With these new observations we tentatively detect these weak vestiges of the outflow of \acp, which may have a structure similar to the one found in similar sources like the Red\,Rectangle.
From the detailed analysis in \app\ref{outflow_acher}, we conclude that the outflow is tentatively detected with an emission $\lsim$\,10\,mJy\,beam$^{-1}$\,km\,s$^{-1}$.

\subsection{\onp}
\label{sec89herobs}

We present NOEMA maps of \on of \doce and \trece in \dosuno emissions. The flux loss in the interferometric observations is discussed in Appendix\,\ref{comp_flujo}. We estimate $\sim$\,30\% of lost flux in \doce\dosuno and $\sim$\,50\% in the wings in \trece \dosuno (\figs\ref{fig:89her_flujo12co} and \ref{fig:89her_flujo13co}). These values exceed the relative calibration error, $\sim$\,30\%. The presence of a certain amount of flux that is filtered out in this case is confirmed by the different profile shapes.

In these maps and PV diagrams (see \figs\ref{fig:89hermapas}, \ref{fig:89her12pv}, and \ref{fig:89her13pv}), we see an extended component and a central clump.
The shape of the CO emission suggests the presence of an extended hourglass-like structure. This shape is also clear in PV diagrams taken along the nebula axis (see the right panels of Figs. 4 and 5).
According to the angular size of the hourglass-like structure, $\sim$\,10$''$, and for a distance of 1000\,pc, we find that the size of the nebula is at least 10\,000\,AU (projected in the sky), 1.5\x\xd{17}cm.

Inspection of the PV diagrams allows us to study the compact component of the CO emission, which seems to be a disk whose projection is elongated in the equatorial direction $PA=150\degree$.
The extent of the disk is not detected and must be $\leq$\,1$''$. Position--velocity diagrams with $PA=150\degree$ (see the left panels of Figs. 4 and 5) suggest that a Keplerian disk with a moderate dispersion of velocity could be responsible for that compact clump.
Asymmetrical velocities are present in both maps because more emission appears at positive velocities than at negative velocities. This kind of phenomenon is often observed in our sources, probably as a result of self-absorption by cold gas in expansion located in front of the disk \citep[e.g.,][]{bujarrabal2018}.
As argued in \sect\ref{observaciones} and \app\ref{dospicos}, we think that there must be a Keplerian disk in the inner region of the nebula of \onp. We find that the line profile from the central component shows the characteristic double peak of rotating disks (see \fig\ref{fig:espectros_dospicos}). However, we stress that these two peaks are barely detected and we cannot accurately discern the emission of the central disk from the emission and absorption from the very inner and dense gas of the outflow.

\subsection{\irasp}
\label{secirasobs}

We present combined NOEMA and 30\,m maps of \iras of \doce\dosuno emission in \fig\ref{fig:iras12mapas} (see \sect\ref{reduccion} for details). Maps of the 30\,m reveals that the source is compact and it presents a smaller size than the beam. The main beam of NOEMA is even bigger than that of the 30\,m,  and so we are sure that there is no flux loss in the combined presented maps and all components of the source are detected.
Due to the small size of the source (see \fig\ref{fig:iras12mapas}), the brightness distribution is barely resolved in channel maps. The extent of the source is better appreciated in PV diagrams (see \fig\ref{fig:iras12pv} \textit{Left} and \textit{Right}), but we must be aware of the synthetic beam. The linear dependence of the position with the velocity suggests a nebula elongated along $PA=50\degree$, which would be the direction of the symmetry axis projection in the plane of the sky, with expansion velocity increasing with distance (a typical field in expanding nebulae around \pagb stars and pPNe).
A velocity gradient is confirmed by our model (see \sect\ref{secirasmodel}). The observed velocity gradient cannot be caused by the rotating disk, because the velocity seems to increase with the distance to the center.
If the observed velocity were attributed to rotation this would imply an extremely large central mass. From the PV diagram in the right panel of \fig\ref{fig:iras12pv} and taking into account the distance (1500\,pc; see \tab\ref{prop}), the supposed central mass would be >\,50\msp. This result is not acceptable. Therefore, the observed velocity gradient must have its origin in the axial expansion of the outflow.
We do not have clear evidence of rotation from the inner region, which is not resolved. However, the moderate dispersion of velocity of the inner region and the characteristic single-dish CO profiles lead us to suggest that there must be a rotating component in the inner part of the nebula.

What is more interesting is the strong emission from the extended component that surrounds the inner region of the nebula (right panel of \fig\ref{fig:iras12pv}). 
The outflow contribution to the total emission is dominant, and its mass could be at least similar to that of the rotating disk.
This is not a surprise because the wings of the single-dish profiles of \iras are very intense in \doce\dosunop. 

In the case of \irasp, and according to \app\ref{dospicos}, there is no sign of the characteristic double peak in the line profile from the central compact component (see \fig\ref{fig:espectros_dospicos}). We are convinced that the inclination of the disk with respect to the line of sight prevents us from seeing that effect. This is expected because \acp \ for example does not present the double peak in its single-dish CO line profiles \citep{bujarrabal2013a}, and this source basically consists in a Keplerian disk and an inclination similar to that of \iras \citep[see  \sects\ref{secachermodel}, \ref{secirasmodel}, and][]{bujarrabal2015}.

\subsection{\rsp}
\label{secrsctobs}

We present combined NOEMA and 30\,m maps of \rs in \doce\dosuno emission in \fig\ref{fig:rs12mapas} (see \sect\ref{reduccion} for details).
This is the source in our sample with the largest angular size. In \fig\ref{fig:rsct12pv}, we show PV diagrams of \doce\dosuno emission, where the presence of an extended component is clearly seen.
The total extent of the \rs nebula is $\sim$\,40$''$, which implies a linear size for the nebula of $\sim$\,40\,000\,AU. We think that the massive outflow contains most of the total mass; see \sect\ref{secrsctmodel}.

Our NOEMA observations were merged with extensive single-dish maps (\sect\ref{reduccion}) and therefore no flux is filtered out in the shown channel maps.

The PV diagram along the equator (\fig\ref{fig:rsct12pv} \textit{Left}) reveals the presence of a central clump in the inner region that probably represents the emission from the Keplerian disk, which remains unresolved in our maps. This interpretation is particularly uncertain in the case of \rsp, because its classification as binary \pagb star is not confirmed (see \sect\ref{secrsctprev}), and the central peak in its single-dish profiles is less prominent than in other sources \citep{bujarrabal2013a}. However, the velocity dispersion of the CO emission from its unresolved central condensation is very similar to those found in other disk-containing \pagb nebulae (see e.g. \citet{bujarrabal2016, bujarrabal2017} and data on other sources in previous sections, including a significant lack of blueshifted emission). This kind of phenomenon is frequently observed in this type of source as a result of self-absorption of inner warmer regions by colder gas in expansion in front of them along the line of sight, as mentioned in \sect\ref{sec89herobs}. We also see (\sect\ref{secrsctmodel}) that models of CO emission from a small disk with very reasonable properties satisfactorily explain the observed emission from the center.
Additionally, we find that the line profile from the central compact component (see \fig\ref{fig:espectros_dospicos}) shows the characteristic double peak of rotating disks (see \sect\ref{observaciones} and \app\ref{dospicos}), and therefore it probably arises from the unresolved Keplerian disk. Nevertheless, we must keep in mind that we cannot discern the emission of the central and rotating component from the emission and absorption from the most inner and dense gas of the outflow.
We therefore conclude that the presence of a Keplerian disk cannot be demonstrated in a conclusive way, but the central condensation found in the innermost region of \rs is probably an unresolved rotating disk.

We must pay attention to the PV diagram along the nebula axis (\fig\ref{fig:rsct12pv} \textit{Right}), with $PA=90\degree$, where the nebular structure is more evident. We can see two big cavities centered at about \mm10$''$. This kind of structure is often present in pPNe, such as M\,1$-$92 \citep{alcolea2007} or M\,2$-$56 \citep{castrocarrizo2002}.

\section{Model fitting of our NOEMA maps}
\label{modelos}

For modeling the sources, we used models and codes similar to those described in our previous mm-wave interferometric works \citep[][see also \sect\ref{introduccion}]{bujarrabal2015,bujarrabal2016,bujarrabal2017,bujarrabal2018}. Our model consists of a rotating disk surrounded by an outer outflow that can present different shapes (no sign of an expanding disk is found in our data). We consider axial and equatorial symmetry for our four nebulae. The model only includes CO-rich regions, because if there is a really extended halo poor in CO, it will not be detectable.
We assume LTE populations for the rotational levels involved in the studied emission. This assumption is reasonable for low-$J$ rotational levels of CO transitions, at least for gas densities over $n \geq$\,10$^{4}$\,cm$^{-3}$, because Einstein coefficients are much smaller than the typical collisional rates \citep{bujarrabalalcolea2013,bujarrabal2016}. Therefore, the characteristic rotational temperature used here is in most relevant cases equal to the kinetic temperature (see \app\ref{trot_tcin} for more details). The use of LTE simplifies the calculations significantly and provides an easier interpretation of the fitting parameters of the model.

The inputs for the model are the nebular shape and distributions of the density, temperature, macroscopic velocity, and local turbulence. We assume constant \doce and \trece abundances.
With these elements, the code calculates the emission and absorption coefficients of the observed lines (\doce \dosuno and \trece in \dosunop). These are computed for a large number of projected velocities and for a large number of elemental cells that fill the nebula. Typically, around 10$^{6}$ cells are used in our models. We solve the radiative transfer equation in the direction of the telescope. 
We then obtain a brightness distribution as a function of the coordinates and of the projected velocity. For this, we take into account the assumed orientation of the nebula axis with respect to the line of sight.
Finally, the derived brightness distribution is numerically convolved with the interferometric clean beam.

Here, we present our best-fit model for the four analyzed sources.
We stress that the nebula models are complex and have a very large number of parameters.
We consider simple laws for the density ($n$) and characteristic rotational temperature ($T$):

\begin{equation}
n=n_{0}\left(\frac{r_{0}}{r}\right)^{\kappa_{n}},
\label{eq:dens_eq}
\end{equation}
\begin{equation}
T=T_{0}\left(\frac{r_{0}}{r}\right)^{\kappa_{T}},
\label{eq:temp_eq}
\end{equation}

\noindent where $r_{0}$ takes the value:
\begin{equation}
  r_{0}=\begin{cases}
      \vspace{1mm}
    \frac{R_{K}}{2}, & \text{if $h \leq h_{K}$ and $r \leq R_{K}$}\\

    \frac{h_{outflow}}{2}, & \text{if $h \textgreater h_{K}$ and $r \textgreater R_{K}$}
  \end{cases}
  ,
\end{equation}

\noindent where $r$ represents the distance to the center, $h$ is the distance to the equator, and $n_{0}$ and $T_{0}$ are the values of the density and temperature at $r=r_{0}$. Here, $\kappa_{n}$ and $\kappa_{T}$ are the values of the slopes in the potential law for density and temperature, respectively, and $R_{K}$ and $h_{K}$ represent the radius and height, respectively, of the disk with Keplerian dynamics. 

We assume pure Keplerian rotation ($V_{rot_{K}}$) in the disk (\eq\ref{eq:vdis_eq}) and radial expansion velocity ($V_{exp}$) in  the outflow (\eq\ref{eq:vout_eq}):

\begin{equation}
V_{rot_{K}} = V_{rot_{K_{0}}}\sqrt{\frac{10^{16}}{r}},
\label{eq:vdis_eq}
\end{equation}
\begin{equation}
V_{exp} = V_{exp_{0}} \frac{r}{10^{16}},
\label{eq:vout_eq}
\end{equation}

\noindent where $V_{rot_{K_{0}}}$ is the tangential velocity of the Keplerian disk at \xd{16}\,cm.
The parameter $V_{exp_{0}}$ represents the expansion velocity of the ouflow at a distance of \xd{16}\,cm.

We assumed values of the relative abundances compatible with previous estimates. In particular, we adopted X(\trecep)\,$\sim$\,2\x\xd{-5} to ease the comparison of our mass estimates with those by \citet{bujarrabal2013a} and a low X(\docep)\,/\,X(\trecep)\,$\sim$\,10, as usually found in those low-mass \pagb nebulae \citep{bujarrabal1990, bujarrabal2013a}.

In \sects\ref{secachermodel}, \ref{sec89hermodel}, \ref{secirasmodel}, and \ref{secrsctmodel} we present the results obtained from the model fitting of the observations for the four sources studied in this paper.

\subsection{\acp}
\label{secachermodel}

\begin{figure*}[h]
        \centering
        \begin{minipage}[b]{0.48\linewidth}
                \includegraphics[width=\sz\linewidth]{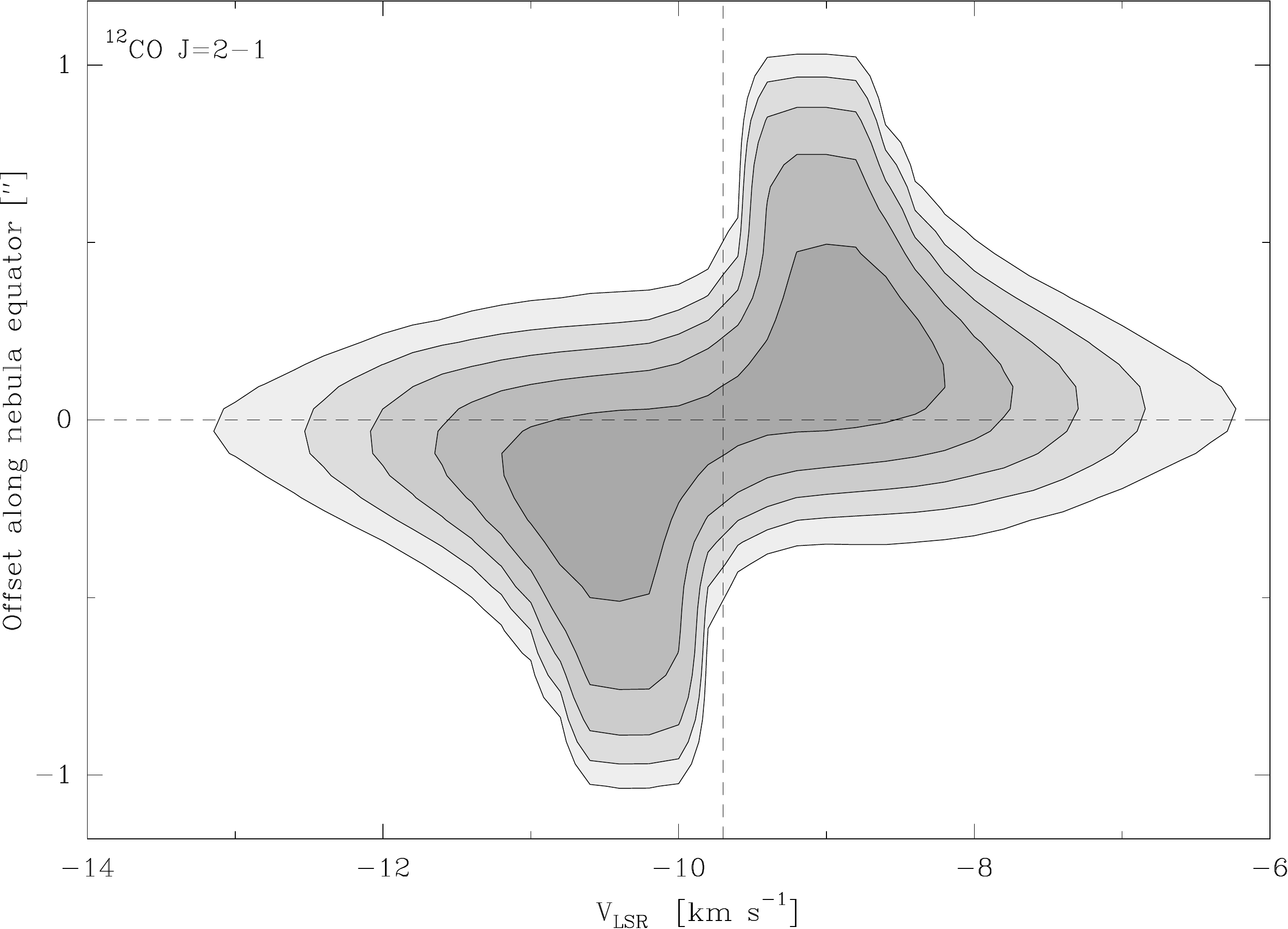}
        \end{minipage}
        \quad
        \begin{minipage}[b]{0.48\linewidth}
                \includegraphics[width=\sz\linewidth]{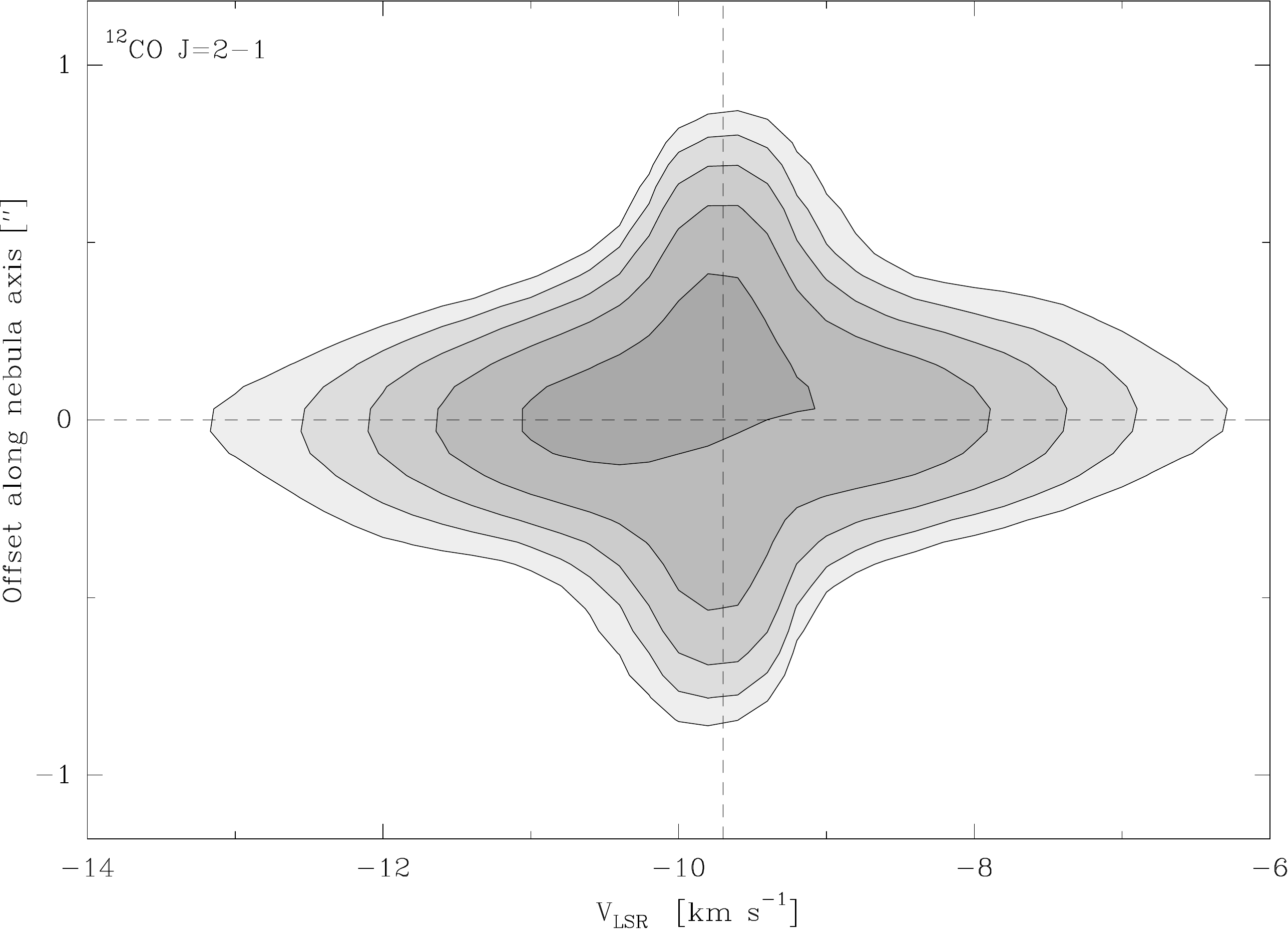}
        \end{minipage}
        \caption{\small \textit{Left:} Synthetic PV diagram from our best-fit model of \doce \dosuno in \ac with an outflow. For ease of comparison with \fig\ref{fig:acher12pv} the scales and contours are the same. \textit{Right}: Same as in \textit{Left} but along $PA=46.1\degree$.}
        \label{fig:acpvmodelodiscooutflow}
\end{figure*}

\begin{figure}[h]
    \centering
                \includegraphics[width=\sz\linewidth]{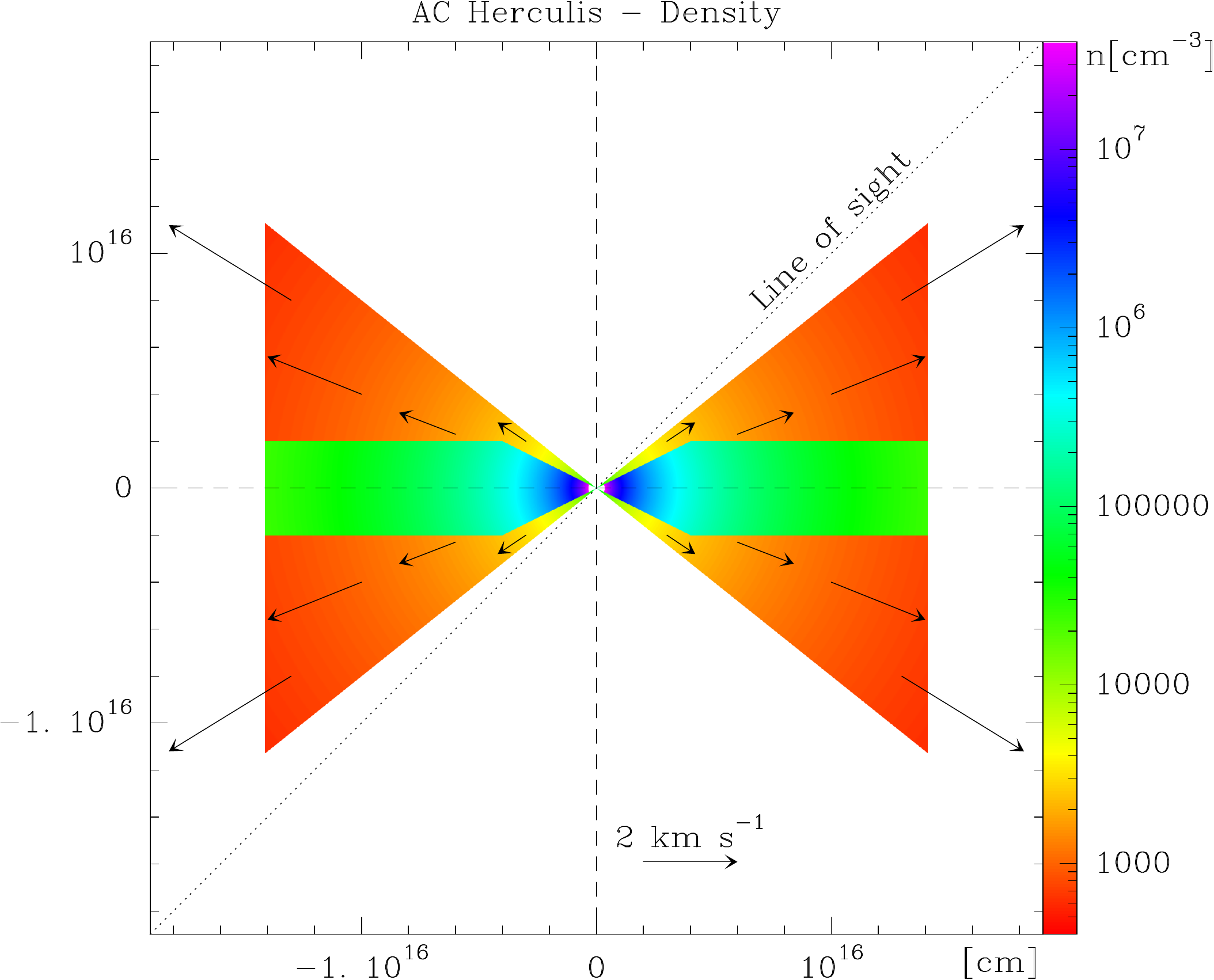}
    \caption{\small Structure and distribution of the density of our best-fit model for the disk and outflow of \acp. The Keplerian disk presents density values $\geq$\xd{5}\,cm$^{-3}$. The expansion velocity is represented with arrows.}
        \label{fig:acherdens}
\end{figure}

The model for the structure of \ac (see \fig\ref{fig:acherdens} and \tab\ref{achermodel}) is similar to the one we developed for the Red\,Rectangle and for IRAS\,08544$-$4431 \citep[see][]{bujarrabal2013b,bujarrabal2016,bujarrabal2018}.
Our maps, and mainly the PV diagram at $PA=136.1\degree$ (\figs\ref{fig:ac12mapas} and \ref{fig:acher12pv}), and following a discussion similar to that in \citet{bujarrabal2015}, confirm that the detected CO emission from \ac comes from an orbiting disk with Keplerian rotation. The presence of an outflow is doubtful, because its emission is very weak (see \sect\ref{secacherobs} and \app\ref{outflow_acher}). However, the slight asymmetry of the emission in the most extreme angular offsets that we see in the right panel of \fig\ref{fig:acher12pv} \textit{} leads us to think that an outflow could be present.  

Our best-fitting model is described in \tab\ref{achermodel}.
We have taken the \doce abundance to be X(\docep)\,=\,2\x\xd{-4}, the same value as in \citet{bujarrabal2013a}.
We assume an inclination for the nebula axis with respect to the line of sight of $45\degree$, in agreement with \citet{bujarrabal2015}.

The density and temperature laws in the disk and outflow are assumed to vary with the distance and they follow potential laws (see \eqs\ref{eq:dens_eq} and \ref{eq:temp_eq}), according to the parameters of \tab\ref{achermodel}. Our model shows densities between \xd{5} and \xd{7}\,cm$^{-3}$ and temperatures between 20 and 200\,K (according to \eqs\ref{eq:dens_eq} and \ref{eq:temp_eq} of \sect\ref{modelos}, and parameters of \tab\ref{achermodel}).
We find that the observations are compatible with Keplerian rotation in the disk (\eq\ref{eq:vdis_eq}) for a central total stellar mass of $\sim$\,1\msp.

A representation of our model nebula and predicted results can be seen in \figs \ref{fig:acpvmodelodiscooutflow}, \ref{fig:acherdens}, and  \ref{fig:acmapasmodelodiscooutflow}.
The main goal of our work is to study the outflow that appears to surround the Keplerian disk.
Our modeling of the outflow is based on a linear law for the density and a constant value for the temperature of 100\,K, which is a value typical of other outflows around Keplerian disks \citep{bujarrabal2016, bujarrabal2017, bujarrabal2018}.
We find that an increase in the outflow density by 50\% yields results that are totally incompatible with observations.
With these premises, we have found an upper limit to the mass of the outflow that is consistent with the observations of $\lsim$\,2.0\x\xd{-5}\msp.
The derived mass for the nebula of \acp, including the extended component, is 8.3\x\xd{-4}\msp, and it is  therefore clearly a disk-dominated \pagb nebula; we find that the mass of outflow represents $\lsim$\,3\% of the total mass.

Additionally, we present predictions from a model for the nebula of \acp, in which only emission of the disk is considered (see \figs\ref{fig:acmapasmodelodisco} and \ref{fig:acher12pvmodelodisco}). As we see, these predictions are almost equal to the ones of our standard model with the outflow emission. This is expected because the outflow contribution is very weak as we see in the observational data (see \sect\ref{secacherobs} and \app\ref{outflow_acher} for more details).

\begin{table}[h]
 \caption{Physical conditions in the molecular disk and outflow of \ac derived from the model fitting of the CO data.}
\small
\vspace{-5mm}
\begin{center}

\begin{tabular*}{\linewidth}{@{\extracolsep{\fill\quad}}lll}
\hline \hline
 \\[-2ex]
Parameter   & Disk   & Outflow \\ \hline
 \\[-2ex]
 \vspace{1mm}
Radius\,[cm] & $1.4\times 10^{16}$ &  $R_{max}=1.4 \times 10^{16}$ \\
\vspace{1mm}
Height\,[cm] & $2.0\times 10^{15}$  & $1.1\times 10^{16}$ \\

\multirow{2}{*}{Density\,[cm$^{-3}$]} & $n_{0}=2.0\times 10^{5}$ &  $n_{0}=1.1\times 10^{3}$ \\
\vspace{1mm}
 & $\kappa_{n}=2.0$   & $\kappa_{n}=1.0$ \\
\multirow{2}{*}{Temperature\,[K]} & $T_{0}=39$ & $T_{0}=100$ \\
\vspace{1mm}
 & $\kappa_{T}=1.0$ & $\kappa_{T}=0$    \\
 \vspace{1mm}
 Rot.\,Vel.\,[km\,s$^{-1}$] & 1.0  & $-$ \\
 \vspace{1mm}
Exp.\,Vel.\,[km\,s$^{-1}$] & $-$ & 2  \\  
\vspace{1mm}
X(\docep) & 2\x10$^{-4}$ & 2\x10$^{-4}$ \\
\vspace{1mm}
\docep\,/\,\trecep & 10 & 10 \\
\vspace{1mm}
Inclination\,[\degree]  & & 45 \\

Position angle\,[\degree] & & 46.1 \\

\hline
 
\end{tabular*}

\end{center}\small
\vspace{-1mm}
\textbf{Notes.} Parameters and their values used in the best-fit model.
$R_{max}$ indicates the maximum radius of the outflow. The density and temperature follow the potential laws of \eq\ref{eq:dens_eq} and \eq\ref{eq:temp_eq}. $V_{rot_{K_{0}}}$ and $V_{exp_{0}}$ are the values of the velocity of the disk and outflow at \xd{16}\,cm in \eqs\ref{eq:vdis_eq} and \ref{eq:vout_eq}. We show the inclination of the nebula symmetry axis with respect to the line of sight and the position angle of its projection on the plane of the sky.
\label{achermodel}
\end{table}



The analysis of the mass of the outflow in \ac is very uncertain, mainly because of the lack of information on its main properties. In an attempt to quantify these uncertainties, we present an alternative model for \ac (see Appendix\,\ref{anexomaps}), where the size of the outflow is somewhat larger than that presented immediately above (but still compatible with our general ideas on such outflows).
This alternative model (see \fig\ref{fig:acherdens12}) presents the same characteristics for the disk. We see in \figs\ref{fig:acmapasmodelodiscooutflow12} and \ref{fig:ac12pvmodelodiscooutflow12} that the increase in the outflow size is still consistent with the observational data.

The total mass derived from our alternative model for the nebula is 8.4\x\xd{-4}\msp, of which the mass of this more extended outflow is  3.2\x\xd{-5}\msp. Therefore, the mass of this alternative and larger extended component represents  just $\lsim$\,4\% of the total mass.
The increase in the percentage of the outflow mass does not change significantly despite this significant increase in the size of the extended component. This alternative model leads us to conclude that the mass of the outflow must be $\lsim$\,5\% of the total mass, but we are aware of the uncertain analysis.

We tried to check the presence of the outflow calculating the derived residual emission from a comparison between observations and predictions of the disk-only model (\app\ref{figuras_adicionales}). Nevertheless, uncertainties in our best-fit models are too large and the residual emission is comparable to the outflow emission we are discussing, because the putative outflow is just $\sim$\,5\% of the maximum emission. Therefore, this method is not useful to prove the existence of the outflow.

Finally, we also tried to estimate the excess emission by comparing PV diagrams obtained for different position angles (see \app\ref{outflow_acher} for more details). We find in this way a tentative detection of the outflow emission, which would contain a mass again  \lsim\,5\% of the total mass.

\begin{figure}[h]
    \centering
                \includegraphics[width=\sz\linewidth]{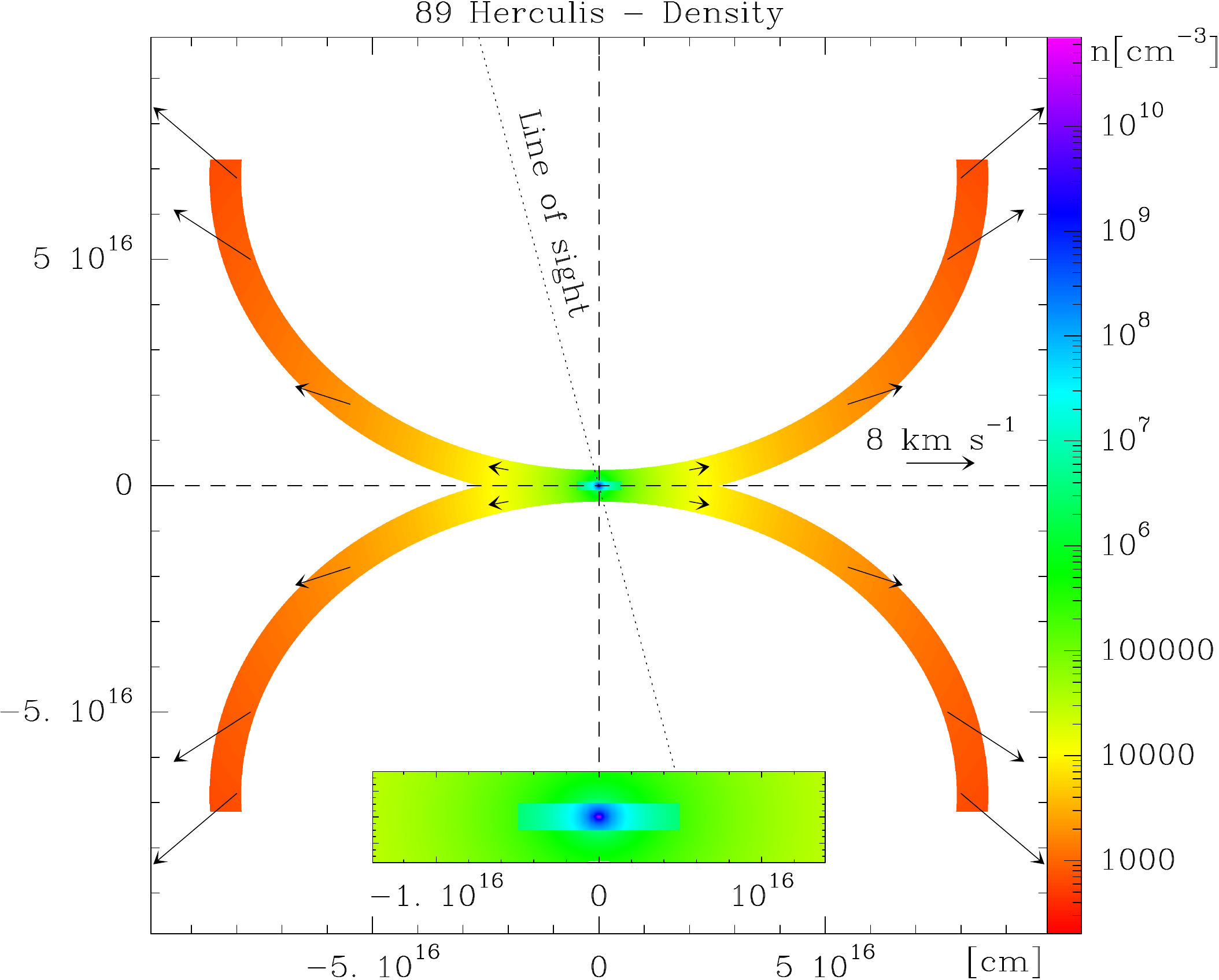}
    \caption{\small Structure and distribution of the density of our best-fit model for the disk and outflow of \onp. The lower inset shows a zoom into the inner region of the nebula where the Keplerian disk presents density values $\geq$\xd{7}\,cm$^{-3}$. The expansion velocity is represented with arrows.}
    \label{fig:89herdens}
\end{figure}

\begin{table}[h]
 \caption{Physical conditions in the molecular disk and outflow of \on derived from the model fitting of the CO data.}
\small
\vspace{-5mm}
\begin{center}

\begin{tabular*}{\linewidth}{@{\extracolsep{\fill\quad}}lll}
\hline \hline
 \\[-2ex]
Parameter   & Disk   & Outflow \\ \hline
 \\[-2ex]
 
\multirow{2}{*}{Radius\,[cm]} & \multirow{2}{*}{$5.0\times 10^{15}$} &  $R_{max}=8.6 \times 10^{16}$\\
 \vspace{1mm}
 & & $W_{o}=0.7$\x\xd{16} \\
  \vspace{1mm}
Height\,[cm] & $1.0\times 10^{15}$  & $7.2\times 10^{16}$ \\
\multirow{2}{*}{Density\,[cm$^{-3}$]} & $n_{0}=2.0\times 10^{7}$ &  $n_{0}=3.0\times 10^{3}$ \\
 \vspace{1mm}
 & $\kappa_{n}=2.0$   & $\kappa_{n}=1.8$ \\
\multirow{2}{*}{Temperature\,[K]} & $T_{0}=75$ & $T_{0}=10$ \\
 \vspace{1mm}
 & $\kappa_{T}=2.5$ & $\kappa_{T}=0$    \\
  \vspace{1mm}
 Rot.\,Vel.\,[km\,s$^{-1}$] & 1.5  & $-$ \\
  \vspace{1mm}
Exp.\,Vel.\,[km\,s$^{-1}$] & $-$ & 1.2  \\  
 \vspace{1mm}
X(\docep) & 2.0\x10$^{-4}$ & 2.0\x10$^{-4}$ \\
 \vspace{1mm}
\docep\,/\,\trecep & 10 & 10 \\
 \vspace{1mm}
Inclination\,[\degree] & & 15 \\

Position angle\,[\degree] & & 60 \\

\hline
 
\end{tabular*}

\end{center}
\small
\vspace{-1mm}
\textbf{Notes.} Parameters and their values used in the best-fit model.
$R_{max}$ indicates the maximum radius of the outflow. $W_{o}$ is the width of the outflow walls. The density and temperature follow the potential laws of \eqs\ref{eq:dens_eq} and \ref{eq:temp_eq}. $V_{rot_{K_{0}}}$ and $V_{exp_{0}}$ are the values of the velocity of the disk and outflow at \xd{16}\,cm in \eqs\ref{eq:vdis_eq} and \ref{eq:vout_eq}. We show the inclination of the nebula symmetry axis with respect to the line of sight and the position angle of its projection on the plane of the sky.
\label{89hermodel}
\end{table}

\subsection{\onp}
\label{sec89hermodel}

\begin{figure*}[h]
        \centering
        \begin{minipage}[b]{0.48\linewidth}
                \includegraphics[width=\sz\linewidth]{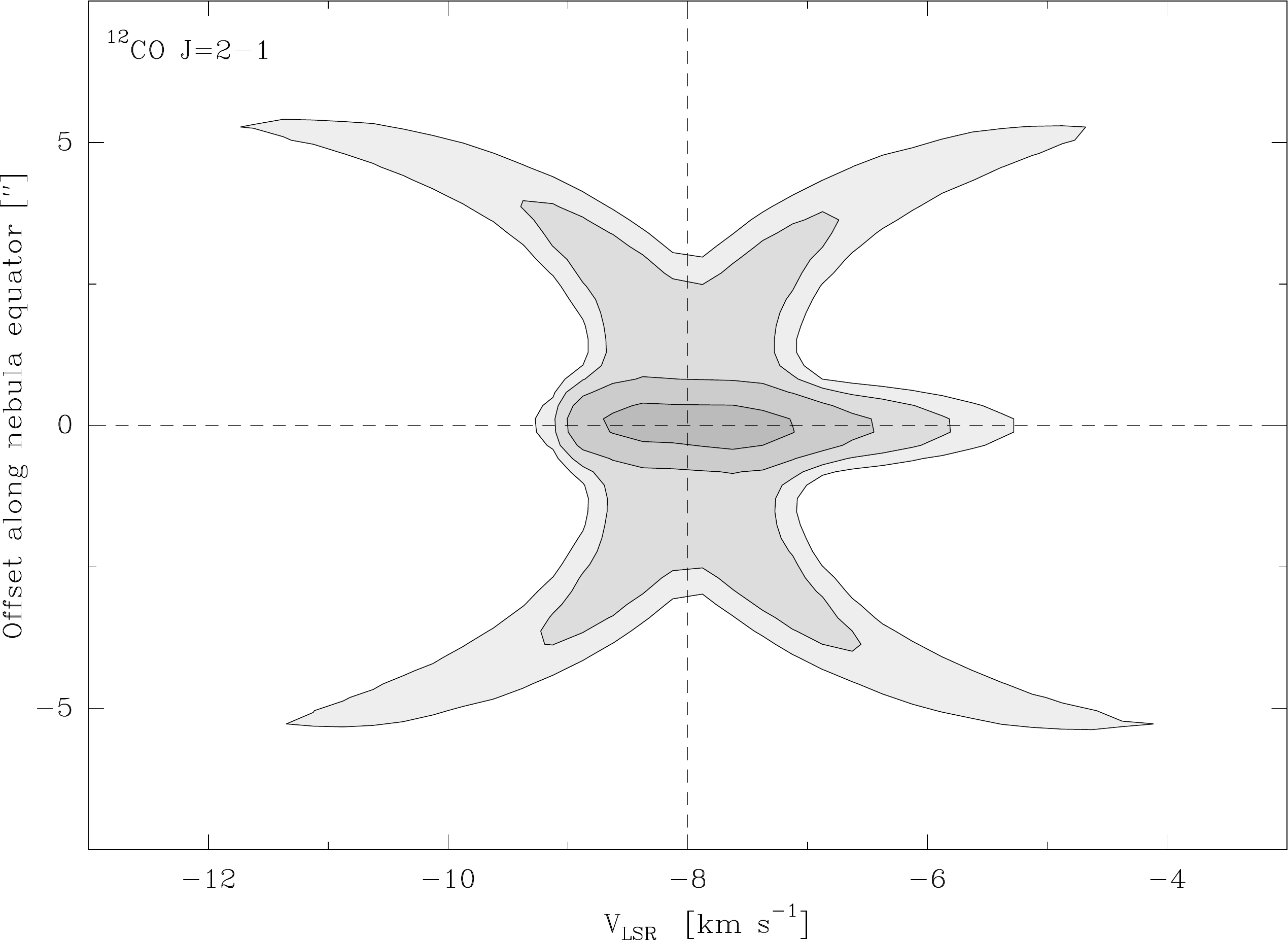}
        \end{minipage}
        \quad
        \begin{minipage}[b]{0.48\linewidth}
                \includegraphics[width=\sz\linewidth]{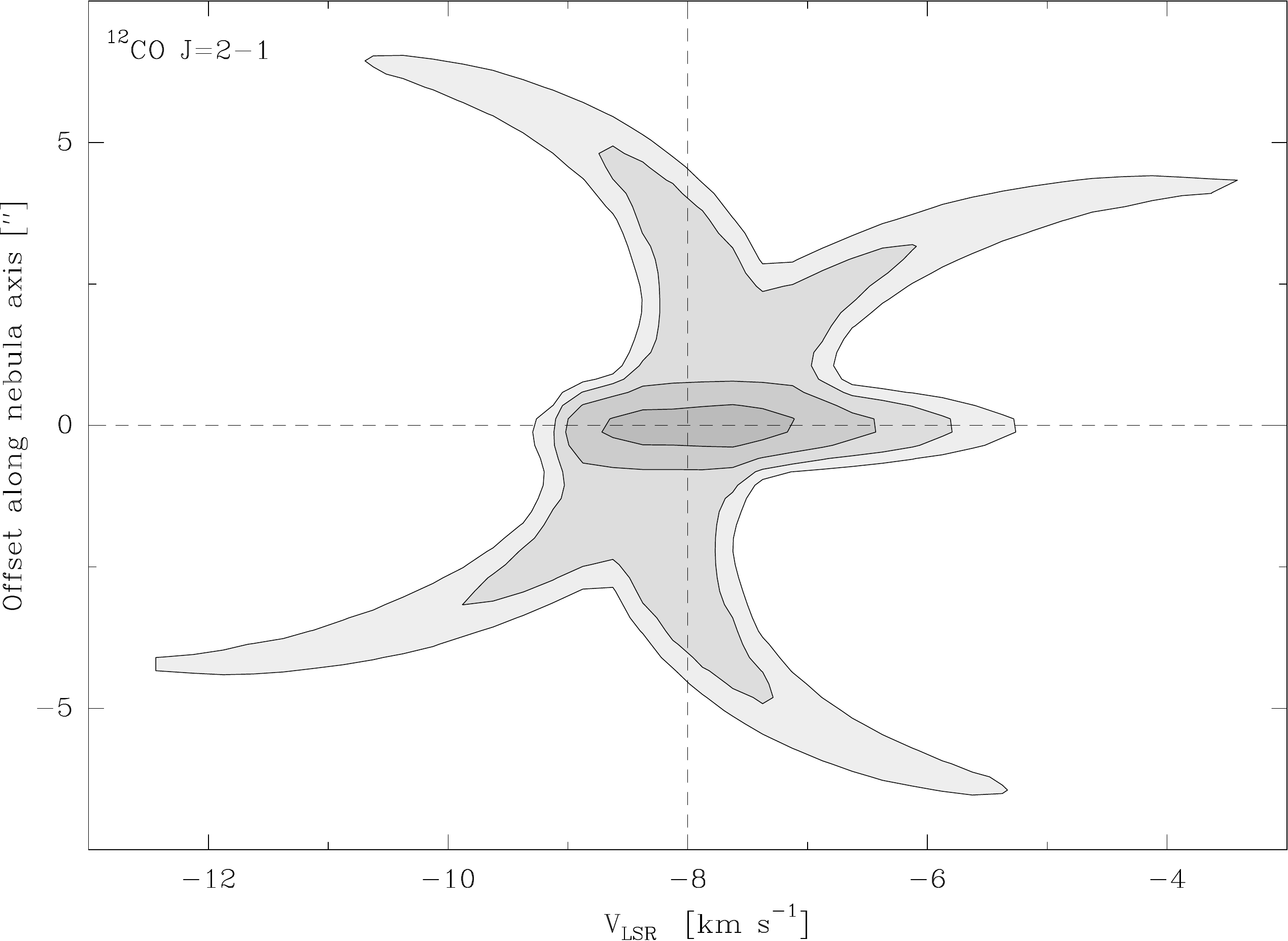}
        \end{minipage}
        \caption{\small \textit{Left:} Synthetic PV diagram from our best-fit model of \doce \dosuno in 89\,Her along the direction $PA=150\degree$. To be compared with the left panel of \fig\ref{fig:89her12pv}, the scales and contours are the same. \textit{Right}: Same as in \textit{Left} but along $PA=60\degree$.}
        \label{fig:89her12pvmodelo}
\end{figure*}

\begin{figure*}[h]
        \centering
        \begin{minipage}[b]{0.48\linewidth}
                \includegraphics[width=\sz\linewidth]{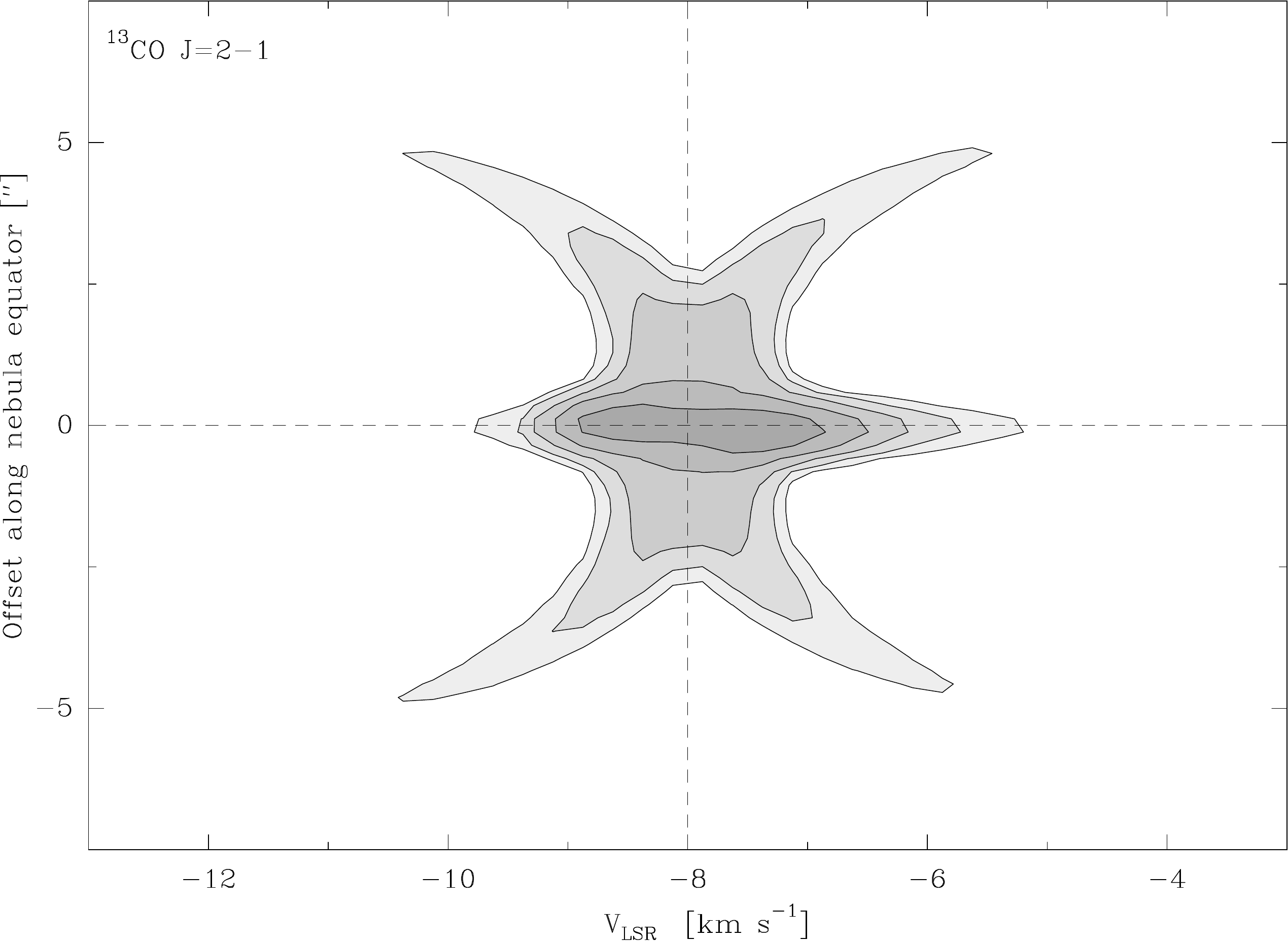}
        \end{minipage}
        \quad
        \begin{minipage}[b]{0.48\linewidth}
                \includegraphics[width=\sz\linewidth]{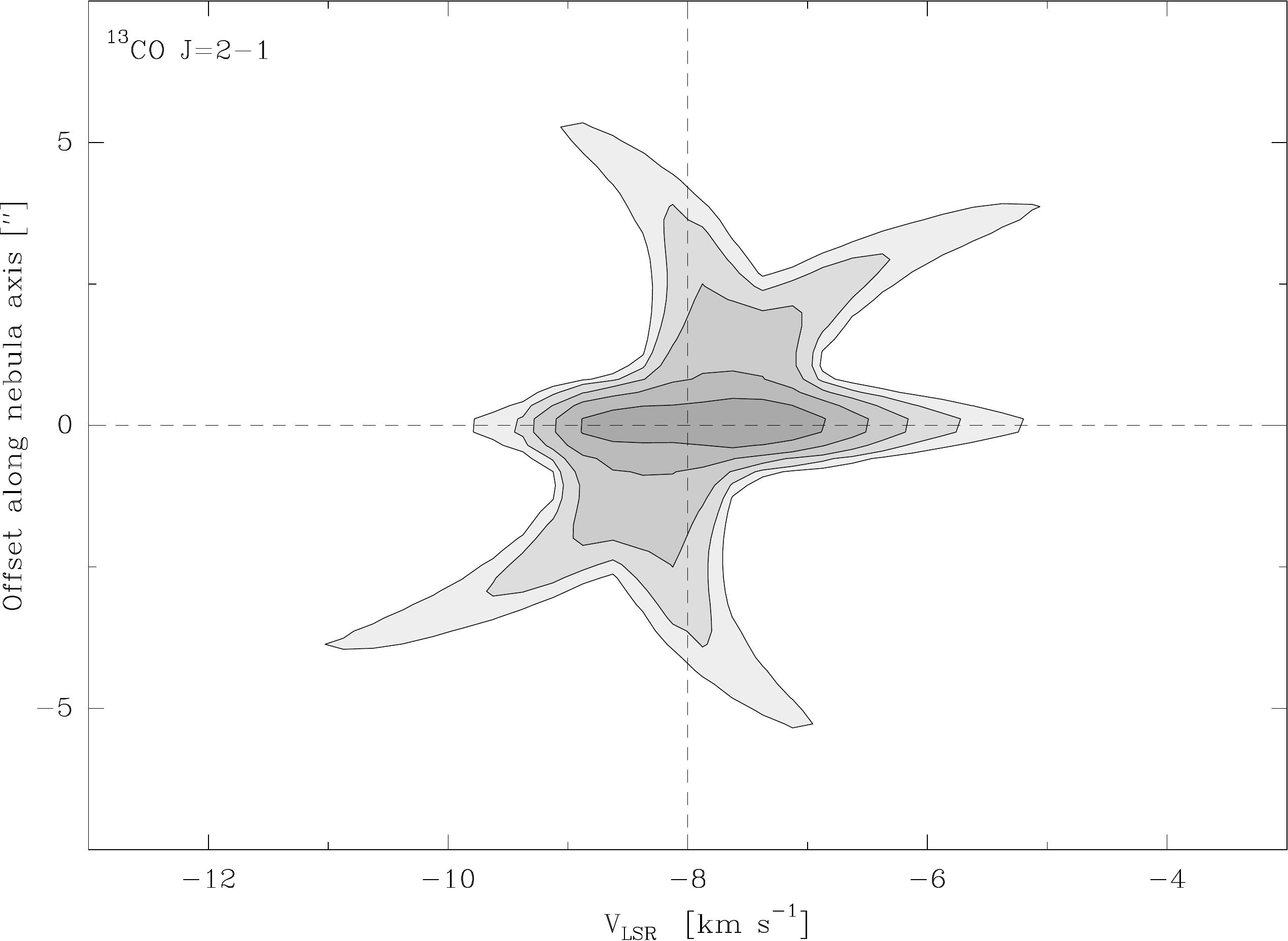}
        \end{minipage}
        \caption{\small Same as in \fig\ref{fig:89her12pvmodelo} but for \trece\dosunop. To be compared with \fig\ref{fig:89her13pv}, the scales and contours are the same.}
        \label{fig:89her13pvmodelo}
\end{figure*}

We adopted a nebula model (see \fig\ref{fig:89herdens} and \tab\ref{89hermodel}) based on reasonable assumptions and predictions and compatible with the observational data and other works of \on \citep[see \sect\ref{sec89herobs} and][]{bujarrabal2007}. We assumed the presence of a compact disk in the central regions of the nebula and a large hourglass-shaped wind. See a representation of our model nebula and predicted results in \figs \ref{fig:89her12pvmodelo}, \ref{fig:89her13pvmodelo}, and \ref{fig:89hermapasmodelo}.

The inclination of the nebular symmetry axis with respect to the line of sight is $15\degree$, with a $PA$ of $40\degree$ ($150\degree$ for the equatorial direction). These results are directly obtained from our observational data and modelling.
Position--velocity diagrams in \figs\ref{fig:89her12pv} and \ref{fig:89her13pv} suggest strong self-absorption effects at negative velocities (\sect\ref{sec89herobs}).

The rotation of the disk is assumed to be Keplerian (\eq\ref{eq:vdis_eq}) and is compatible with a central stellar mass of 1.7\msp. 
The density and temperature of the outflow are assumed to vary with the distance to the center of the nebula following potential laws; see \eqs\ref{eq:dens_eq} and \ref{eq:temp_eq}.
We find reliable results for density and temperature laws with high slope values. We also find expansion velocity in the outflow, according to \eq\ref{eq:vout_eq}.
We can see the predictions from these considerations and the model parameters in the synthetic velocity maps and PV diagrams in \figs\ref{fig:89her12pvmodelo}, \ref{fig:89her13pvmodelo}, and \ref{fig:89hermapasmodelo}.

Our model reproduces the NOEMA maps and yields a total mass for the nebula of $\sim$\,1.1\x\xd{-2}\ms with a disk mass of $\sim$\,60\%.
This standard model includes the clear missed flux of the interferometric process with respect to the single-dish flux (see Appendix\,\ref{comp_flujo}). This implies that the derived mass is a moderate lower limit, and because of the observed geometry could mostly apply to the extended outflow. To quantify this effect, we considered that the spatially extended emission that is filtered out by the interferometer partially fills the shells, as in the cases of outflows in young stellar sources \citep[e.g.,][]{gueth1996}. 
In this alternative model, the width of the outflow walls is increased by $\sim$\,70\% and yields a good fit of the \trece\dosuno single-dish profile in \fig\ref{fig:89her_flujo13co_mo_chetado}. 
With this, we derive a total mass of the nebula of $\sim$\,1.4\x\xd{-2}\msp, of which $\sim$\,6.4\x\xd{-3}\ms corresponds to the disk mass. We take these values as the most probable ones.
We do not develop this topic in more detail, because we are discussing parts of the outflow structure whose emission is not detected. In the future, we plan to perform on-the-fly observations using the 30\,m\,IRAM telescope and new observations with NOEMA.

We conclude that 41\,$-$\,53\% of the total nebular mass is contained in the outflow.
We consider the last value to be the most probable, because this result takes into account the lost flux, which very probably has its origin in the hourglass-shaped extended outflow.
Therefore, it is an intermediate case between disk- and outflow-dominated \pagb nebulae.


\subsection{\irasp}
\label{secirasmodel}


\begin{figure*}[h]
        \centering
        \begin{minipage}[b]{0.48\linewidth}
                \includegraphics[width=\sz\linewidth]{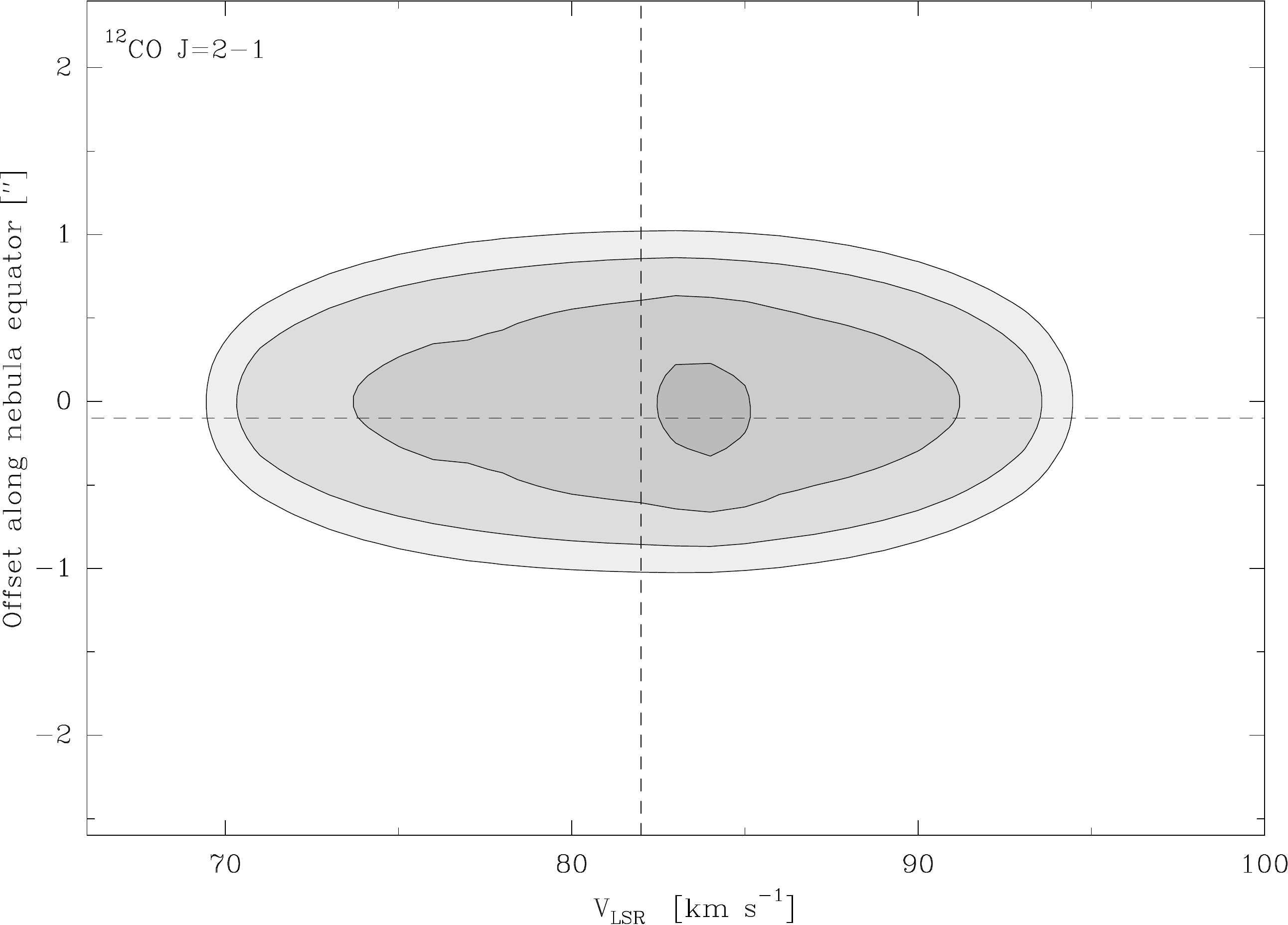}
        \end{minipage}
        \quad
        \begin{minipage}[b]{0.48\linewidth}
                \includegraphics[width=\sz\linewidth]{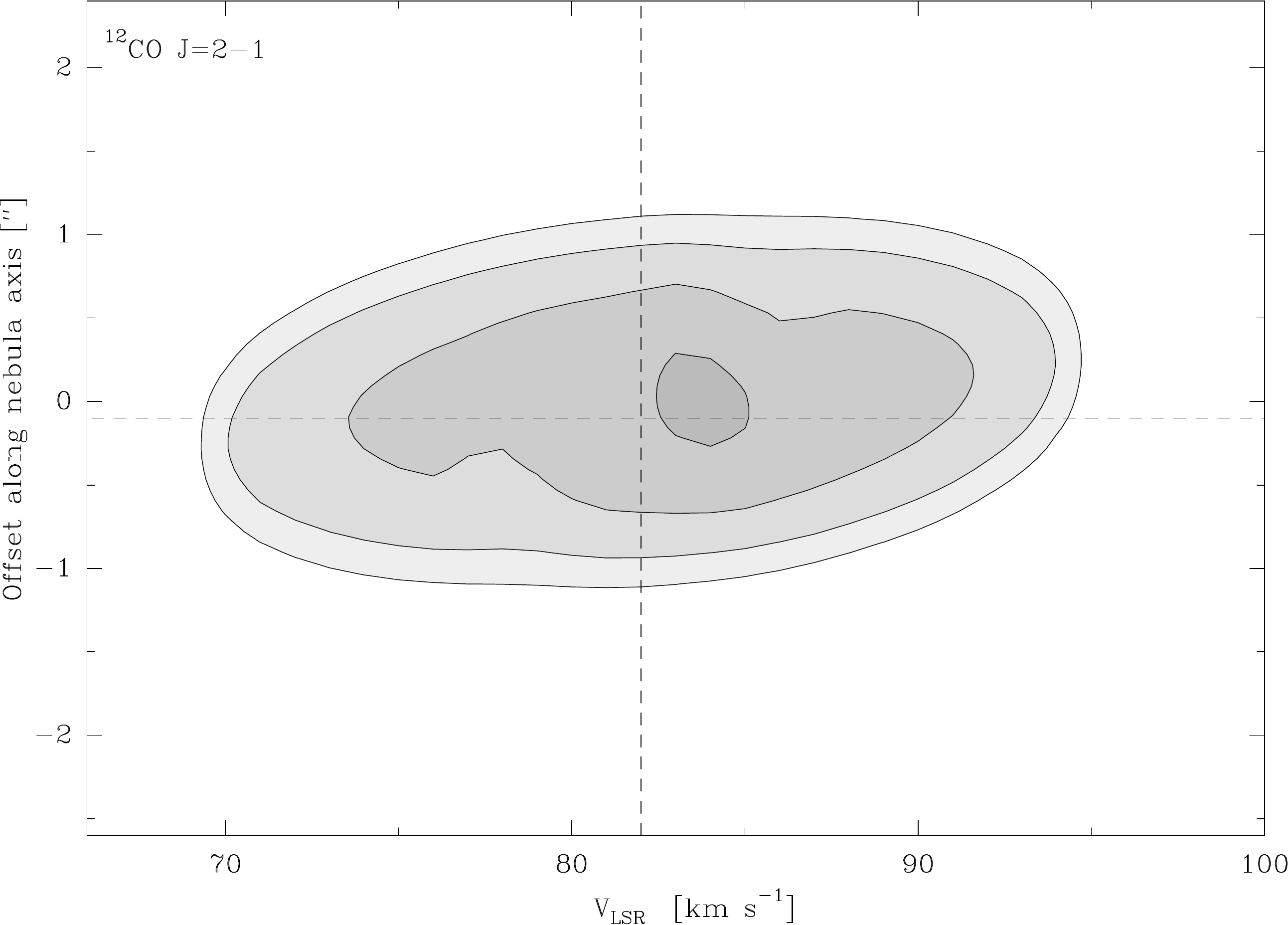}
        \end{minipage}
        \caption{\small \textit{Left:} Synthetic PV diagram from our best-fit model of \doce \dosuno in \iras along the direction $PA=-40\degree$. To be compared with the left panel of \fig\ref{fig:iras12pv}, the scales and contours are the same. \textit{Right}: Same as in \textit{Left} but along $PA=50\degree$.}
        \label{fig:iras12pvmodelo}
\end{figure*}

\begin{figure}[h]
    \centering
                \includegraphics[width=\sz\linewidth]{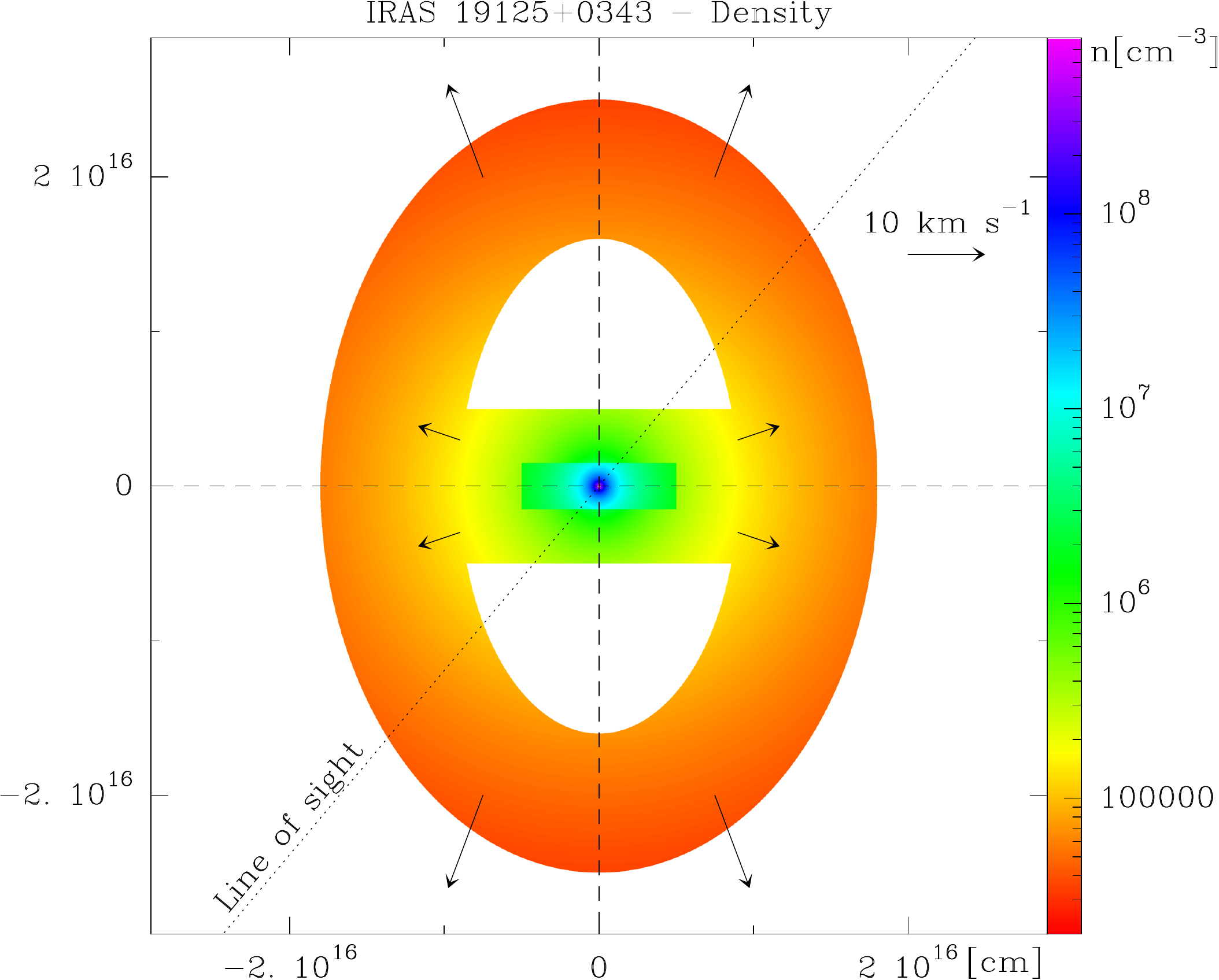}
    \caption{\small Structure and distribution of the density of our best-fit model for the disk and outflow of \irasp. The Keplerian disk presents density values $\geq$\xd{6}\,cm$^{-3}$. The expansion velocity is represented with arrows.}
        \label{fig:irasdens}
\end{figure}

We present our best nebula model for \iras in \fig\ref{fig:irasdens} and \tab\ref{irasmodel}. As the angular resolution is relatively poor (0\secp 7\x0\secp 7), we cannot resolve the disk, but we do know some features of the extended component. We propose a model consisting of a rotating disk and an outflow with cavities along the axis.
These cavities are parameterized with variables $h_{o}$ and $W_{o}$, where $h_{o}$ describes the height of the central region to the cavity and $W_{o}$ describes the width of the outflow walls.
This kind of shape is present in many pPNe (see \sect\ref{secrsctobs})

As the disk is not resolved, our main goal for this source is to study the outflow. For that purpose, the PV diagram along the nebula axis is the most relevant result.

The final nebula model yields results that are absolutely compatible with the observations, as we can see in \figs\ref{fig:iras12mapas} and \ref{fig:iras12pv}.
An inclination for the nebula axis with respect to the line of sight of $40\degree$ is compatible with the data.
We analyzed PV diagrams along different $PA$, and we find that the PV diagram along $PA=50\degree$ is the best to show the velocity position gradient characteristic of the expansion dynamics of the \pagb and pPNe outflows. This fact implies that the PV diagram along $PA=-40\degree$ should show the rotation of the equatorial disk (\figs \ref{fig:iras12pv} and \ref{fig:iras12pvmodelo}); due to the relatively low resolution and the confusion with the intense extended component, the presumed Keplerian rotation of the disk is not detectable. However, in view of the CO line profiles compatible with those of well-identified disks, we think that there must be a rotating disk surrounded by an outflow.

The density and temperature laws of the disk and outflow are assumed to vary with the distance to the center of the nebula following potential laws (see \eqs\ref{eq:dens_eq} and \ref{eq:temp_eq}). 
The velocity law of the disk is Keplerian (\eq\ref{eq:vdis_eq}) and compatible with a central stellar mass of 1.1\msp. In the  outflowing component, we assume radial velocity with a modulus linearly increasing with the distance to the center (\eq\ref{eq:vout_eq}).

We find a total mass value of 1.1\x\xd{-2}\ms of which 7.9\x\xd{-3}\ms corresponds to the outflow mass. This means that \iras has a Keplerian disk surrounded by an outflow, the mass of which constitutes $\sim$\,71\% of the total mass.


\begin{table}[h]
 \caption{Physical conditions in the molecular disk and outflow of \iras derived from the model fitting of the CO data.}
\small
\vspace{-5mm}
\begin{center}
\begin{tabular*}{\linewidth}{@{\extracolsep{\fill\quad}}lll}
\hline \hline
 \\[-2ex]
Parameter  & Disk    & Outflow \\ \hline
 \\[-2ex]
 
\multirow{2}{*}{Radius\,[cm]} & \multirow{2}{*}{ $5.0\times 10^{15}$ }  & $R_{max}=1.8\times 10^{16}$ \\
 \vspace{1mm}
 & & $W_{o}=9.0\times 10^{15}$ \\
 
\multirow{2}{*}{Height\,[cm]} & \multirow{2}{*}{ $1.5\times 10^{15}$ } & $2.5\times 10^{16}$ \\
 \vspace{1mm}
 &  & $h_{o}=5.0 \times 10^{15}$ \\
\multirow{2}{*}{Density\,[cm$^{-3}$]} & $n_{0}=1.0\times 10^{7}$ &  $n_{0}=2.0\times 10^{5}$ \\
 \vspace{1mm}
 & $\kappa_{n}=1.5$   & $\kappa_{n}=1.5$ \\

\multirow{2}{*}{Temperature\,[K]} & $T_{0}=100$ & $T_{0}=14$ \\
 \vspace{1mm}
 & $\kappa_{n}=0.5$ & $\kappa_{T}=0.4$  \\ 
 \vspace{1mm}
Rot.\,Vel.\,[km\,s$^{-1}$] & 1.2   & $-$ \\
 \vspace{1mm}
Exp.\,Vel.\,[km\,s$^{-1}$] & $-$  & 5.6  \\ 
 \vspace{1mm}
X(\docep) & 2\x10$^{-4}$  & 2\x10$^{-4}$\\
 \vspace{1mm}
\docep\,/\,\trecep & 10  & 10 \\
 \vspace{1mm}
Inclination\,[\degree] & & 40 \\

Position angle\,[\degree] & & 50 \\

\hline
 
\end{tabular*}

\end{center}\small
\vspace{-1mm}
\textbf{Notes.} Parameters and their values used in the best-fit model.
$R_{max}$ indicates the maximum radius of the  outflow. The density and temperature follow the potential laws of \eqs\ref{eq:dens_eq} and \ref{eq:temp_eq}. $V_{rot_{K_{0}}}$ and $V_{exp_{0}}$ are the values of the velocity of the disk and outflow at \xd{16}\,cm in \eqs\ref{eq:vdis_eq} and \ref{eq:vout_eq}. We show the inclination of the nebula symmetry axis with respect to the line of sight and the position angle of its projection on the plane of the sky.
\label{irasmodel}
\end{table}

\begin{figure*}[h]
        \centering
        \begin{minipage}[b]{0.48\linewidth}
                \includegraphics[width=\sz\linewidth]{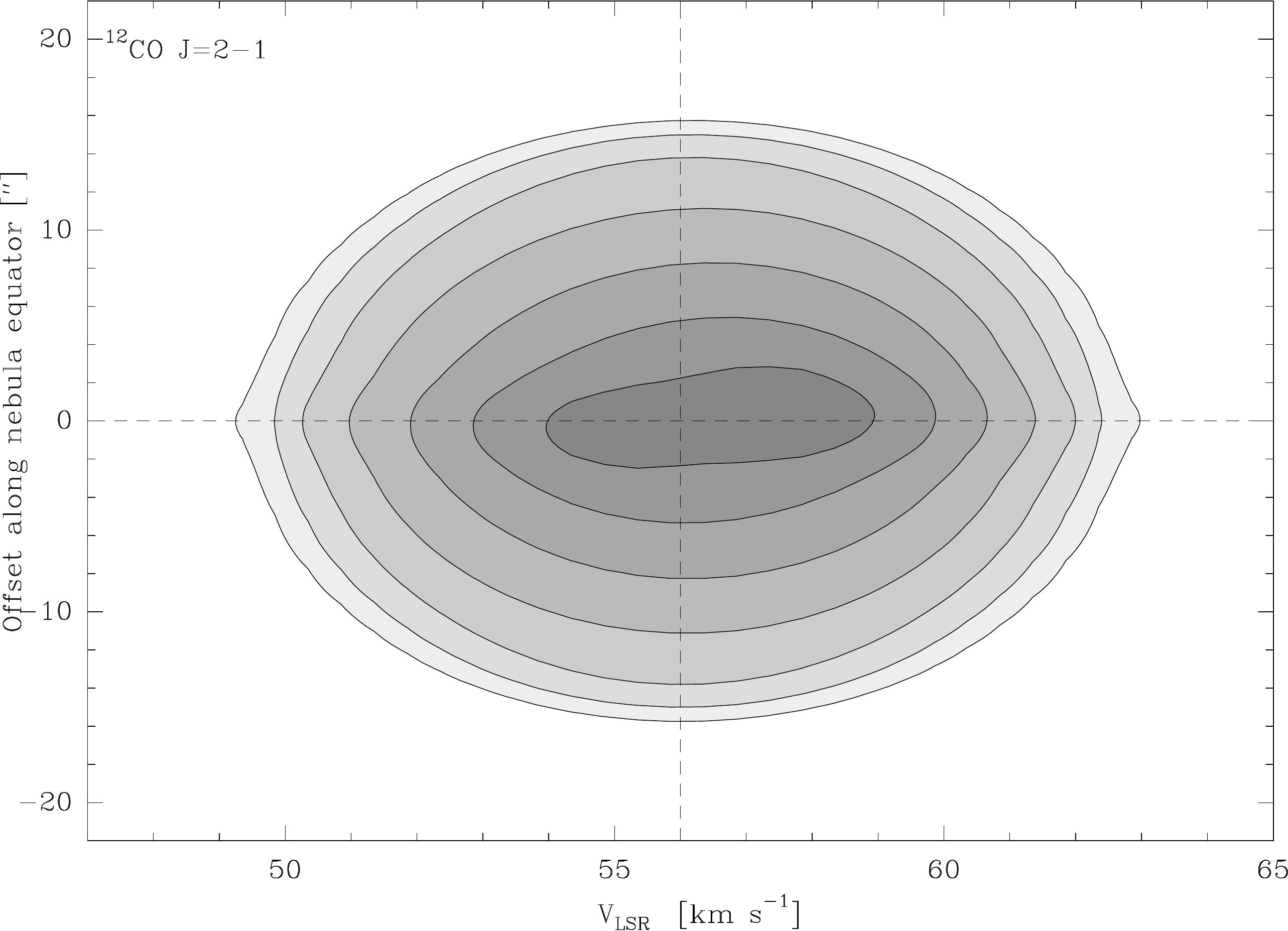}
        \end{minipage}
        \quad
        \begin{minipage}[b]{0.48\linewidth}
                \includegraphics[width=\sz\linewidth]{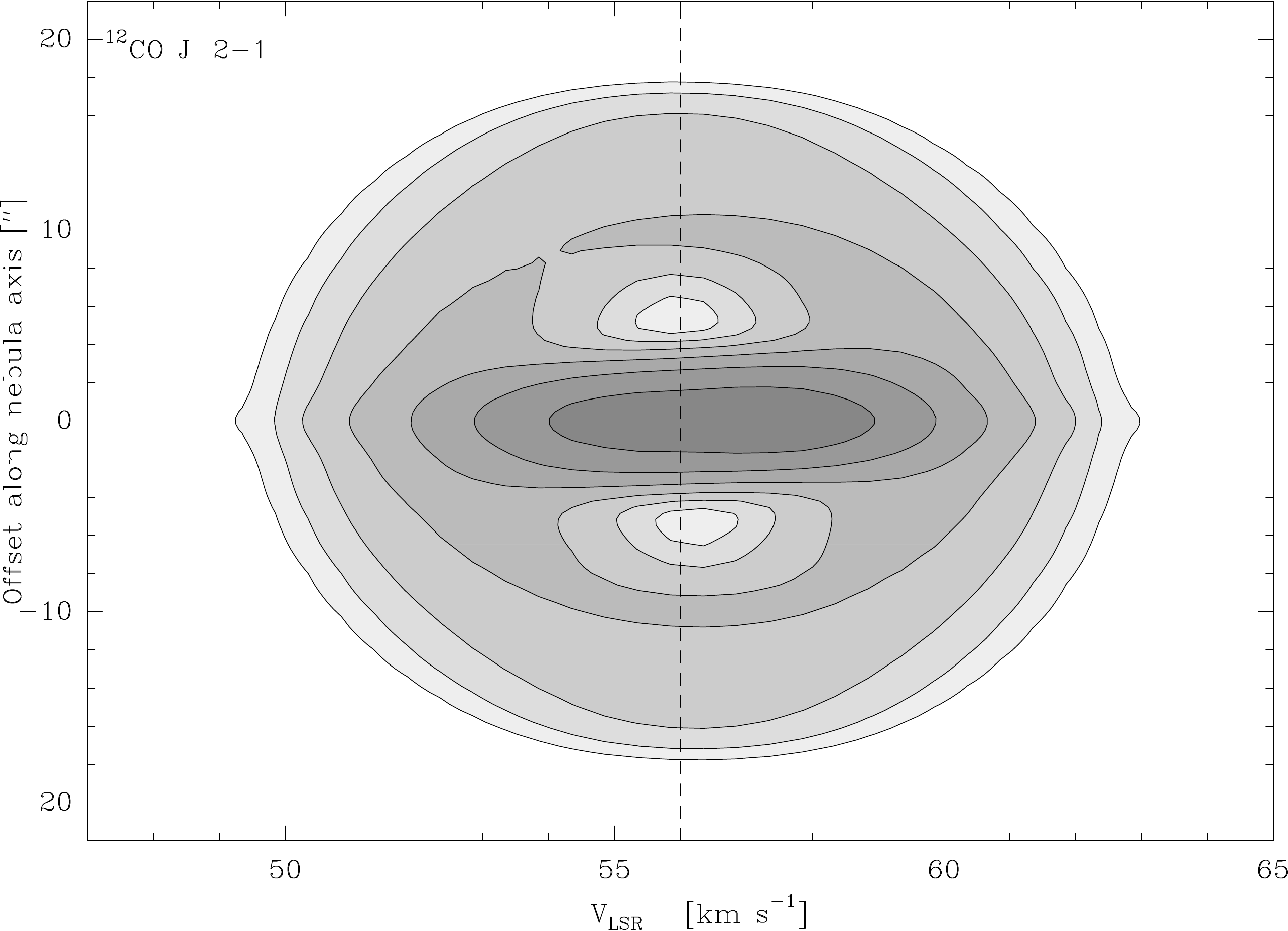}
        \end{minipage}
        \caption{\small \textit{Left:} Synthetic PV diagram from our best-fit model of \doce\dosuno in R\,Sct along the direction $PA=0\degree$. To be compared with the left panel of \fig\ref{fig:rsct12pv}, the scales and contours are the same. \textit{Right}: Same as in \textit{Left} but along $PA=90\degree$.}
        \label{fig:rsct12pvmodelo}
\end{figure*}


We checked that \iras cannot present an hourglass-shaped extended component similar to the one in \on (see \fig\ref{fig:89herdens}) because predictions are incompatible with the maps. However, we cannot exclude that cavities were smaller.

Accordingly, we propose an alternative model to the extended component of \irasp, in order to check the validity of our best model. We present an outflow without extended cavities in the nebula axis (see \fig\ref{fig:irasdensst}). 
We show the synthetic velocity maps (\fig\ref{fig:irasmapasmodelost}) and PV diagrams along the nebula axis and along the nebula equator (\fig\ref{fig:iras12pvmodelost}) from our alternative model. The predictions from this model are also compatible with the observational data of \sect\ref{secirasobs}. The shape of the third level of emission at $\sim$\,78\kms and $\sim$\,85\kms at $\sim$\,$\pm$\,0\secp 4 in the right panel of \fig\ref{fig:iras12pv} can only be reproduced after including cavities in the nebula model. We do not exclude models with small cavities or without them, but predictions of our alternative model are slightly worse than those from our best-fit model. This alternative model also serves to demonstrate our uncertainties.



\subsection{\rsp}
\label{secrsctmodel}

\begin{figure}[h]
    \centering
                \includegraphics[width=\sz\linewidth]{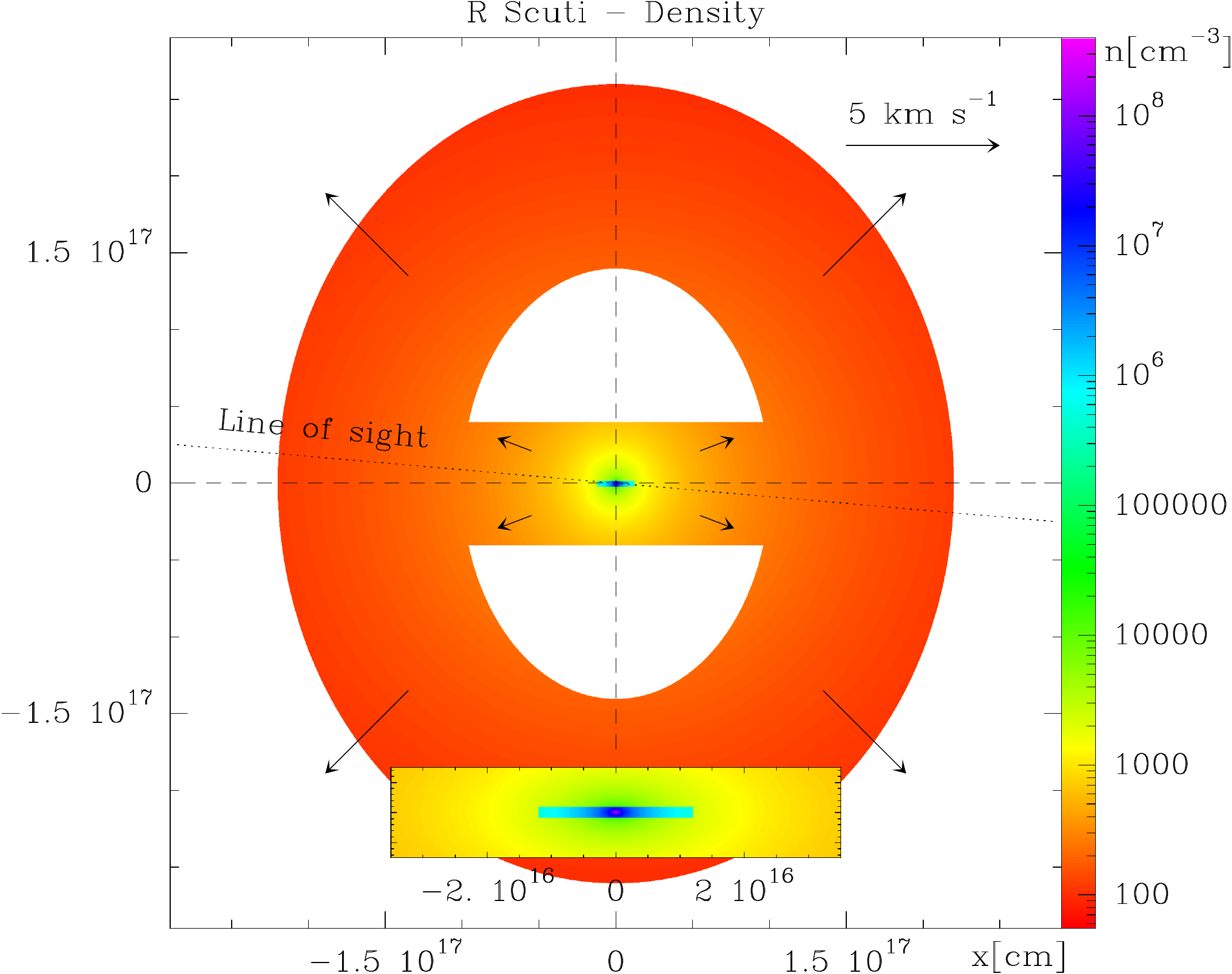}
    \caption{\small Structure and distribution of the density of our best-fit model for the disk and outflow of \rsp. The lower inset shows a zoom into the inner region of the nebula where the Keplerian disk presents density values $\geq$\xd{6}\,cm$^{-3}$. The expansion velocity is represented with arrows.}
        \label{fig:rsctdens}
\end{figure}

 Figure 18 and \tab\ref{rsctmodel} present our best-fit model for \rsp.
According to the PV diagram shown in the right panel of \fig\ref{fig:rsct12pv}, the structure of \rs is extended and contains two big cavities along the axis at about \mm10$''$ (see \fig\ref{fig:rsctdens}). Similar structures often appear in other similar objects (\sects \ref{irasmodel} and \ref{secrsctobs}).

The parameters that describe our best-fit model are shown in \tab\ref{rsctmodel}. We adopted a relative abundance with respect to the total number of particles of X(\docep)\,=\,\xd{-4} and X(\trecep)\,=\,2\x\xd{-5}. We find a low [$^{12}$C]\,/\,[$^{13}$C] abundance ratio for this source ($\sim$\,5), the same as that found by \citet{bujarrabal1990}.

An inclination for the nebula axis with respect to the line of sight of $85\degree$ is compatible with the data.
We can see the predicted PV diagrams along the equator and along the nebula axis in \fig\ref{fig:rsct12pvmodelo}. The agreement between the observations and predictions is reasonable. We note the good fitting of the self-absorption \citep[also present in the line profiles][see \sect\ref{secrsctobs}]{bujarrabal2013a}, but we stress that the central disk is barely detected due to the angular resolution (3\secp 12\x 2\secp 19).
In view of the satisfactory model fitting and following the arguments already presented in \sect\ref{secrsctobs} (see also \app\ref{dospicos}), we think that there is probably a rotating disk in the center of the \rs nebula.

The outflow cavities are also present in our predictions, as we can see in the velocity maps and in the PV diagram along the nebula axis (\fig\ref{fig:rsct12pvmodelo} \textit{Right}). The outflow shape is parameterized with variables $W_{o}$ and $h_{o}$, where $W_{o}$ describes the width of the outflow walls, and $h_{o}$ describes the height of the central region to the cavity (see \tab\ref{rsctmodel} and \fig\ref{fig:rsctdens}).

As we see, we predict a very diffuse and cold outflow. The predicted temperatures may appear very low, but as we see in \app\ref{trot_tcin}, the kinetic temperature ($T_{K}$) is expected to be higher than excitation temperature ($T_{ex}$) for these low densities because of the expected underpopulations of relevant levels. In reality, we expect $T_{K}$\,$\gsim$\,15\,K in the outer regions of \rsp.

The Keplerian velocity of the disk (\eq\ref{eq:vdis_eq}) is compatible with a central stellar mass of 1.7\msp. 
The total mass derived from our best-fit model is $\sim$\,3.2\x\xd{-2}\ms with a disk mass of $\sim$\,8.5\x\xd{-3}\msp. 

We considered other options for the shape of the outflow, for instance similar to the adopted model for \on (\sect\ref{sec89hermodel}). However, we ruled out this option because regardless of the orientation  we choose for the hourglass, the result is not consistent with the observational data. Significant changes in the size of the empty regions in the outflow lobes also lead to unacceptable predictions.

\begin{table}[h]
 \caption{Physical conditions in the molecular disk and outflow of \rs derived from the model fitting of the CO data.}
 \small
\vspace{-5mm}
\begin{center}

\begin{tabular*}{\linewidth}{@{\extracolsep{\fill\quad}}lll}
\hline \hline
 \\[-2ex]
Parameter  & Disk    & Outflow \\ \hline
 \\[-2ex]
\multirow{2}{*}{Radius\,[cm]} & \multirow{2}{*}{$1.1\times 10^{16}$}  & $R_{max}=2.2\times 10^{17}$ \\
 \vspace{1mm}
 & & $W_{o}=1.2\times 10^{17}$ \\
\multirow{2}{*}{Height\,[cm]} & \multirow{2}{*}{$1.9\times 10^{15}$} & $2.6\times 10^{17}$ \\
 \vspace{1mm}
 &  & $h_{o}=4.0 \times 10^{16}$ \\
\multirow{2}{*}{Density\,[cm$^{-3}$]} & $n_{0}=3\times 10^{6}$ &  $n_{0}=4\times 10^{2}$ \\
 \vspace{1mm}
 & $\kappa_{n}=1.5$   & $\kappa_{n}=1.0$ \\
\multirow{2}{*}{Temperature\,[K]} & $T_{0}=150$ & $T_{0}=7$ \\
 \vspace{1mm}
 & $\kappa_{n}=0.2$ & $\kappa_{T}=1.0$  \\
  \vspace{1mm}
Rot.\,Vel.\,[km\,s$^{-1}$] & 1.5   & $-$ \\
 \vspace{1mm}
Exp.\,Vel.\,[km\,s$^{-1}$] & $-$  & 0.2  \\ 
 \vspace{1mm}
X(\docep) & 1\x10$^{-4}$  & 1\x10$^{-4}$\\
 \vspace{1mm}
\docep\,/\,\trecep & 5  & 5 \\
 \vspace{1mm}
Inclination\,[\degree] & & 85 \\

Position angle\,[\degree] &  & 90 \\

\hline
 
\end{tabular*}

\end{center}
\small
\vspace{-1mm}
\textbf{Notes.} Parameters and their values used in the best-fit model.
$R_{max}$ indicates the maximum radius of the ellipsoidal outflow walls. $W_{o}$ is the width of the outflow walls.  The density and temperature follow the potential laws of \eqs\ref{eq:dens_eq} and \ref{eq:temp_eq}. $V_{rot_{K_{0}}}$ and $V_{exp_{0}}$ are the values of the velocity of the disk and outflow at \xd{16}\,cm in \eqs\ref{eq:vdis_eq} and \ref{eq:vout_eq}.
We show the inclination of the nebula symmetry axis with respect to the line of sight and the position angle of its projection on the plane of the sky.
\label{rsctmodel}
\end{table}


\subsection{Uncertainties of the model parameters}
\label{incertidumbres}

Due to the relatively low angular resolution of the observational data, some of the nebula properties are not precisely determined. In particular, owing to the insufficient angular resolution and the inclination of the disk with respect to the plane of the sky, the height of the rotating disk is not well determined in all cases,  the angular resolution being the basic estimate of the uncertainty for these parameters. On the contrary, the radius of the disk is better measured. This parameter is derived from the observational data and is confirmed with the best-fit model. The width of the disk (twice the height) is limited by the angular resolution and, often more restrictively, by the disk radius, because the disk width \textcolor[rgb]{0,0,0}{in real cases is} significantly smaller than its equatorial size. For instance, we know that the disk width of \on must be $\lsim$\,3\x\xd{15}\,cm (equivalent to 0\secp 2, smaller than the angular resolution).

The structure of the outflow is well determined in the case of \onp, because the hourglass-shaped structure is absolutely clear in the observational data. In the case of \rsp, its extended outflow structure with cavities is clear according to the velocity maps.
The maps of \iras do not allow us to determine the shape of the outflow  in detail, and only the extent along the symmetry axis is actually measured. We present a simple model to explain the observations of this source, as well as an alternative model similar to those used in \rs and many pPNe which is also compatible with the data and serves to demonstrate the uncertainties in the structure of \irasp. The case of the outflow of \ac is complex. We detect weak emission of the extended component in the PV diagrams, but the shape of the outflow is not well determined. For this reason we assume properties and shape similar to those of the Red\,Rectangle as they are similar objects. These assumptions clearly increase the uncertainty of our results and we can only give an approximate upper limit to the mass of the outflow.

Other uncertainties in the modeling also affect our main results, and particularly crucial for our discussion is the determination of the mass of the various components. From a comparison of several alternative models, we see that the assumed nebular gas distribution affects the derived mass values  only slightly. The derived density distribution and mass are inversely proportional to the assumed CO abundances. The values we adopted, X(\docep)\,$\sim$\,1\,$-$\,2\x\xd{-4} and X(\trecep)\,$\sim$\,2\x\xd{-5}, have been extensively studied in previous works on those \pagb sources \citep{bujarrabal1990,bujarrabal2013a,bujarrabal2016} and are accurate to within less than a factor of two. The dependence of the mass on the gas temperature is lower, less than proportional \citep[see detailed discussion in][]{bujarrabal2013a}. Our temperature distributions are again very similar to those usually found in these objects and are constrained by the observed brightness in optically thick areas (where the brightness is almost equal to the gas temperature after correcting for dilution within the beam). The effects of the temperature on the mass determination uncertainty are therefore a minor contribution.

The distances of these objects are very uncertain and affect other parameters of the models. The size of the nebula determined by the model scales linearly with distance $d$. The density varies with $d^{-1}$ because the column density must be conserved to yield the same optical depth in all lines of sight. A change of distance or density implies that the volume of the nebula also changes, which implies that the mass of the nebula varies. The mass scales with $d^{2}$.
According to the model, the typical values of temperature $T$ and velocity field are not affected by a change in the distance. These dependencies on the distance are those usually encountered when modeling nebulae.

\begin{table*}[t]
\caption{Mass and size values for the disk and outflow of our \pagb stars sample.}
\vspace{-5mm}
\begin{center}
\setlength{\tabcolsep}{0pt} 

\begin{tabular*}{\textwidth}{@{\extracolsep{\fill}\quad}lccccccc}


\hline \hline
\noalign{\smallskip}
 \\[-2ex]
\multirow{2}{*}{Source} & Total mass & Disk mass  & Outflow mass  & $\frac{Disk}{Total}$ & $\frac{Outflow}{Total}$ & Radius disk & Outflow size \\ 
 & [M$_{\odot}$]  & [M$_{\odot}$] & [M$_{\odot}$] & [\%]  & [\%] & [cm] & [cm] \\
\hline
\\[-2ex]
\vspace{1mm} 
AC\,Herculis & 8.3\x\xd{-4} & 8.1\x\xd{-4} & $\lsim$\,2.0\x\xd{-5} & $\gsim$\,95 & $\lsim$\,5 & 1.4\x\xd{16} & $-$\\
\vspace{1mm} 
89\,Herculis & 1.4\x\xd{-2} & 6.4\x\xd{-3} & 7.1\x\xd{-3}  & 48 & 53 & $\leq$\,6.0\x\xd{15} & 1.5\x\xd{17} \\
\vspace{1mm} 
\irasp  & 1.1\x\xd{-2} & 3.3\x\xd{-3}  & 7.9\x\xd{-3} & 29 & 71 & $\leq$\,6.0\x\xd{15} & 5.0\x\xd{16} \\

R\,Scuti & 3.2\x\xd{-2} & 8.5\x\xd{-3} & 2.4\x\xd{-2} &  27 & 73 & $\leq$\,1.5\x\xd{16} & 5.2\x\xd{17} \\

\hline
 
\end{tabular*}
 \\[1ex]

\end{center}
\small
\vspace{-1mm}
\textbf{Notes.} Masses calculated for the disk and outflow. These results have been calculated based on the distances of \tab\ref{prop}. Together with the masses, we show the value of the Keplerian disk radius and the outflow size in the axial direction.
\label{tablamasas}
\end{table*}

\section{Conclusions}
 \label{conclusiones}

We present interferometric NOEMA maps of \doce and \trece \dosuno in 89\,Her and \doce \dosuno in \acp, \irasp, and \rsp. These objects belong to the binary \pagb stars sample with \nirp,  and are thought to be surrounded by rotating disks and outflowing disk winds that show axial and equatorial symmetry (see \sect\ref{introduccion}). We carefully modeled our observations and find that they are always compatible with this paradigm.
The masses and sizes derived from our model fitting are given in \tab\ref{tablamasas}.

\begin{itemize}[label={$\cdot$},leftmargin=*,topsep=2.5pt]\itemsep0.5em

\item \acp: Our maps are more sensitive than previously published ones \citep{bujarrabal2015}. The goal of this work is to study the presence of an extended outflowing component whose presence is not obvious in the observation.
The PV diagram along  $PA = 136.1 \degree$ (nebula equator) is exactly coincident with Keplerian dynamics.
To study the presence of the outflow we analyzed the observational data in detail and tentatively detect its emission: $\lsim$\,10\,mJy\,beam$^{-1}$\,km\,s$^{-1}$.
The disk radius is 1.4\x\xd{16}\,cm. We find densities between \xd{5} and \xd{7}\,cm$^{-3}$ and temperatures between 20 and 200\,K. The rotational velocity field is purely Keplerian in the disk and is compatible with a central (total) stellar mass of $\sim$\,1\msp.
The mass of the disk is 8.1\x\xd{-4}\msp, the same value as in previous works: $\sim$\,8\x\xd{-4}\ms from single-dish observations \citep{bujarrabal2013a} with $d=1100$\,pc (same as in this work); 1.5\x\xd{-3}\ms from mm-wave interferometric observations \citep{bujarrabal2015} with $d=1600$\,pc (which implies $\sim$\,8\x\xd{-4}\ms rescaling to the distance of this work).
We modeled an extended structure that surrounds the disk, assuming a structure similar to the one found in the Red\,Rectangle. We find that the mass of the outflow must be $\lsim$\,5\% of the total mass.
We conclude that \ac is clearly a binary \pagb star surrounded by a disk-dominated nebula, because $\gsim$\,95\% of the total mass corresponds to the disk.
However, even more sensitive maps will be necessary to confirm our model of the low-mass and extended component of \acp.

\item \onp: We can see an extended hourglass-like structure in the velocity maps and PV diagrams. 
The detected outflow emission corresponds to a size of 1.5\x\xd{17}\,cm.
We can also see a central clump of strong emission with relatively low dispersion of velocity, which we think that a Keplerian disk must be responsible for. Due to the limited spatial resolution, we cannot resolve the inner structure of the disk.
We find that the line profile from the most central compact component, which arises from the unresolved Keplerian disk and very inner outflow, shows the characteristic double-peak shape of rotating disks (see \sect\ref{sec89herobs} and \app\ref{dospicos}).
The PV diagram along the nebula equator, $PA \sim 150\degree$, is compatible with characteristics of Keplerian dynamics, and we derive the main properties of the rotating disk directly from the model fitting. We find that the disk radius must be $\leq$\,6\x\xd{15}\,cm and Keplerian rotation that is compatible with a central stellar mass of 1.7\msp.
The nebula of \on contains a total mass of 1.4\x\xd{-2}\msp, of which the outflow is responsible for 41\,$-$\,53\% (see \sect\ref{sec89hermodel}).

\item \irasp: We cannot resolve the disk structure in our data. We developed a model with a rotating disk with Keplerian dynamics surrounded by an outflow $\sim$\,5.0\x\xd{16}\,cm in size in the axial direction, which is completely consistent with the observations. We find that the disk radius must be $\leq$\,6.0\x\xd{15}\,cm.
The Keplerian rotation of \iras is compatible with a central stellar mass of 1.1\msp.
However, we note that the disk structure and dynamics are particularly difficult to study from our observations and that the results on this object are less reliable; the existence of an inner rotating disk is mostly based on our experience in the analysis of single-dish profiles and the satisfactory model fitting (see \sect\ref{secirasmodel}). Observations with higher resolution are required in order to confirm these results.
The total mass of the nebula of \iras is 1.1\x\xd{-2}\ms (1.3\x\xd{-2}\ms in the alternative model), of which 3.3\x\xd{-3}\ms corresponds to the Keplerian disk mass, meaning that $\sim$\,71\% ($\sim$\,74\%) of the total mass conforms the outflow escaping from the Keplerian disk.
Therefore, \iras is an outflow-dominated \pagb nebula.

\item \rsp: Velocity maps and PV diagrams show strong emission from an inner region, which has a relatively low velocity dispersion. This central clump is not resolved in our maps and the interpretation of the central emission in this source is less reliable than for our other sources; see \sect\ref{secrsctobs}. Nevertheless, we think that this condensation is probably an unresolved rotating disk, as also argued in \sect\ref{secrsctobs}, according to the velocity dispersion and slight redshift, comparable to those observed in similar sources, and that our models easily reproduce its observational properties. 
In addition, we find a line profile coming from the very central region of the nebula that shows the characteristic double peak shape of rotating disks (see \sect\ref{secrsctobs} and \app\ref{dospicos}).
The nature of \rs is not yet clear (\sect\ref{secrsctprev}), but our observations strongly suggest that \rs is also a \pagb star surrounded by a Keplerian disk (and by a high-mass extended outflow). Therefore, we suspect a binary nature for the central star of \rsp.
We present the model composed of a compact rotating disk with Keplerian dynamics and an outflow. The outflow shows two cavities that are clearly identified in the observations and are very extended, $\sim$\,5.2\x\xd{17}\,cm.
The main properties of the disk derived from the model are necessarily uncertain. The disk radius is $\leq$\,1.5\x\xd{16}\,cm and the Keplerian rotation is compatible with a central stellar mass of 1.7\msp.
The total mass of the nebula is $\sim$\,3.2\x\xd{-2}\msp. Taking into account that the disk represents 26.5\% of the total mass and the large size of the outflow (compared with other similar objects), it is clear that \rs is an outflow-dominated \pagb nebula.
Although the presented results are consistent with the observational data, it would be necessary to observe \rs with higher resolution to resolve the disk and firmly conclude on the nature of the source.

\end{itemize}

\subsection*{General conclusions}

Based on the presented results for \onp, \irasp, and \rsp, we conclude that there is a new subclass of binary \pagb stars with \nirp: the outflow-dominated nebulae. These present massive outflows, even more massive than their disks. These outflows are mostly composed of cold gas. There are other single-dish observed sources, such as AI\,CMi and IRAS\,20056+1834, that present narrow CO line profiles with strong wings. These could also belong to this new subclass.

We also observed \acp, where the outflow is barely detected. We estimate an upper limit to its mass of $\lsim$\,5\% of the total mass, which is even smaller than those of the disk-dominated subclass (the Red\,Rectangle, IW\,Carinae, and IRAS\,08544$-$4431).

\on is an intermediate source in between disk-dominated and outflow-dominated nebula.

\begin{acknowledgements}
We are grateful to the anonymous referee for the relevant recommendations and comments.
This work is based on observations of IRAM telescopes. IRAM is supported by INSU/CNRS (France), MPG (Germany), and IGN (Spain). 
This work is part of the AxiN and EVENTs\,/\,NEBULEA\,WEB research programs supported by Spanish AEI grants AYA\,2016-78994-P and PID2019-105203GB-C21.
IGC acknowledges Spanish MICIN the funding support of BES2017-080616.

\end{acknowledgements}

\bibliographystyle{aa}
\bibliography{referencias.bib}


\appendix

\section{Flux comparison}
\label{comp_flujo}

\subsection{NOEMA vs. 30\,m}

\begin{figure}[h]
\includegraphics[width=\sz\linewidth]{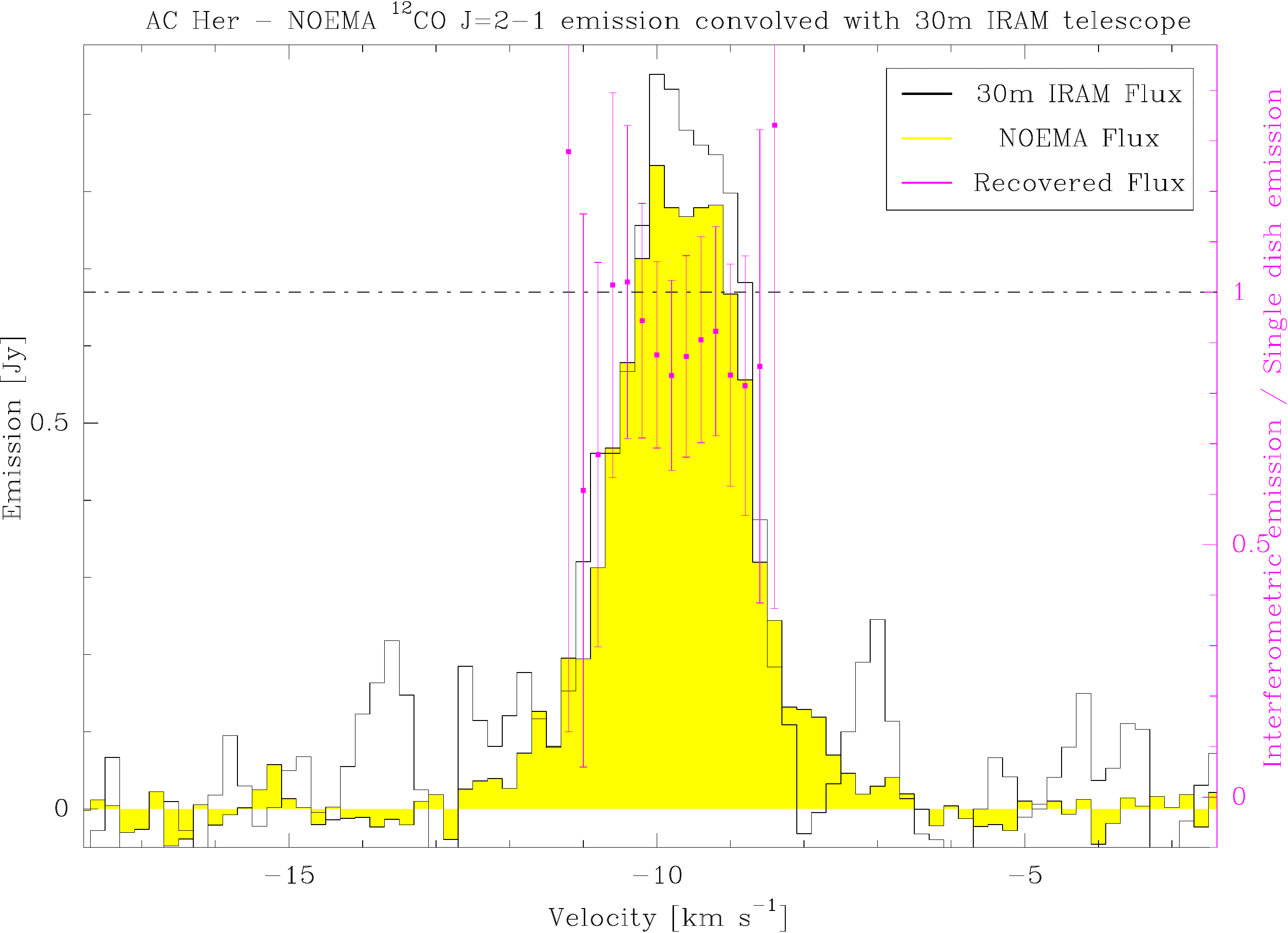}
\caption{\small Comparison of the NOEMA flux (yellow histogram) of \doce\dosuno with 30\,m observation (white histogram) for \acp. The points show the recovered flux across velocity channels (right vertical axis), with error bars including uncertainties in the calibration of both instruments.}
    \label{fig:acher_flujo12co}  
\end{figure}

\begin{figure}[h]
\includegraphics[width=\sz\linewidth]{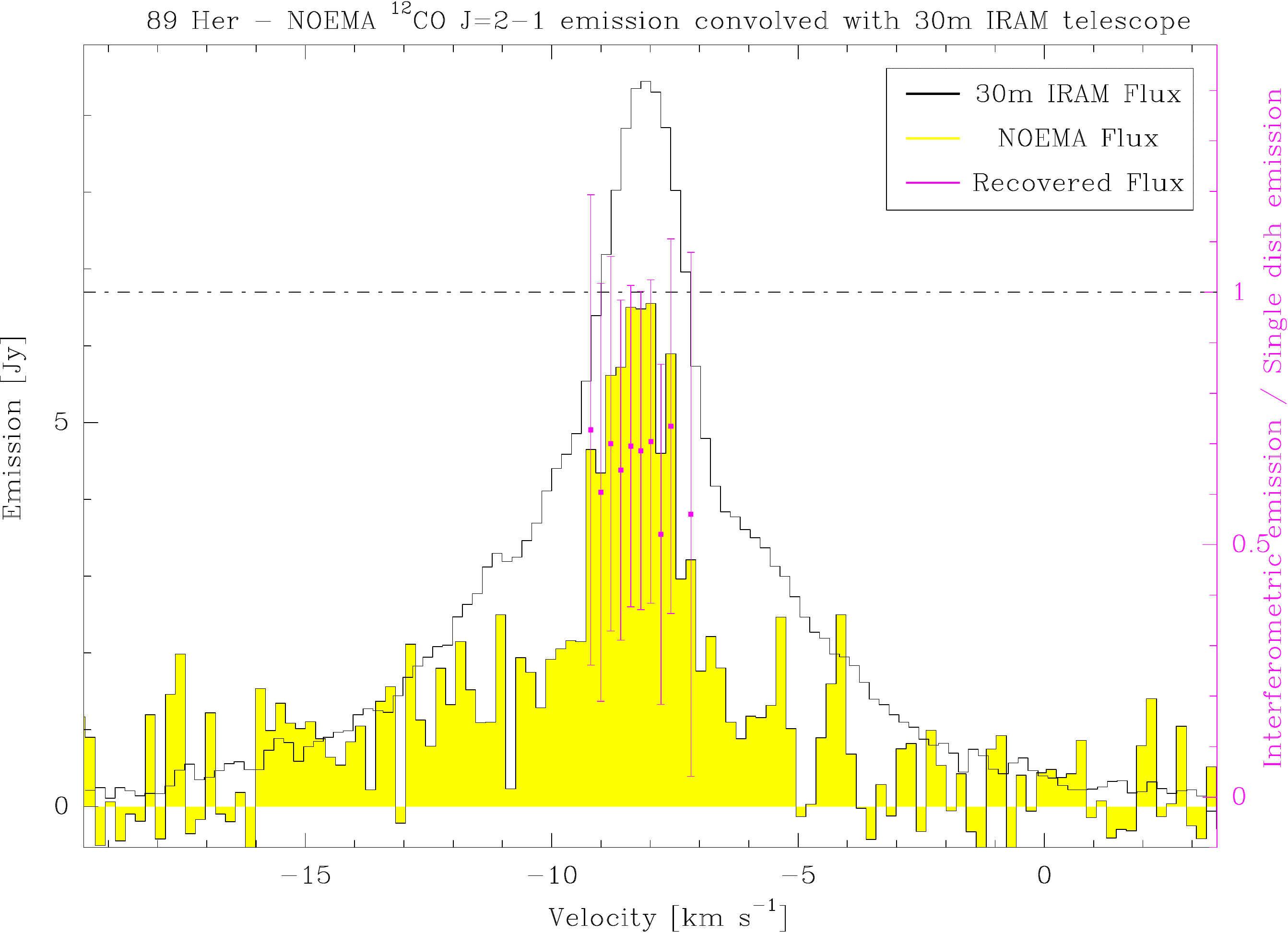}
\caption{\small Comparison of the NOEMA flux (yellow histogram) of \doce\dosuno with 30\,m observation (white histogram) for \onp. The points show the recovered flux across velocity channels (right vertical axis), with error bars including uncertainties in the calibration of both instruments.}
    \label{fig:89her_flujo12co}  
\end{figure}

\begin{figure}[h]
\includegraphics[width=\sz\linewidth]{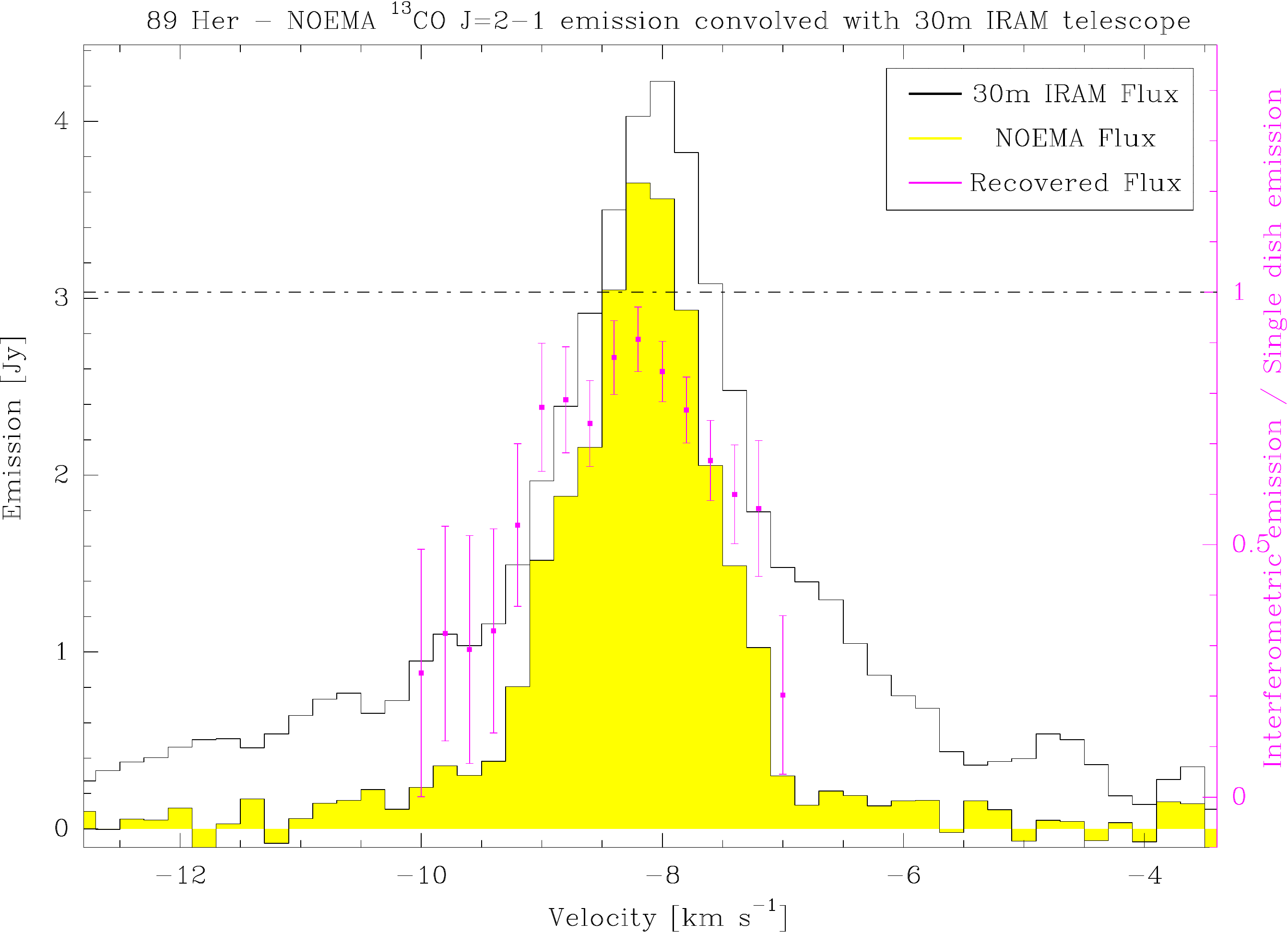}
\caption{\small Comparison of the NOEMA flux (yellow histogram) of \trece\dosuno with 30\,m observation (white histogram) for \onp. The points show the recovered flux across velocity channels (right vertical axis), with error bars including uncertainties in the calibration of both instruments.}
    \label{fig:89her_flujo13co}  
\end{figure}

Here, we show the comparison between the interferometric flux and the single-dish integrated flux \citep[data taken from][]{bujarrabal2013a}. In the case of \acp, no significant amount of flux was filtered out in the interferometric data (\fig\ref{fig:acher_flujo12co}). The flux loss in \doce\dosuno is $\sim$\,30\% and $\sim$\,50\% in the wings of \trece\dosuno for \on (\figs\ref{fig:89her_flujo12co} and \ref{fig:89her_flujo13co}, respectively). 
In the case of \irasp, the interferometric visibilities were merged with zero-spacing data obtained with the 30\,m\,IRAM telescope, which guarantees that there is not lost flux in final maps.
There is no flux loss for \rsp, because our NOEMA observations were merged with large single-dish maps.

\subsection{Flux loss in \onp: implications for the mass value}
\label{89her13co_masa} 

\begin{figure}[h]
\includegraphics[width=\sz\linewidth]{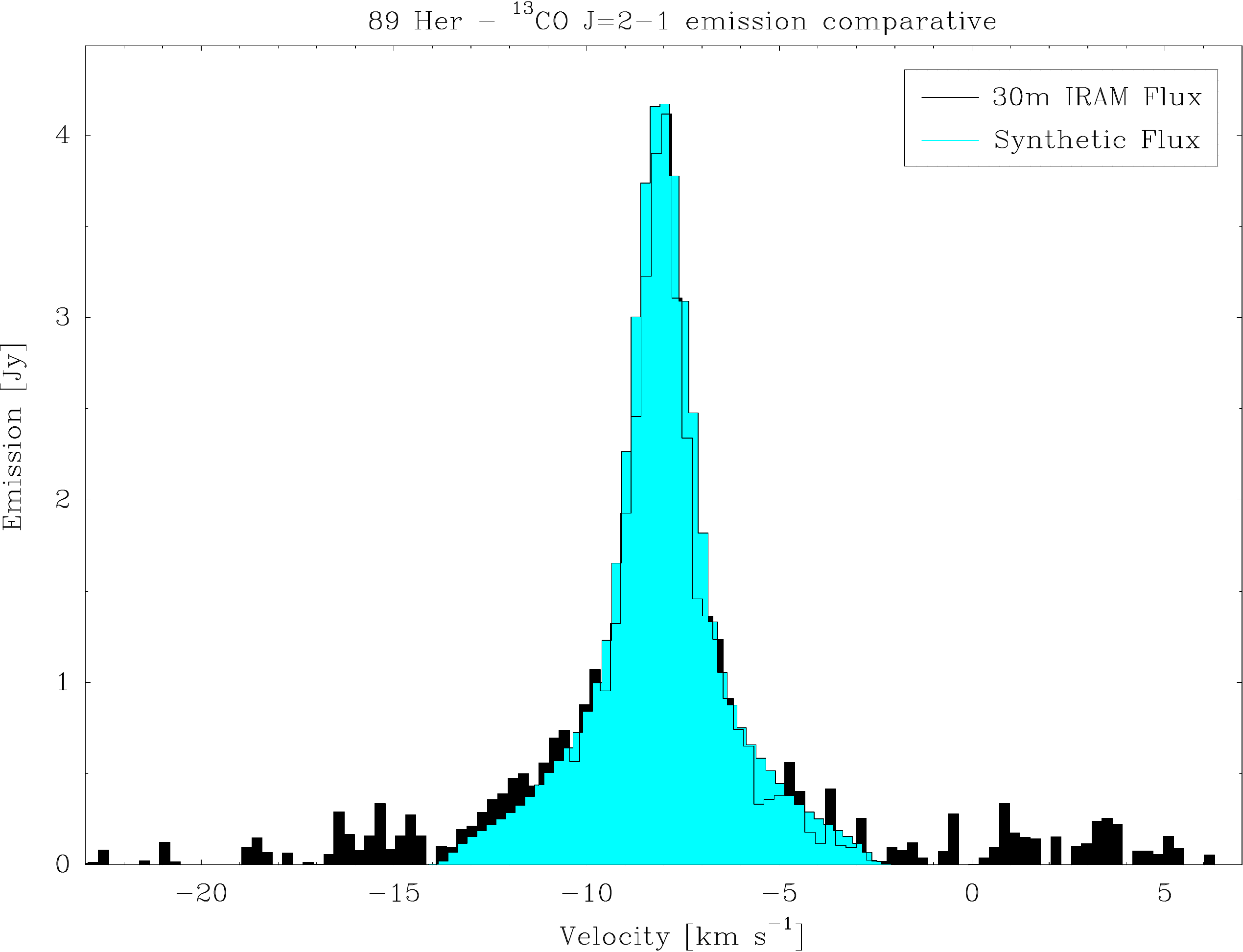}
\caption{\small Comparison between the 30\,m\,IRAM \trece\dosuno line profile (black histogram) and the synthetic one (cyan histogram). The synthetic line profile corresponds to the model after increasing the width of the hourglass of the standard model.}
    \label{fig:89her_flujo13co_mo_chetado}  
\end{figure}

In this Appendix, we present the synthetic \trece\dosuno emission convolved with the 30\,m\,IRAM telescope for \on (see \fig\ref{fig:89her_flujo13co_mo_chetado}) compared with the observational data \citep[obtained from][]{bujarrabal2013a}.
The flux derived of the standard model (see \tab\ref{89hermodel}) is undervalued. We focus our analysis on \trece\dosunop, which is a transition that is more sensitive to density than \doce\dosunop.
After increasing the width of the outflow walls of our standard model by $\sim$\,70\%, we see that we can recover the lost flux. Therefore, we conclude that the mass derived from this modified model, 1.4\x\xd{-2}\msp, is realistic (see \sect\ref{sec89hermodel} and \tab\ref{89hermodel}).

\section{Analyzing the inner region of the nebulae}
\label{dospicos}
\begin{figure}[h]
\center
\includegraphics[width=\sz\linewidth]{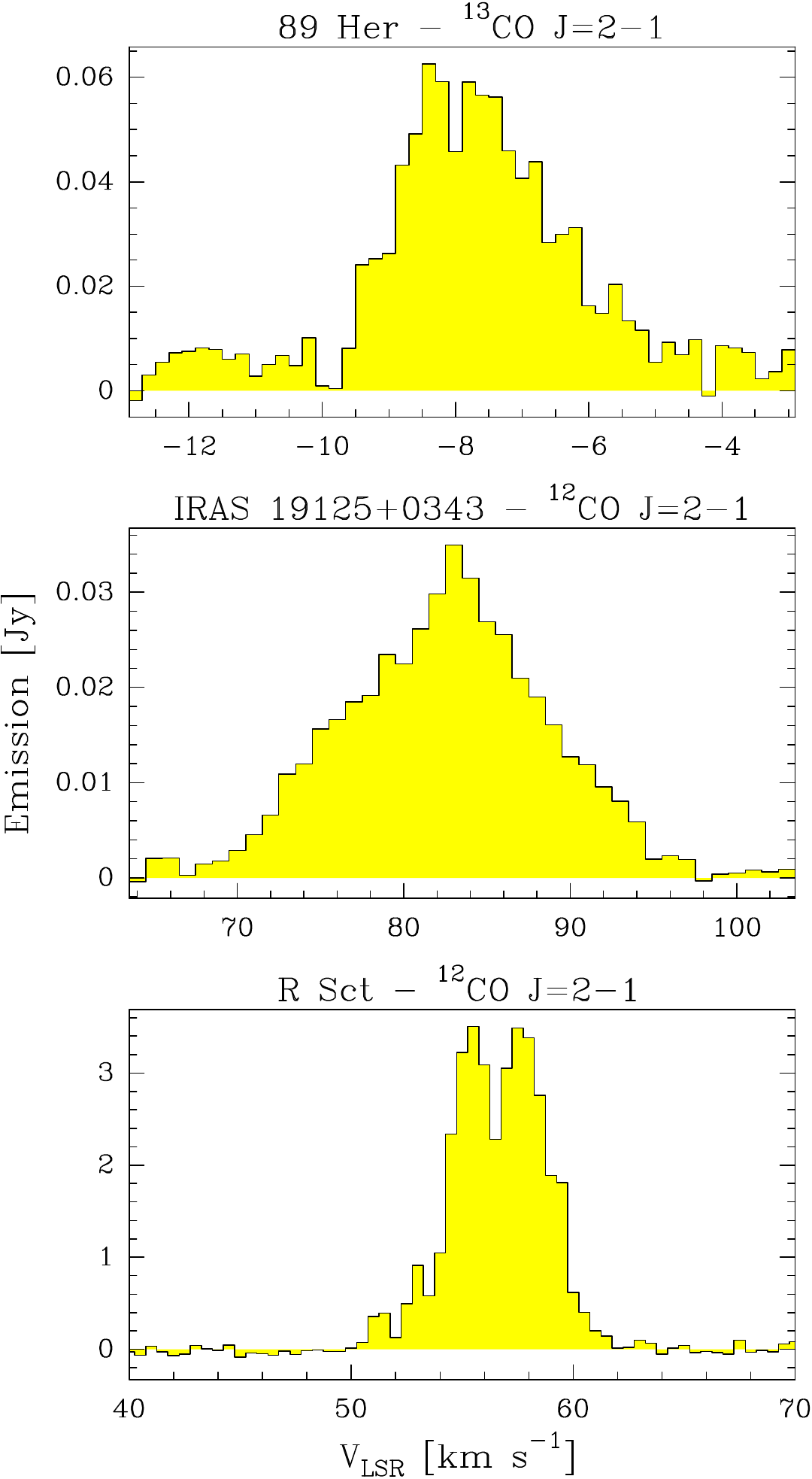}
\caption{\small Observational line profiles of \on (\textit{Top}), \iras (\textit{Middle}), and \rs (\textit{Bottom}) from the innermost region of each nebulae. See main text for details.}
    \label{fig:espectros_dospicos}  
\end{figure}

In this Appendix, we show the observational line profile from the central component of the nebula of \onp, \irasp, and \rs (see \fig\ref{fig:espectros_dospicos}). 
The lines of \on and \iras come from maps in which longer baselines are favored. The line of \rs come from interferometric maps without the 30\,m\,IRAM telescope contribution. 
In addition, all these selected spectra come from the most inner region because we select a central region with the size of the beam.
For these reasons, only a fraction of the total flux can be seen.
We tentatively see the double peak line profile in the case of \on and \rsp. 
We must clarify that it is impossible to distinguish the disk emission from emission and absorption of the innermost and dense outflow gas, which limits the reach of our discussion.
In the case of \iras we cannot see the double-peak shape in the line profile. We think that the inclination of the disk with the line of sight (130$\degree$, see \sect\ref{secirasmodel}) must be a key factor to explain the absence of that double peak in this case because this profile shape tends to disappear for face-on disks. We remind the reader that \acp, for instance, clearly presents a disk with Keplerian rotation and there is no sign of that effect in its single-dish CO line profiles \citep[see][]{bujarrabal2013a}.

\section{Detailed analysis of \ac maps}
\label{outflow_acher}

\begin{figure*}[h]
\centering
\includegraphics[width=\sz\textwidth]{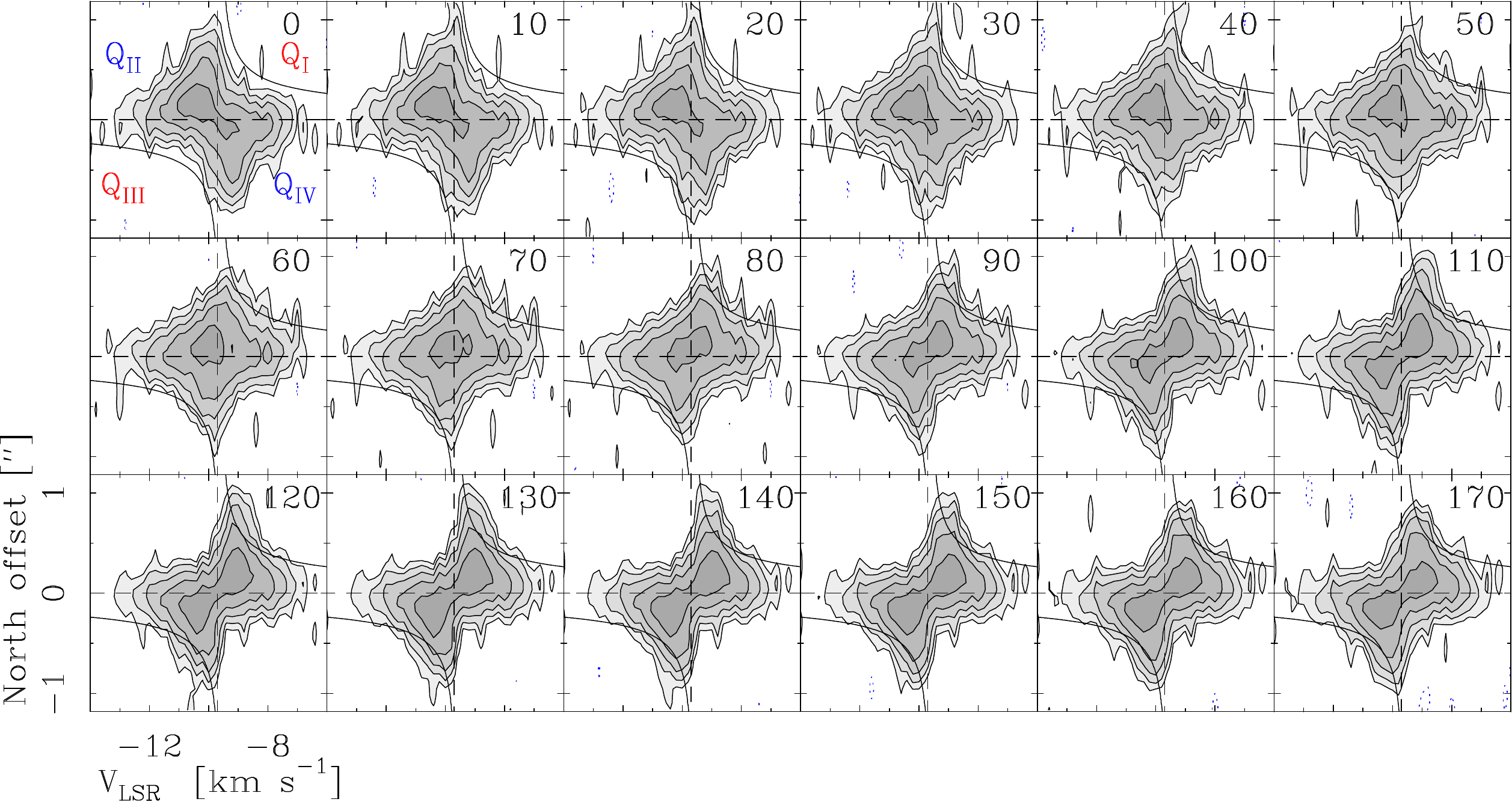}
\caption{\small \ac position--velocity diagrams found along different position angles from 0\degree\ to 170\degree\ with a step of 10\degree. The $PA$ is indicated in each panel in the top right corner. The name of each quadrant is indicated in the  panel showing the first velocity channel. Contours are the same as in the corresponding channel maps (\fig\ref{fig:ac12mapas}). To help in the identification of the Keplerian dynamics, we show hyperbolic functions in each panel.}
    \label{fig:cortes_acher}  
\end{figure*}

This Appendix explains how we estimate the exact position angle that best reveals the Keplerian dynamics of the rotating disk. We also calculate an upper limit to the outflow emission through the observational data. We developed a code that allows us to analyze the emission of a rotating disk and  is adapted to the case of disk-dominated pPNe such as \ac (see \sect\ref{secacherobs} and previous works).

 Figure C.1 shows PV diagrams of \ac along different $PA$. The $PA$ is indicated in the top right corner of each panel. We have drawn hyperbolic functions ($r^{-1/2}$) in each panel to help in the visual inspection. Additionally, we divided each panel in four quadrants ($Q_{i}$), which are enumerated in order (counter clockwise starting from the upper right quadrant).

\begin{figure}[h]
\includegraphics[width=\sz\linewidth]{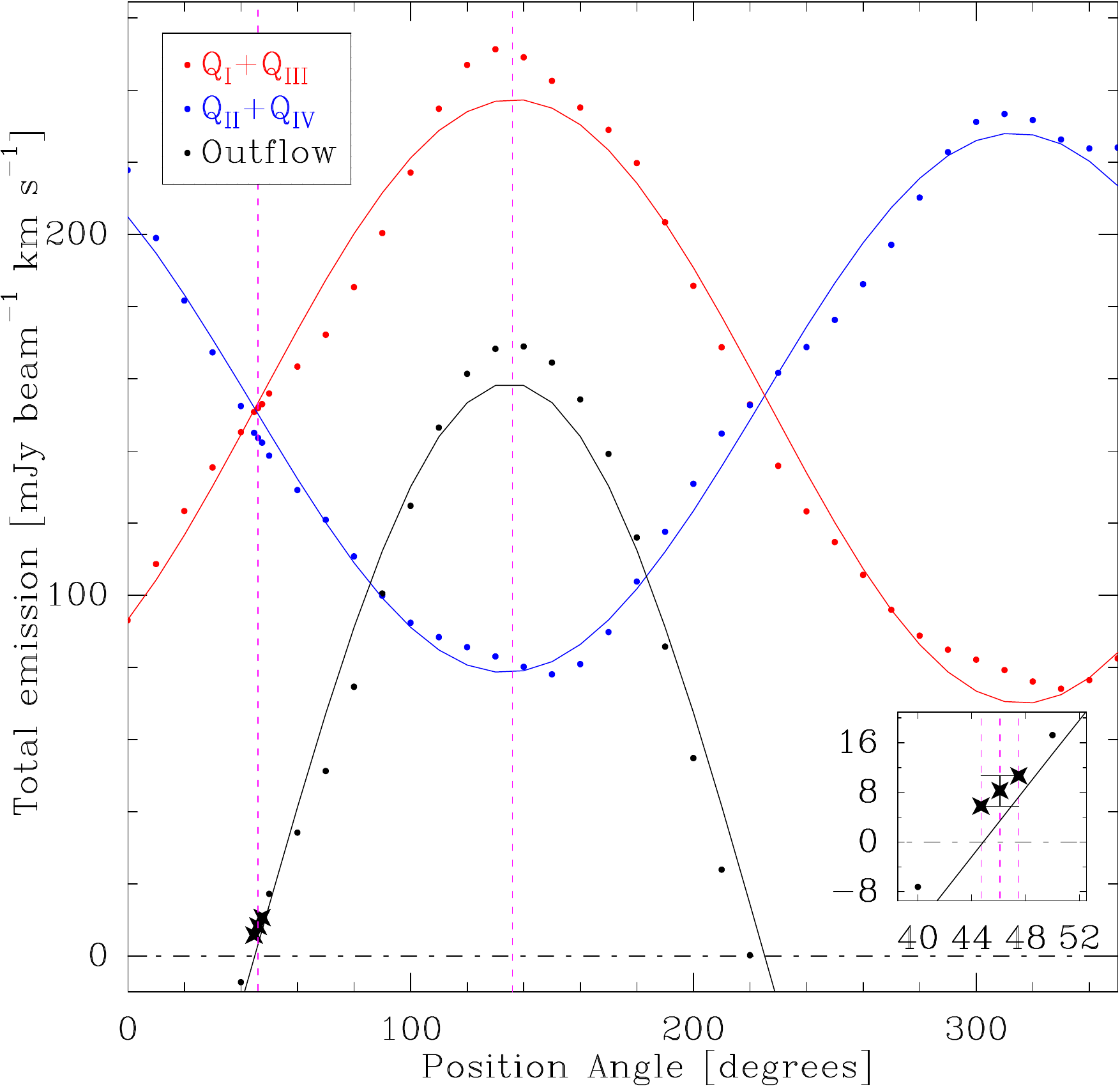}
\caption{\small Variation of the emission of the first and third quadrants (Q$_{I}$ and Q$_{III}$ in red dots) and the second and fourth quadrants (Q$_{II}$ and Q$_{IV}$ in blue dots) with the $PA$ for our observation of \ac (see \fig\ref{fig:cortes_acher}). 
The maximum emission of the mean of Q$_{I}$ and Q$_{III}$ indicates the $PA$, for which the rotational effect of the Keplerian dynamics along the nebula equator are best detected. The angle found is: $PA=136.1\degree$.
The subtraction of what we name Q$_{I}$\,+\,Q$_{III}$ and Q$_{II}$\,+\,Q$_{IV}$ shows the excess emission of the putative outflow along the nebula axis (black dots). We use the same colors as in \fig\ref{fig:cortes_acher}.
The magenta dotted lines represents $PA=46.1\degree$ and $PA=136.1\degree$. These $PA$ values represent the cut along the nebula axis and equator, respectively.
The emission of the putative outflow is represented as a star at $PA=46.1\degree \pm 1.4\degree$. The error bar is the same for each point. The emission at the $PA$ along the nebula axis is shown in the inset of the right corner.}
    \label{fig:grafica_acher}  
\end{figure}

A disk with Keplerian dynamics must show the typical and well-known ``butterfly'' shape in the PV diagram along the equatorial direction.
Therefore, in an ideal situation (pure Keplerian rotation and infinite spatial resolution), the emission should only appear in two opposite quadrants. The same applies for the case of a bipolar outflow with linear velocity gradient, but when we take the PV diagram along the symmetry axis. On the contrary, in an isotropically expanding  envelope, the emission should be the same in all four quadrants regardless of the orientation. In our case here, a source dominated by the central rotating disk, we can use these properties to infer the orientation of the major axes of the nebula, equatorial and axial, by finding the $PA$ that maximizes the emission in two opposite quadrants, $Q_{I}$ and $Q_{III}$ for instance, with respect to the other two.
The emission in each quadrant $Q_{i}$ has been spatially averaged and added in velocity.
This emission along different $PA$ is shown in \fig\ref{fig:grafica_acher} with red dots ($Q_{I} + Q_{III}$ emission). We derived this $PA$ by the numerical fitting of a sinusoidal function. The optimal value is found to be $136.1\degree \pm 1.4\degree$.

The presence of a putative outflow should be present in the PV diagram along the nebula axis. The $PA$ along the nebula axis will be $46.1\degree$ because the $PA$ along the equator disk is $136.1\degree$. The theoretical PV diagram along the nebula axis in the presence of a rotating disk must show emission with a form similar to a rhombus with equal emission in all four quadrants. 
The subtraction of $Q_{I} + Q_{III}$ emission (red dots) from $Q_{II} + Q_{IV}$ emission (blue dots) removes the disk contribution emission (``Outflow'' in black dots in \fig\ref{fig:grafica_acher}) and reveals excess emission. The emission of an outflow will be the one of ``Outflow'' at $PA=46.1\degree$. The excess emission at this $PA$ is 8.3\,mJy\,beam$^{-1}$\,km\,s$^{-1}$.
Taking into account the uncertainty of the $PA$, the different values of $PA=44.7\degree$ and $PA=47.5\degree$ are measurements of the derived outflow intensity. Thus, we find that the excess emission that we attribute to the outflow emission is 8.3$^{+2.4}_{-2.5}$\,mJy\,beam$^{-1}$\,km\,s$^{-1}$. We must highlight that the three different $PAs$ present positive emission.
In addition, the formal error for the trigonometrical fitting yields an uncertainty of 2.4\,mJy\,beam$^{-1}$\,km\,s$^{-1}$, which is highly consistent with the previous calculated uncertainty.

There is another option to estimate the emission of the putative outflow through the PV diagram along the nebula axis. We see how the emission at central velocities is inclined in the right panel of \fig\ref{fig:acher12pv}. 
We see an excess emission in quadrants $Q_{I}$ and $Q_{III}$, with central velocities of $\pm\,$0.5\kmsp and extreme offsets of $\pm$\,0\secp8.
This excess emission could be explained by the presence of an outflow. Taking into account the emission ($\sim$\,10\,mJy\,beam$^{-1}$) and the width of that emission in terms of velocity ($\sim$\,0.75\,km\,s$^{-1}$), the putative outflow presents an emission of $\sim$\,10\,mJy\,beam$^{-1}$\,km\,s$^{-1}$.
The uncertainty for this value, $\Delta E$, is:
\begin{equation}
\Delta E = rms \frac{\sqrt{n_{V}}}{\sqrt{n_{Y}}},
\label{error_acher}
\end{equation}
where $rms$ is the noise of the map (4.8\,mJy\,beam$^{-1}$), $n_{V}$ is the number of velocity channels used in that region ($\sim$\,0.75\,km\,s$^{-1}$\,/\,0.2\,km\,s$^{-1}$\,=\,3.8), and $n_{Y}$ is the number of position channels in that region (0\secp5\,/\,0\secp07\,=\,7.1). We find $\Delta E$\,=\,3.5\,mJy\,beam$^{-1}$\,km\,s$^{-1}$. Therefore, we conclude that the emission of the outflow is 10\,$\pm$\,3.5\,mJy\,beam$^{-1}$\,km\,s$^{-1}$.

We obtained very close values for the outflow emission through two very different methods, which reinforces our conclusions about the presence of an extended outflow in \acp. Therefore, we consider the outflow tentatively detected with an emission of $\lsim$\,10\,mJy\,beam$^{-1}$\,km\,s$^{-1}$.

\section{Outflow kinetic and rotational temperatures}
\label{trot_tcin}

The sources studied in this work  present relatively massive and extended outflows, which show low rotational temperatures (see \sect\ref{sec89hermodel}, \ref{secirasmodel}, and \ref{secrsctmodel}). When densities are very low, these rotational temperatures may significantly depart from the kinetic ones. 
We must remember that  ``excitation temperature'' is always defined as an equivalent temperature for a given line, while ``rotational temperature'' is defined as the typical value or average of the excitation temperatures of the relevant rotational lines (in our case, low-$J$ transitions including the observed one), and is used to approximately calculate the partition function and the absorption and emission coefficients. In the limit of thermalization (LTE), which is attained for sufficiently high densities, all excitation temperatures are the same, and equal to the rotational and kinetic temperatures.

\begin{figure}[h]
    \centering
                \includegraphics[width=\sz\linewidth]{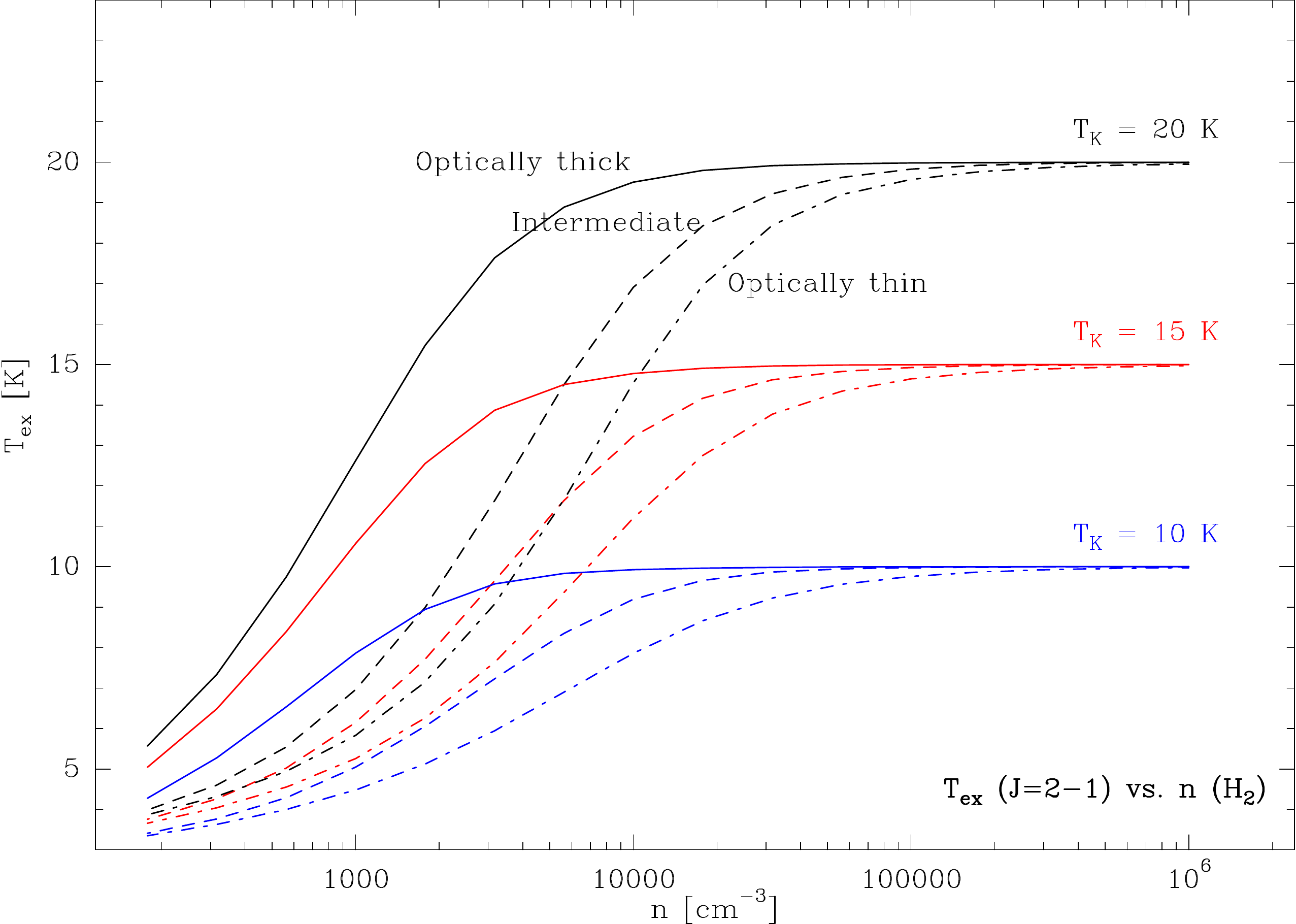}
                \quad
                \includegraphics[width=\sz\linewidth]{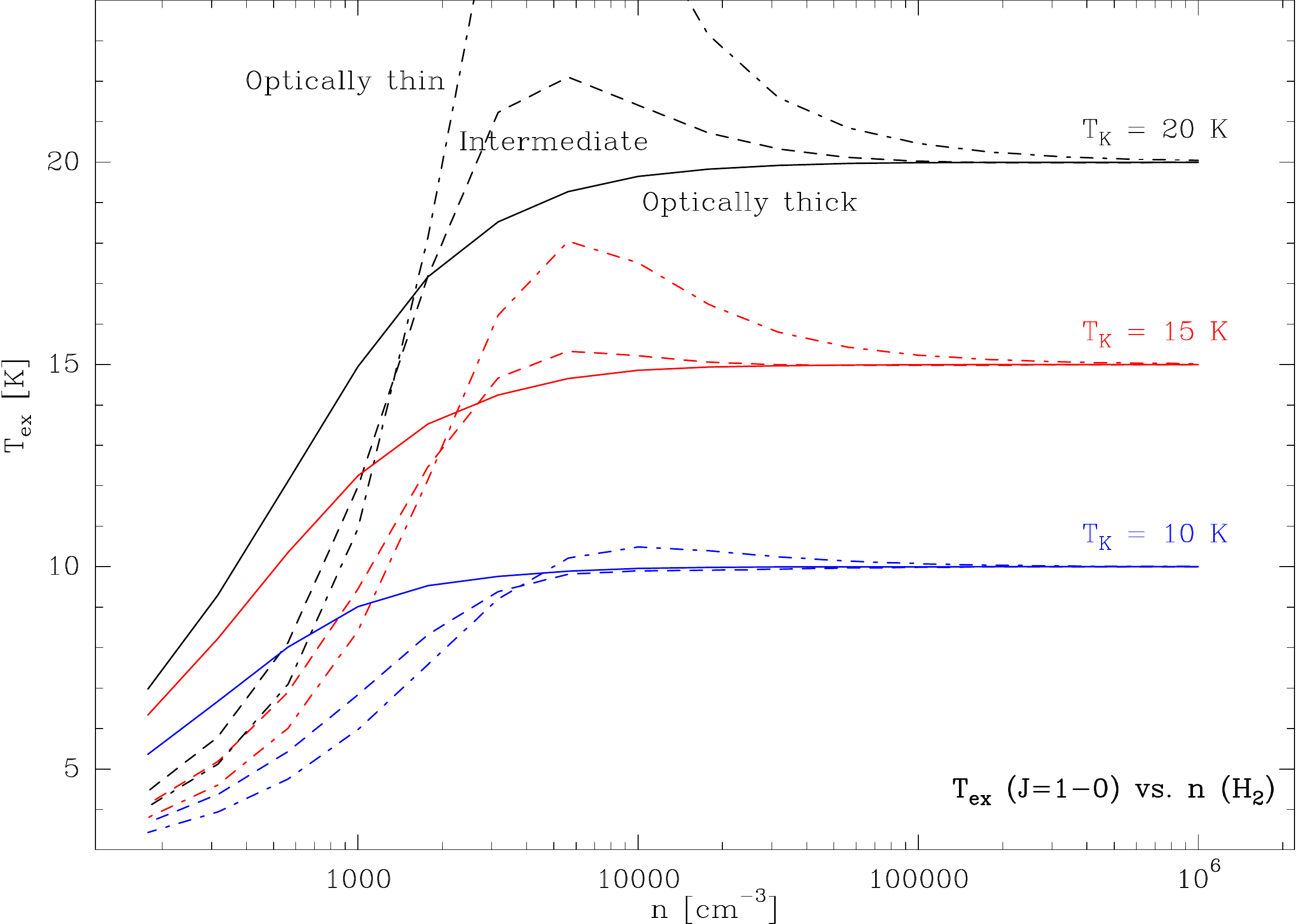}
    \caption{\small Estimates of the equivalent excitation temperature as a function of the kinetic temperature and the density for \dosuno (\textit{Top}) and \unocero (\textit{Bottom}) transitions.}
        \label{fig:temp_21}
\end{figure}

In \fig\ref{fig:temp_21} we present theoretical estimates of the equivalence between kinetic temperatures ($T_{K}$) and excitation temperatures ($T_{ex}$) for relevant lines as a function of density. (In the case of very low excitation temperatures, <\,10\,K, only the three lowest levels of CO are significantly populated.)
To perform these calculations, we used a standard LVG code very similar to that described for example by \citet{bujarrabalalcolea2013}, with the simplest treatment of the velocity field, taking logarithmic velocity gradient to be equal to 1. Three cases are considered, very low optical depths, optically thick case, and calculations with intermediate optical depths. In the optically thick and intermediate opacity cases, we varied the characteristic lengths of the considered regions at the same time as the density, in order to keep opacities of $\sim$\,10 and 1, respectively.

As we can see, for larger density values we find that $T_{ex}=T_{K}$. However, for lower density values we find  $T_{ex} \textless T_{K}$, as expected. In some cases, the difference is significant; for the very diffuse extended layers around \rsp, a rotational temperature of $\sim$\,7\,K  corresponds to kinetic temperatures of over $\sim$\,15\,K.

\section{Additional figures of model calculation} \label{anexomaps}
\label{figuras_adicionales}

In this Appendix, we show the maps and PV diagrams derived from models for the four analyzed sources.
We present the synthetic velocity maps of \ac for \doce\dosuno (see \fig\ref{fig:acmapasmodelodisco}) for our best-fit model (\sect\ref{secachermodel}). Additionally, we show synthetic velocity maps and PV diagrams along the equator and nebula axis for a disk-only model (the standard model of \tab\ref{achermodel} and \fig\ref{fig:acherdens} without the outflow) in \figs\ref{fig:acmapasmodelodisco} and \ref{fig:acher12pvmodelodisco}. We also show an alternative model of \ac where the size of the outflow is somewhat larger than the standard one (see \figs\ref{fig:acherdens12}, \ref{fig:acmapasmodelodiscooutflow12}, and \ref{fig:ac12pvmodelodiscooutflow12}), both cases being compatible with observations.

We present synthetic velocity maps of \on for \doce and \trece \dosuno emission for our best-fit model (see \sect\ref{sec89hermodel} and \fig\ref{fig:89hermapasmodelo}). The hourglass-shaped structure is present in the synthetic velocity maps for \doce and \trece \dosunop.
We show velocity maps of \iras for \doce\dosuno (see \fig\ref{fig:irasmapasmodelo}) of our best-fit model (see \sect\ref{secirasmodel}). Additionally, we present synthetic velocity maps for the alternative model of \iras and PV diagrams along the equator disk and nebula axis (see \fig\ref{fig:irasmapasmodelost}, \ref{fig:iras12pvmodelost}, and \ref{fig:irasdensst}).

We present synthetic velocity maps of \rs for \doce\dosuno for our best-fit model (see \fig\ref{fig:rs12mapasmodelo} and \sect\ref{secrsctmodel}). These maps are similar to those of \fig\ref{fig:rs12mapas}. We note that cavities of the extended outflow are also present.


\begin{figure*}[h]
\centering
\includegraphics[width=\sz\textwidth]{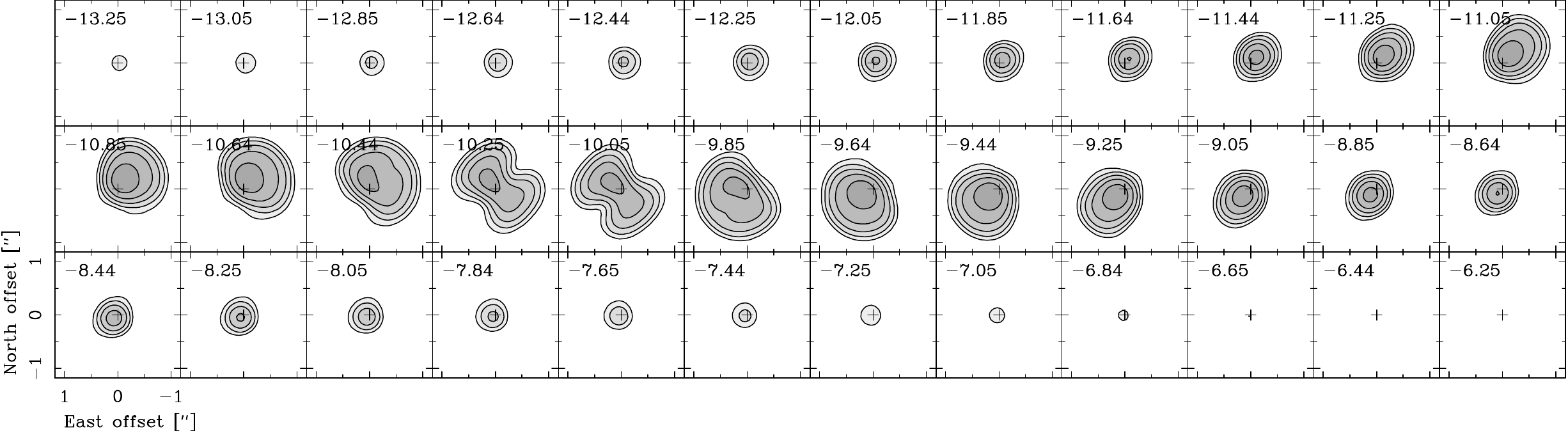}

\caption{\small Synthetic maps predicted by the alternative model of the \doce\dosuno line emission for the nebula around AC\,Her. To be compared with \fig\ref{fig:ac12mapas}, the scales and contours are the same.}
    \label{fig:acmapasmodelodiscooutflow}  
\end{figure*}

\begin{figure*}[h]
\centering
\includegraphics[width=\sz\textwidth]{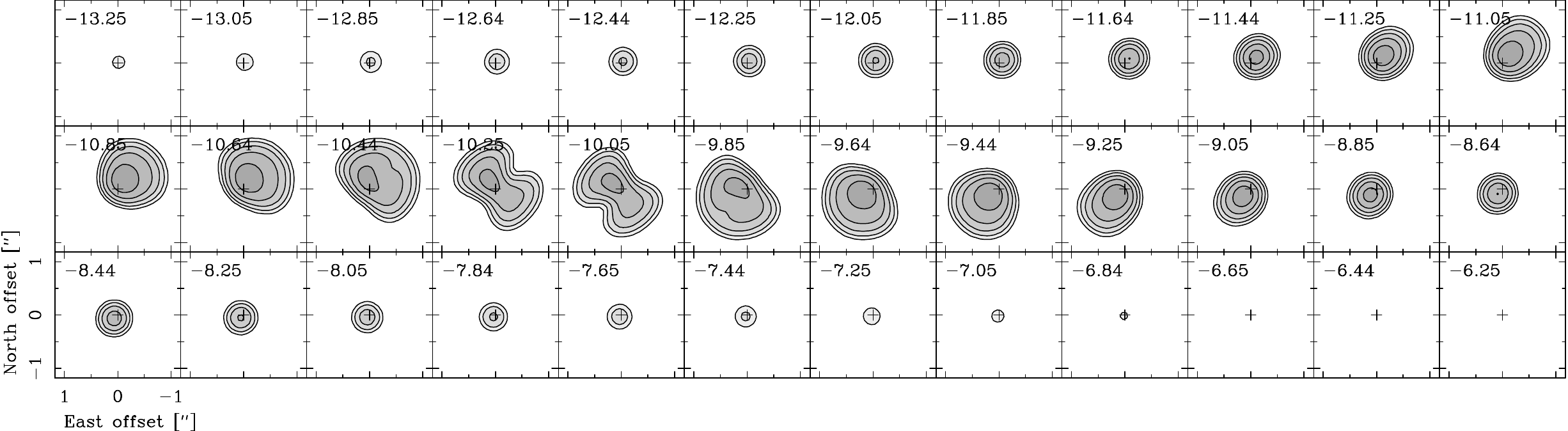}

\caption{\small Synthetic maps predicted by the disk-only model of the \doce\dosuno line emission for the nebula around AC\,Her. To be compared with \fig\ref{fig:ac12mapas}, the scales and contours are the same.}
    \label{fig:acmapasmodelodisco}  
\end{figure*}

\begin{figure*}[h]
        \centering
        \begin{minipage}[b]{0.48\linewidth}
                \includegraphics[width=\sz\linewidth]{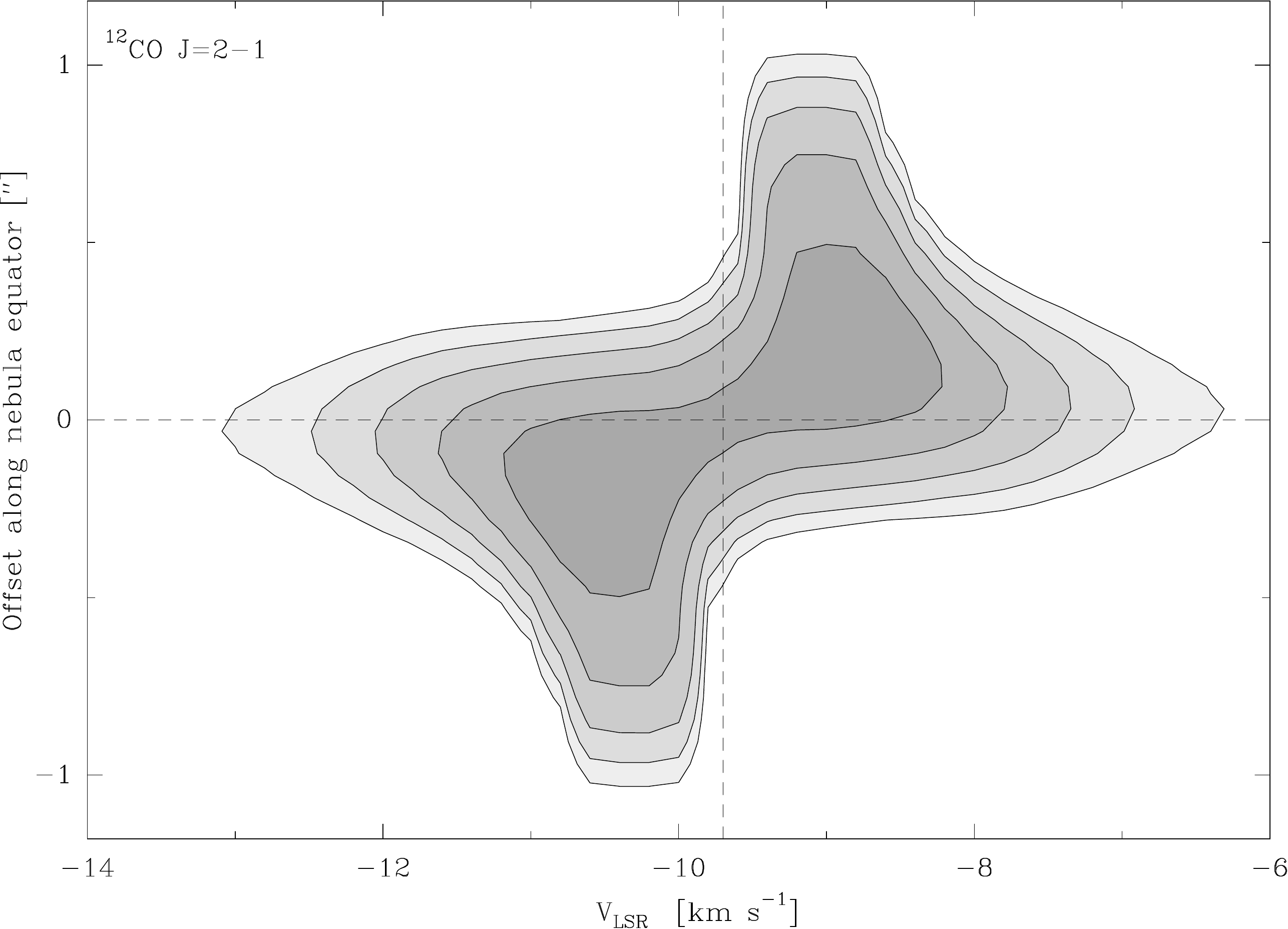}
        \end{minipage}
        \quad
        \begin{minipage}[b]{0.48\linewidth}
                \includegraphics[width=\sz\linewidth]{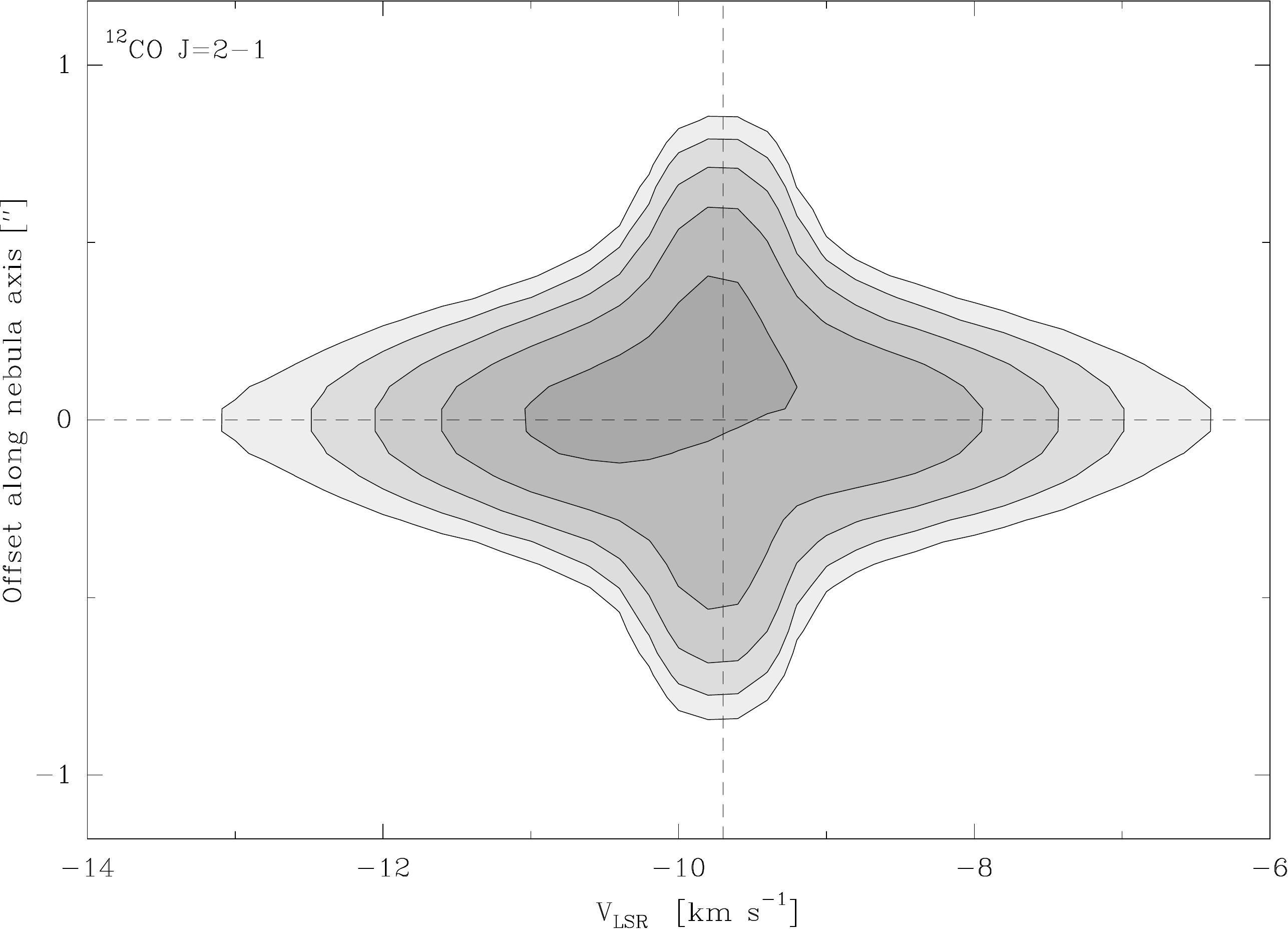}
        \end{minipage}
        \caption{\small \textit{Left:} Synthetic position-velocity diagram from our disk-only best-fit model of \doce \dosuno in AC\,Her along the direction $PA=136.1\degree$. To be compared with the left panel of \fig\ref{fig:acher12pv}, the scales and contours are the same. \textit{Right}: Same as in \textit{Left} but along $PA=46.1\degree$.}
        \label{fig:acher12pvmodelodisco}
\end{figure*}

\begin{figure*}[h]
\centering
\includegraphics[width=\sz\textwidth]{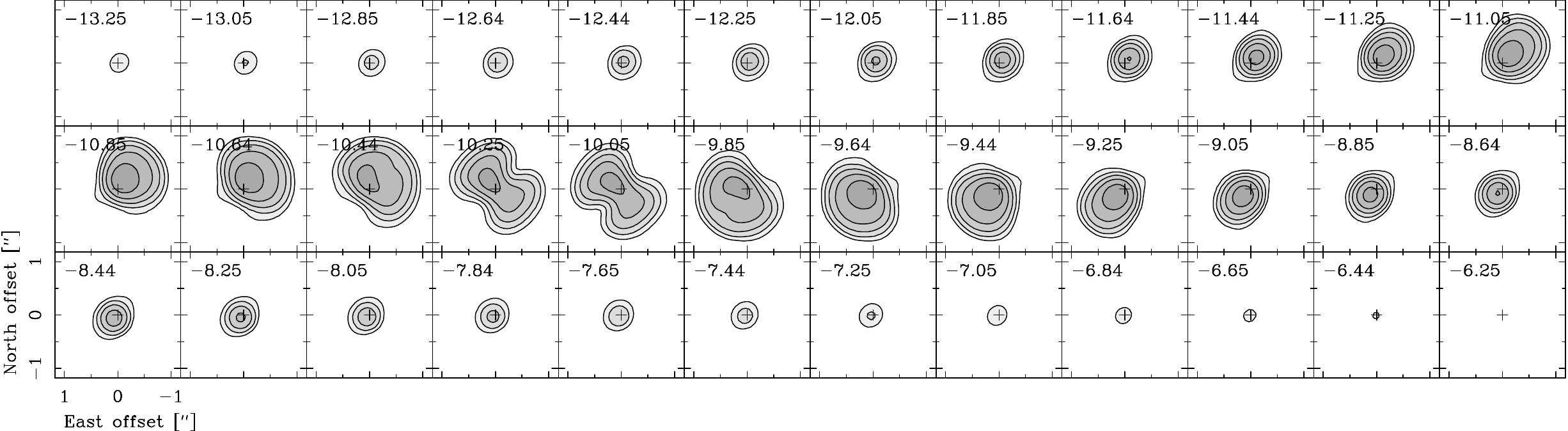}

\caption{\small Synthetic maps predicted by the alternative model of the \doce\dosuno line emission for the nebula around AC\,Her. To be compared with \fig\ref{fig:ac12mapas}, the scales and contours are the same.}
    \label{fig:acmapasmodelodiscooutflow12}  
\end{figure*}

\begin{figure*}[h]
        \centering
        \begin{minipage}[b]{0.48\linewidth}
                \includegraphics[width=\sz\linewidth]{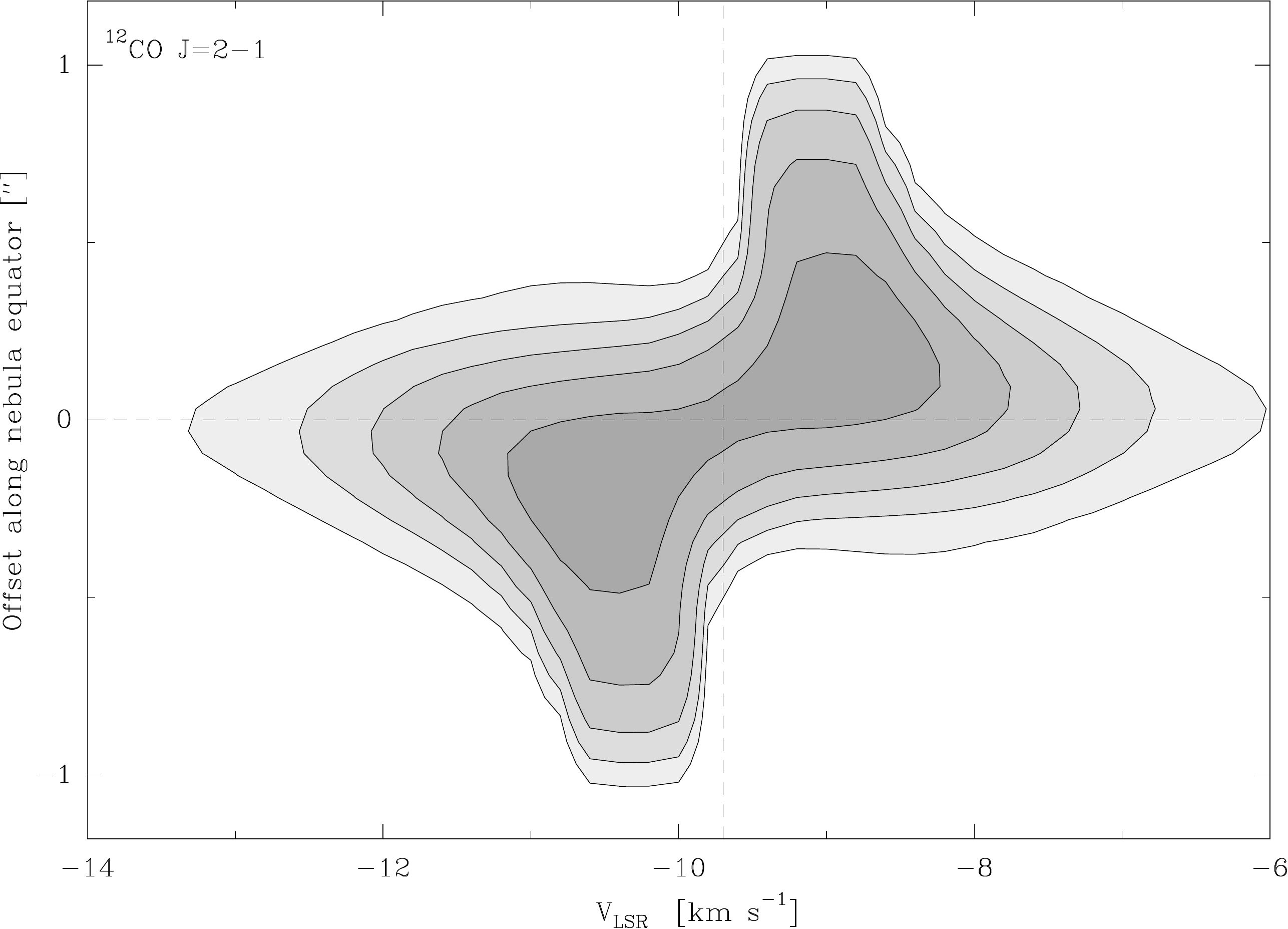}
        \end{minipage}
        \quad
        \begin{minipage}[b]{0.48\linewidth}
                \includegraphics[width=\sz\linewidth]{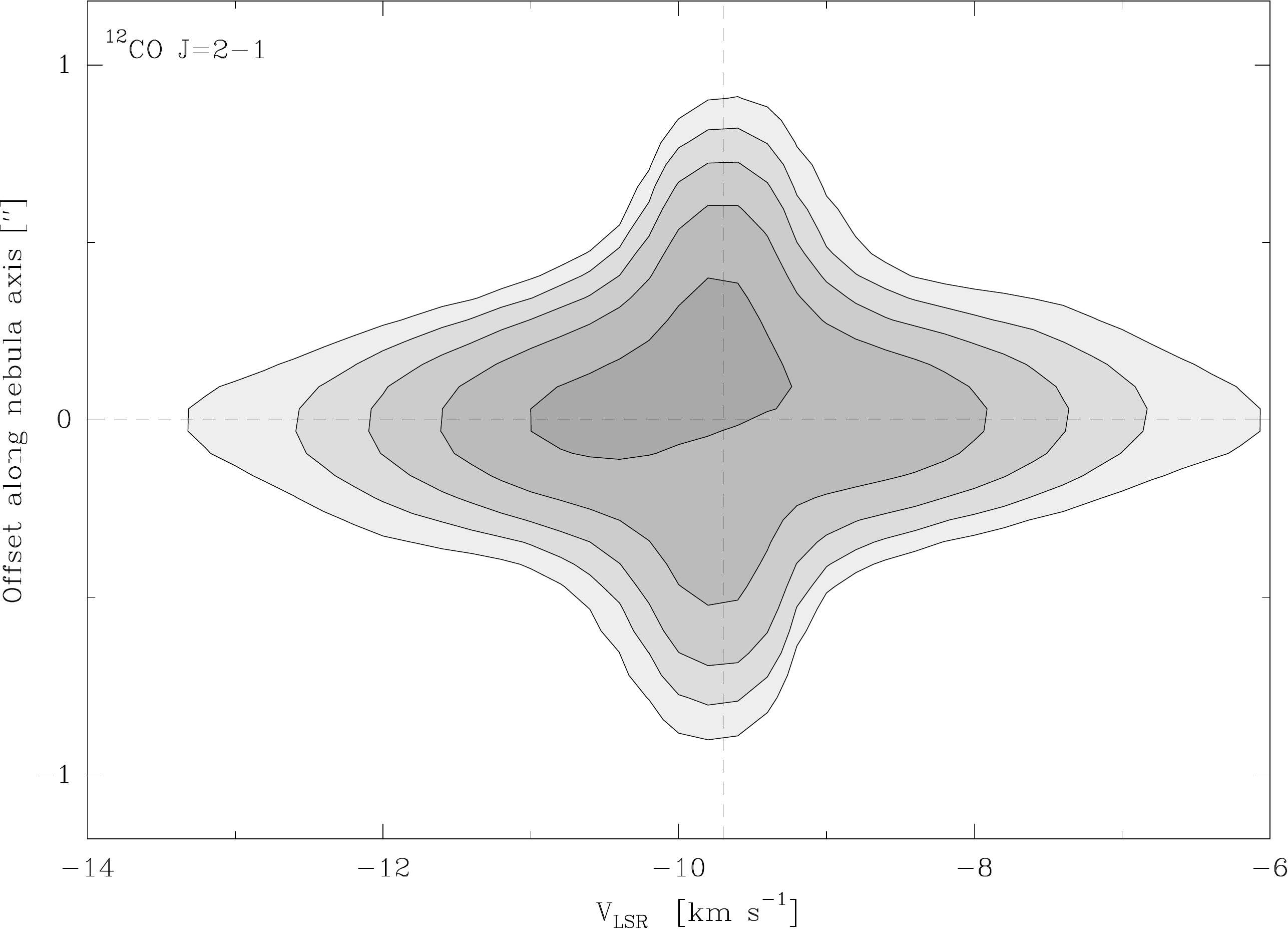}
        \end{minipage}
        \caption{\small \textit{Left:} Synthetic position-velocity diagram from our best-fit alternative model of \doce \dosuno in \ac with an outflow. To be compared with \fig\ref{fig:acher12pv}, the scales and contours are the same. \textit{Right}: Same as in \textit{Left} but along $PA=46.1\degree$.}
        \label{fig:ac12pvmodelodiscooutflow12}
\end{figure*}

\begin{figure*}[h]
\centering
\includegraphics[width=\textwidth]{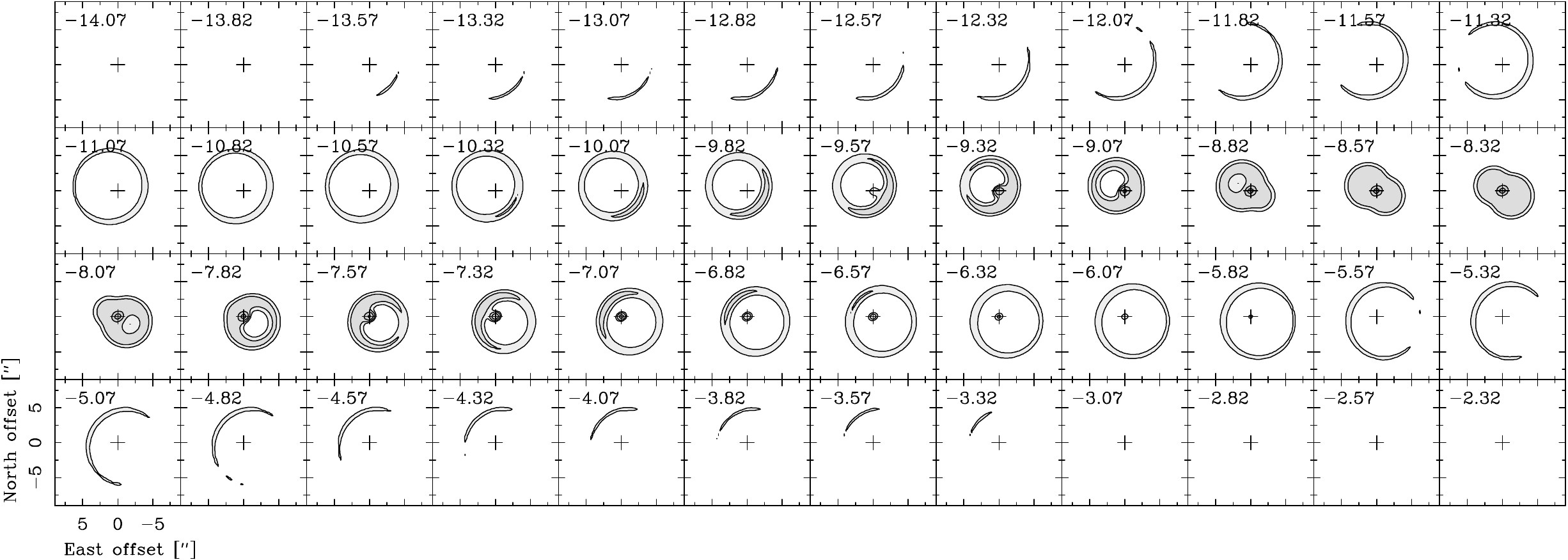}
 \includegraphics[width=\textwidth]{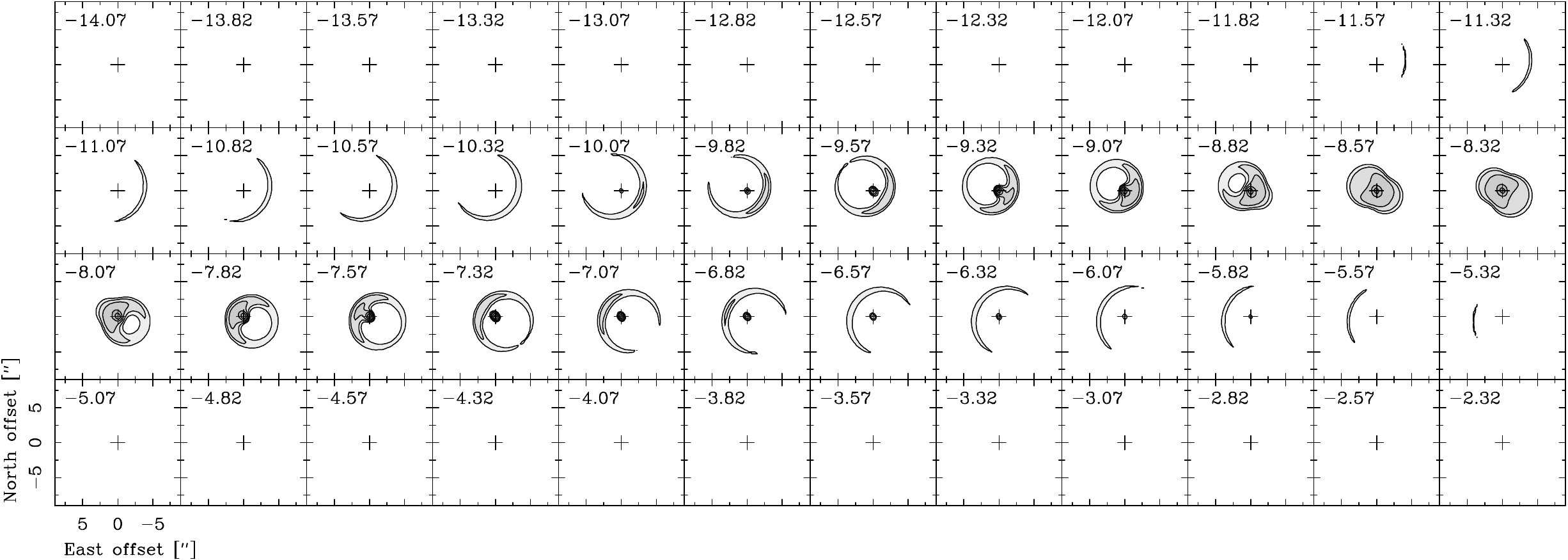}
\caption{\small Synthetic maps predicted by the model of the \doce\dosuno line emission (\textit{Top}) and \trece\dosuno (\textit{Bottom}) for the nebula around \onp. To be compared with \fig\ref{fig:89hermapas}, the scales and contours are the same.}
    \label{fig:89hermapasmodelo}  
\end{figure*}

\begin{figure*}[h]
\centering
\includegraphics[width=\sz\textwidth]{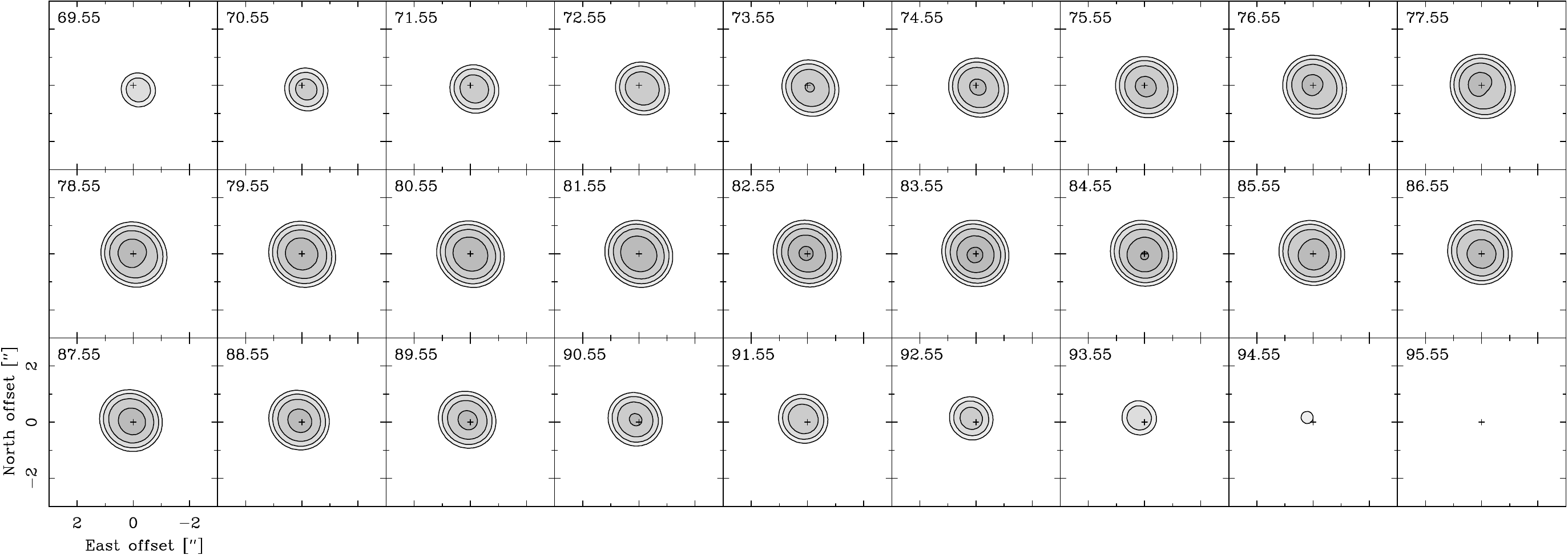}

\caption{\small Synthetic maps predicted by the model of the \doce\dosuno line emission for the nebula around \irasp. To be compared with \fig\ref{fig:iras12mapas}, the scales and contours are the same.}
    \label{fig:irasmapasmodelo}  
\end{figure*}

\begin{figure*}[h]
\centering
\includegraphics[width=\sz\textwidth]{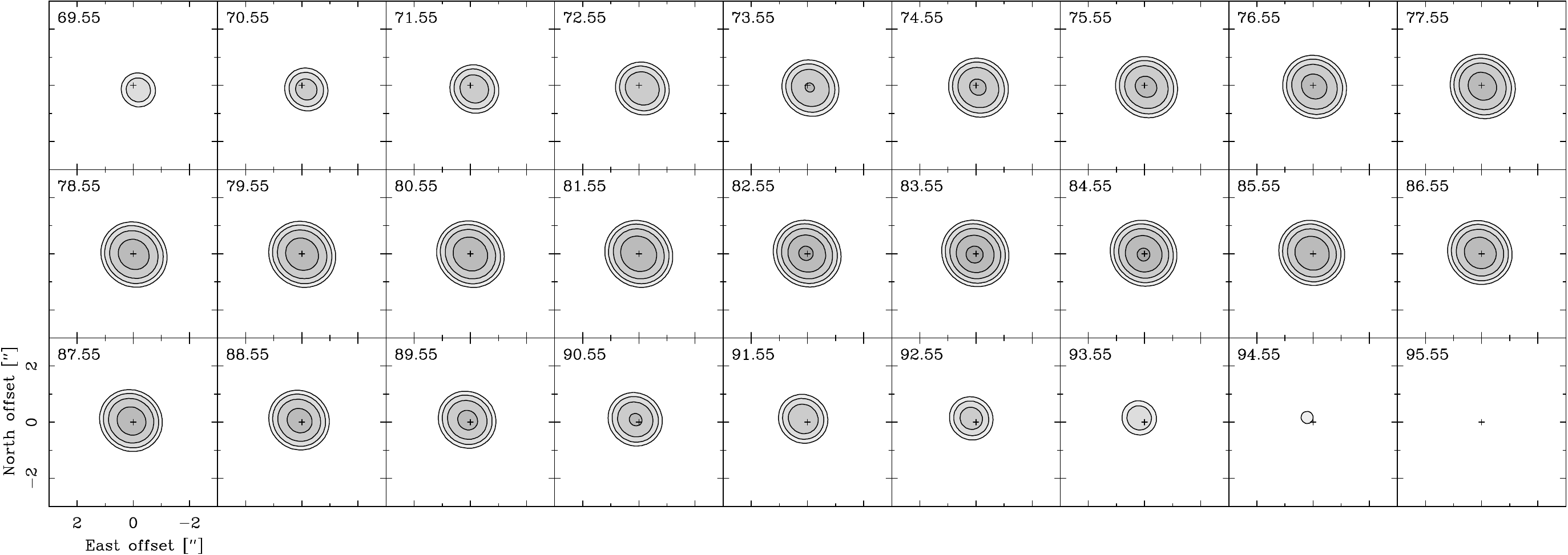}
\caption{\small Same as in \fig\ref{fig:irasmapasmodelo} but for the alternative model. To be compared with \fig\ref{fig:iras12mapas}, the scales and contours are the same.}
\label{fig:irasmapasmodelost}
\end{figure*}

\begin{figure*}[h]
        \centering
        \begin{minipage}[b]{0.48\linewidth}
                \includegraphics[width=\sz\linewidth]{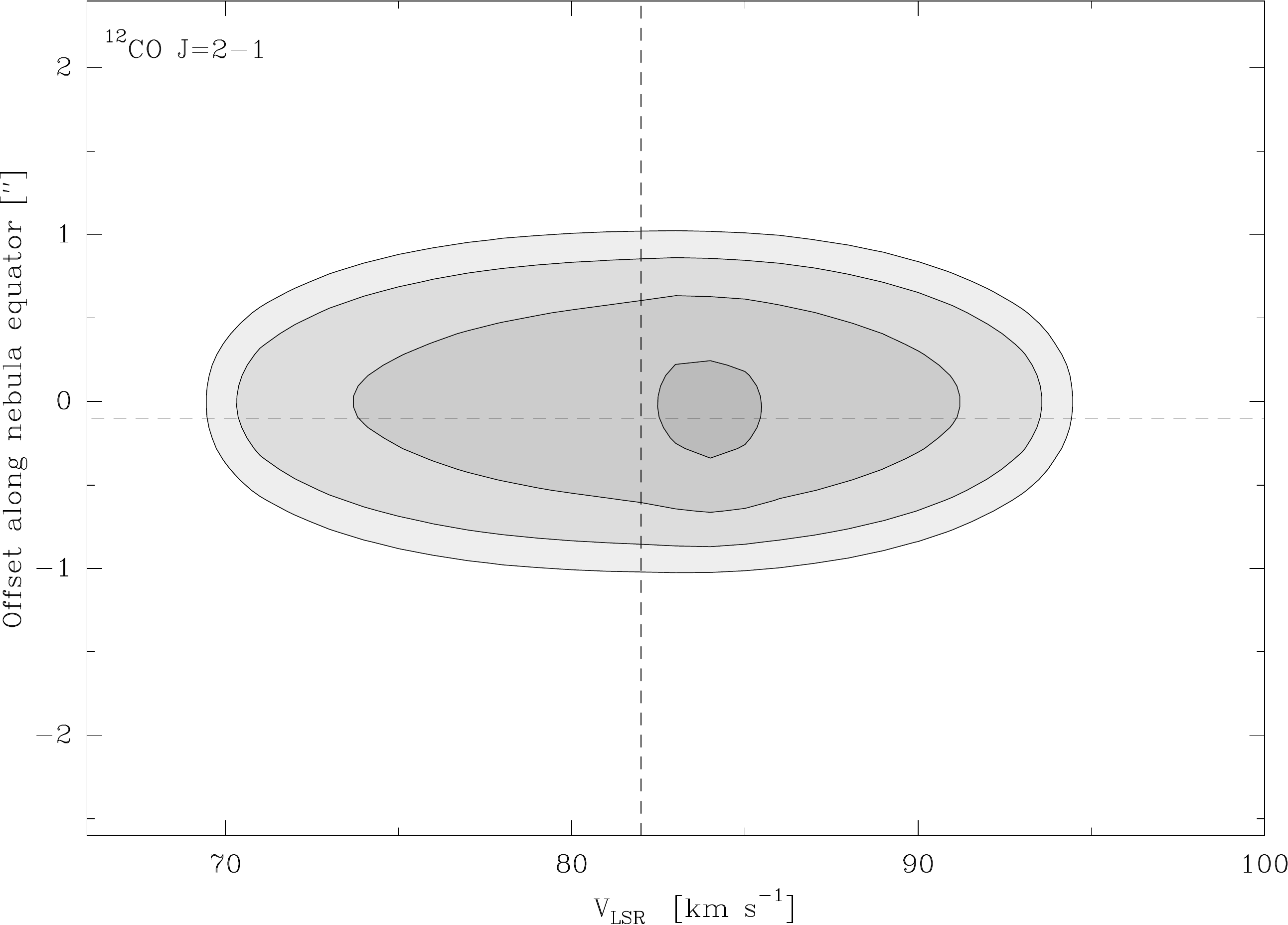}
        \end{minipage}
        \quad
        \begin{minipage}[b]{0.48\linewidth}
                \includegraphics[width=\sz\linewidth]{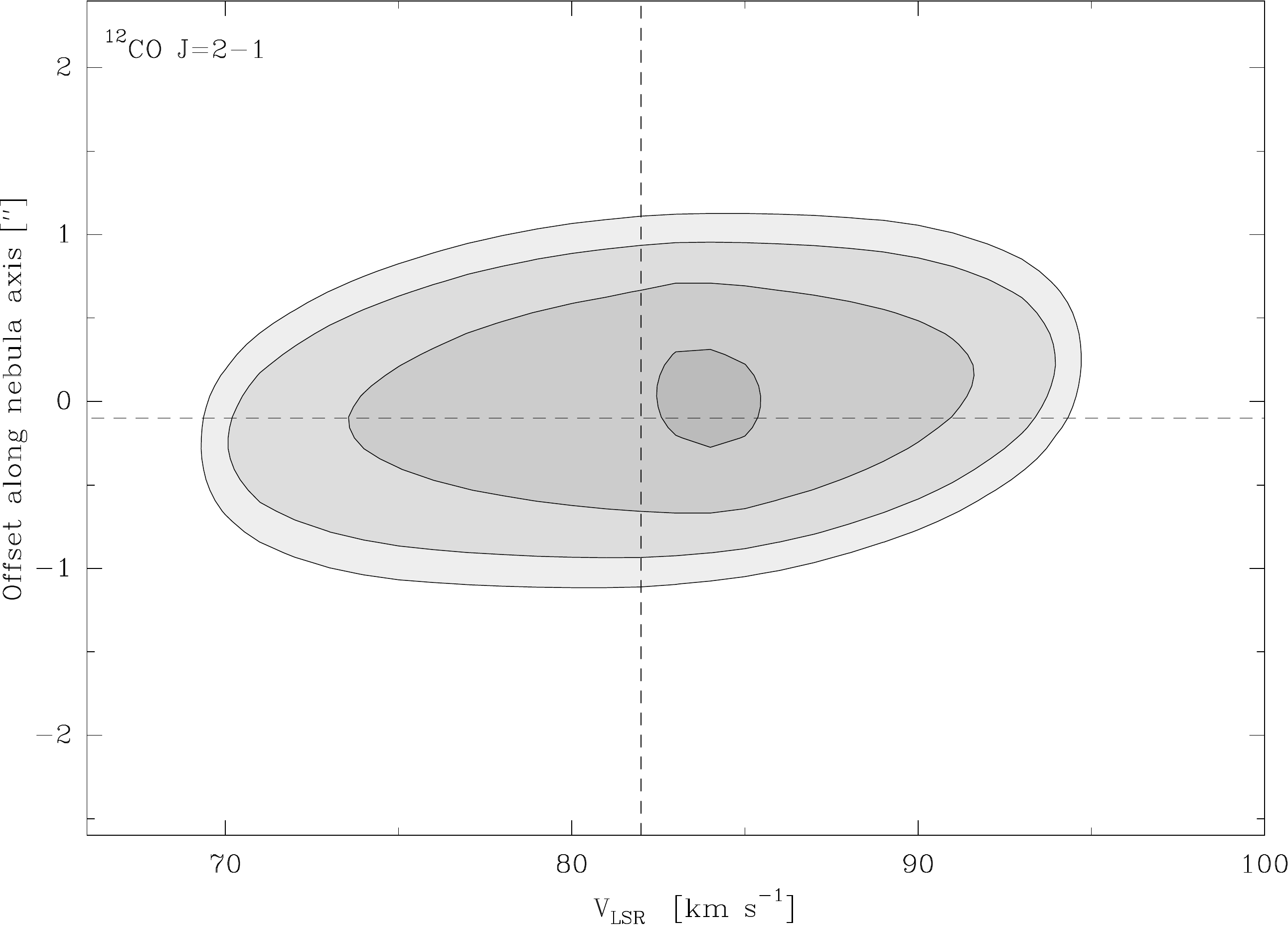}
        \end{minipage}
        \caption{\small \textit{Left:} Synthetic position-velocity diagram from our alternative best-fit model of \doce \dosuno in \iras along the direction $PA=-40\degree$. To be compared with the left panel of \fig\ref{fig:iras12pv}, the scales and contours are the same. \textit{Right}: Same as in \textit{Left} but along $PA=50\degree$.}
        \label{fig:iras12pvmodelost}
\end{figure*}

\begin{figure*}[h]
\includegraphics[width=\sz\textwidth]{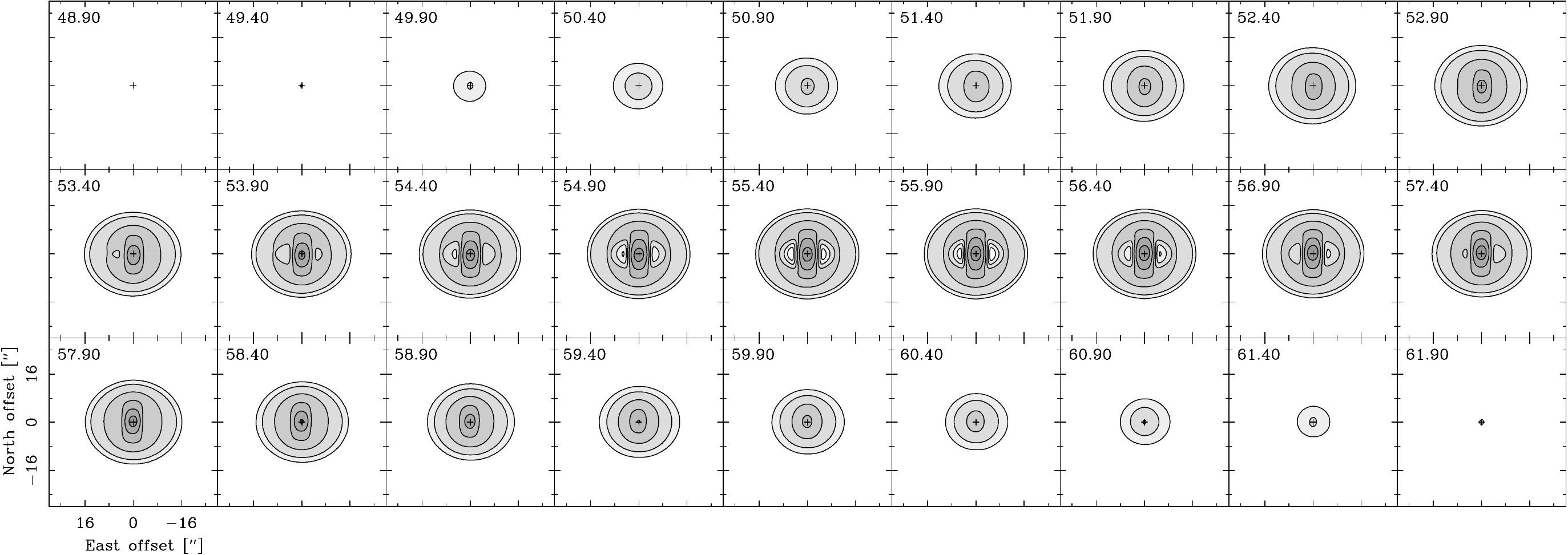}
 
\caption{\small Synthetic maps predicted by the model of the \doce\dosuno line emission for the nebula around \rsp. To be compared with \fig\ref{fig:rs12mapas}, the scales and contours and units are the same.}
    \label{fig:rs12mapasmodelo}  
\end{figure*}

\begin{figure}[h]
    \centering
                \includegraphics[width=\sz\linewidth]{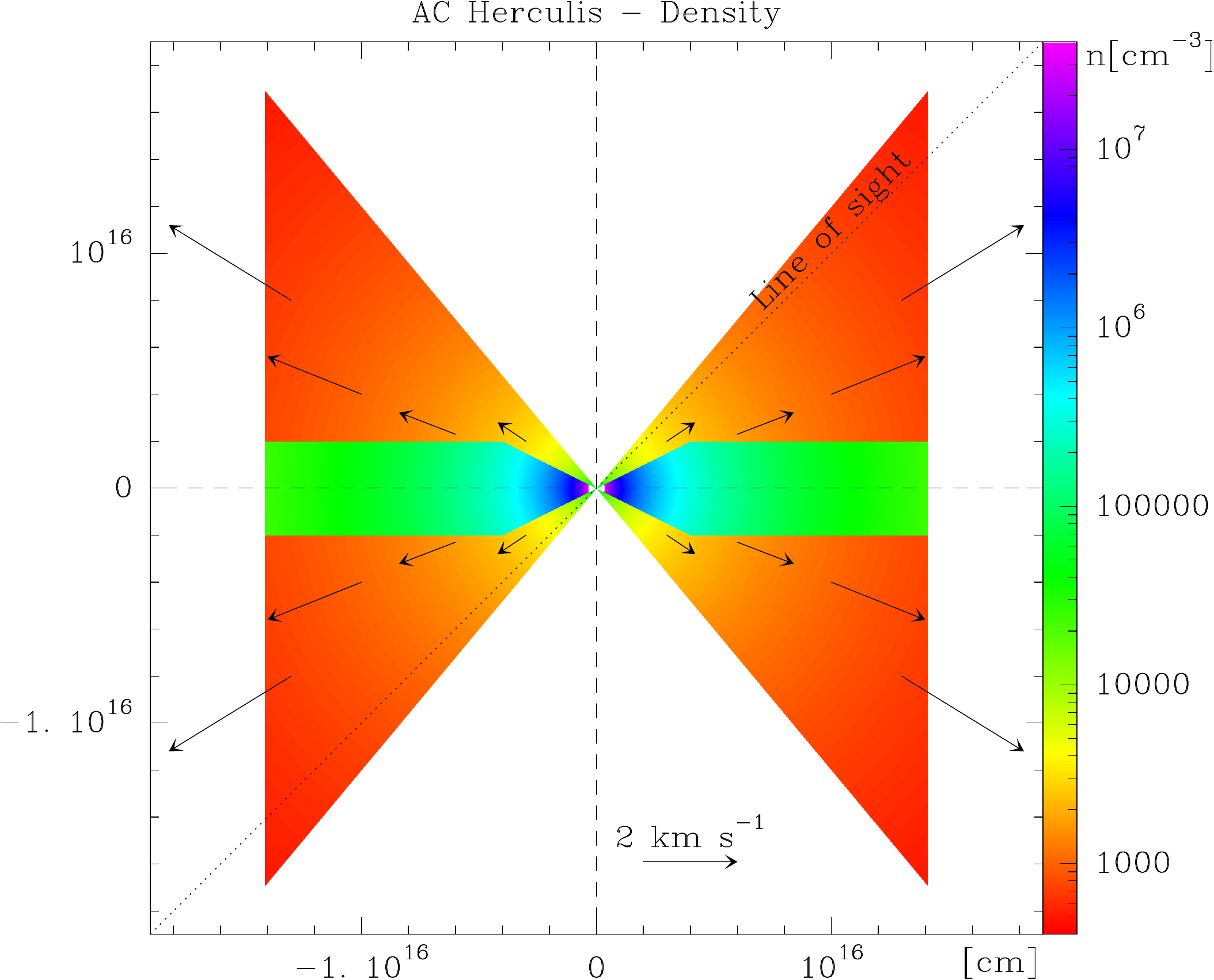}
    \caption{\small Structure and distribution of the density of our alternative best-fit model for the disk and outflow of \acp. The Keplerian disk presents density values $\geq$\xd{5}\,cm$^{-3}$. The expansion velocity is represented with arrows.}
        \label{fig:acherdens12}
\end{figure}

\begin{figure}[h]
    \centering
                \includegraphics[width=\sz\linewidth]{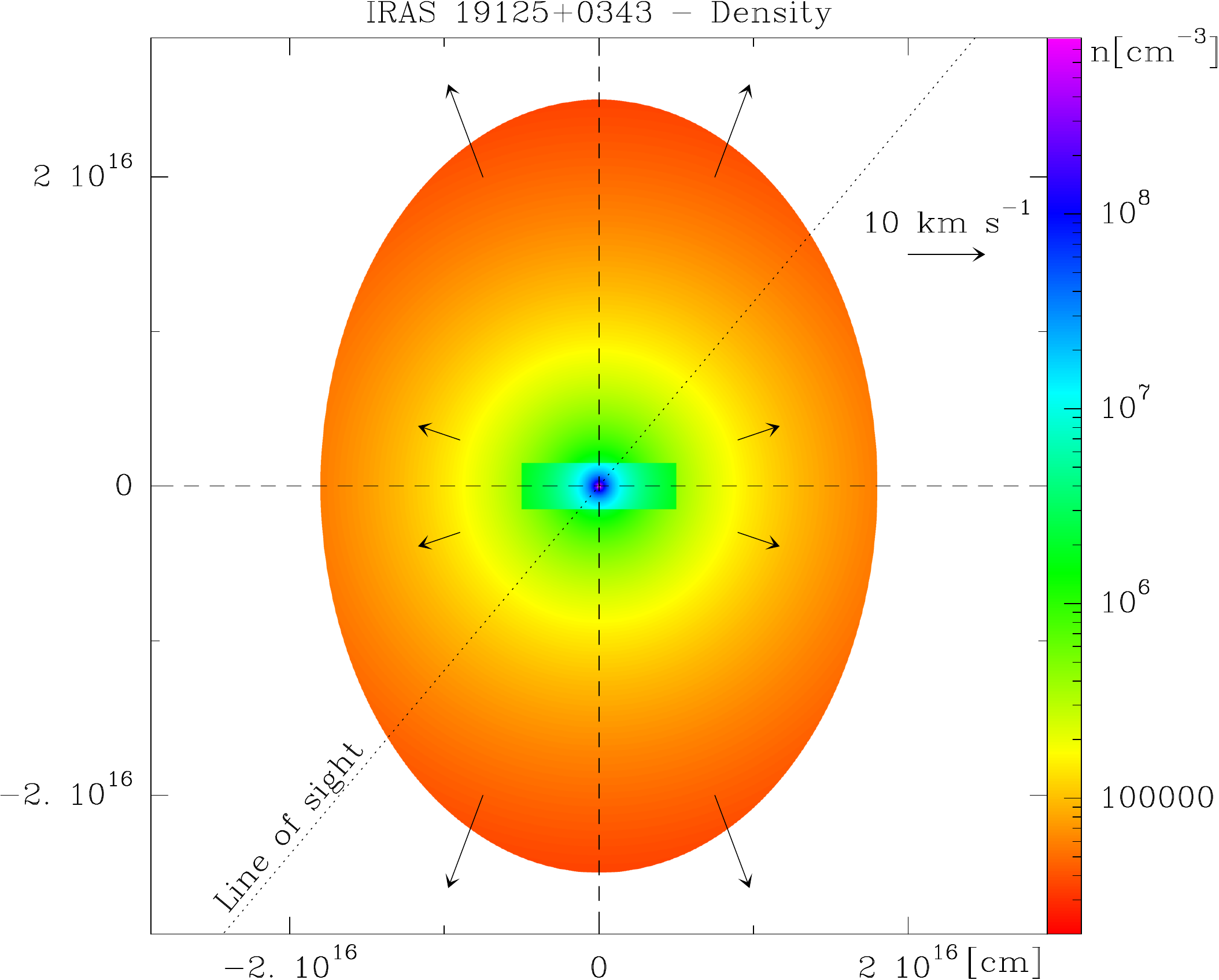}
    \caption{\small Structure and distribution of the density of our best-fit alternative model for the disk and outflow of \irasp. The Keplerian disk presents density values $\geq$\xd{6}\,cm$^{-3}$. The expansion velocity is represented with arrows.}
    \label{fig:irasdensst}
\end{figure}

\end{document}